\documentclass[10pt,letterpaper,final]{iopart}
\usepackage{bbm,times,gerrymath,amssymb,amsbsy,subfigure}
\usepackage[dvips]{graphicx}

\newcommand{\arcsinh}{{\rm arcsinh}}
\newcommand{\op}[1]{\hat{#1}}
\newcommand{\adj}[1]{{#1}^{\dag}}
\newcommand{\comm}[2]{\left[#1,#2\right]}
\newcommand{\hh}{\mathcal{H}}
\newcommand{\ro}{\varrho}
\newcommand{\diag}{\,{\rm diag}\,}
\newcommand{\R}{\mathbbm{R}}

\newcommand{\rr}{\mathbbm{R}}

\newcommand{\id}{\mathbbm{1}}

\newcommand{\sy}[1]{Sp_{(#1,\R)}}

\renewcommand{\tr}{{\rm Tr}\,}
\renewcommand{\det}{{\rm Det}\,}
\newcommand{\gr}[1]{\boldsymbol{#1}}
\newcommand{\be}{\begin{equation}}
\newcommand{\ee}{\end{equation}}
\newcommand{\bea}{\begin{eqnarray}}
\newcommand{\eea}{\end{eqnarray}}
\newcommand{\ket}[1]{|#1\rangle}
\newcommand{\bra}[1]{\langle#1|}
\newcommand{\ketbra}[2]{\vert #1 \rangle \! \langle #2 \vert}

\newcommand{\N}{{\cal N}}

\newcommand{\sig}{\gr{\sigma}}
\newcommand{\eps}{\gr{\varepsilon}}
\newcommand{\bet}{\gr{\beta}}
\newcommand{\alp}{\gr{\alpha}}
\newcommand{\abs}[1]{\left\vert#1\right\vert}
\newcommand{\eq}[1]{Eq.~(\ref{#1})}
\newcommand{\ineq}[1]{Ineq.~(\ref{#1})}

\newcommand{\pref}[1]{(\ref{#1})}
\newcommand{\eg}{\emph{e.g.}~}
\newcommand{\ie}{\emph{i.e.}~}

\newcommand{\T}{^{\sf T}}
\newcommand{\PT}[1]{^{{\sf T\!}_#1}}
\newcommand{\ps}{\ket{\psi}}
\newcommand{\s}{\mathcal{S}}
\newcommand{\set}[1]{\left\{#1\right\}}
\newcommand{\ph}{\varphi}
\newcommand{\W}{\op{\mathcal{W}}}
\newcommand{\dd}{\mathcal{D}}
\newcommand{\norm}[1]{\left\Vert#1\right\Vert}
\newcommand{\ptr}[2]{{\rm Tr}_{#1}\, #2}

\begin{document}
\topical{Entanglement in continuous variable systems: \\
         Recent advances and current perspectives}
\author{Gerardo Adesso$^{1,2}$ and Fabrizio
Illuminati$^{2,3}$\footnote{Home page:
{\texttt{www.sa.infn.it/quantumtheory}}}}

\address{$^1$Dipartimento di Fisica, Universit\`a di Roma ``La
Sapienza'', Piazzale Aldo Moro 5, 00185 Roma, Italy. \\
$^2$Dipartimento di Matematica e Informatica, Universit\`a degli
Studi di Salerno; CNR-INFM {\it``Coherentia''}; and INFN, Sezione di
Napoli-Gruppo Collegato di Salerno,
Via Ponte don Melillo, I-84084 Fisciano (SA), Italy.\\
$^3$ISI Foundation for Scientific Interchange Viale S. Severo 65,
I-10133 Torino, Italy}

\begin{abstract}
We review the theory of continuous-variable entanglement with special
emphasis on foundational aspects, conceptual structures, and mathematical
methods. Much attention is devoted to the discussion of separability
criteria and entanglement properties of Gaussian states, for their
great practical relevance in applications to quantum optics and quantum
information, as well as for the very clean framework that they allow for
the study of the structure of nonlocal correlations. We give a self-contained
introduction to phase-space and symplectic methods in the study of Gaussian
states of infinite-dimensional bosonic systems. We review the most important
results on the separability and distillability of Gaussian states and discuss
the main properties of bipartite entanglement. These include the extremal
entanglement, minimal and maximal, of two-mode mixed Gaussian states, the
ordering of two-mode Gaussian states according to different measures of
entanglement, the unitary (reversible) localization, and the scaling of bipartite
entanglement in multimode Gaussian states. We then discuss recent advances in
the understanding
of entanglement sharing in multimode Gaussian states, including the proof
of the monogamy inequality of distributed entanglement for all Gaussian states.
Multipartite entanglement of Gaussian states is reviewed by discussing its
qualification by different classes of separability, and the main consequences
of the monogamy inequality, such as the quantification of genuine tripartite
entanglement
in three-mode Gaussian states, the promiscuous nature of entanglement
sharing in symmetric Gaussian states, and the possible coexistence of unlimited
bipartite and multipartite entanglement. We finally review recent advances and
discuss possible perspectives on the qualification and quantification of
entanglement in non Gaussian states, a field of
research that is to a large extent yet to be explored.
\end{abstract}


\newpage

\setcounter{tocdepth}{2}

\tableofcontents

\title[Entanglement in continuous variable systems]{}

\maketitle

\newpage

\section{Prologue.}

About eighty years after their inception, quantum mechanics and
quantum theory are still an endless source of new and precious
knowledge on the physical world and at the same time keep evolving
in their mathematical structures, conceptual foundations, and
intellectual and cultural implications. This is one of the reasons
why quantum physics is still so specially fascinating to all those
that approach it for the first time and never ceases to be so for
those that are professionally involved with it. In particular, since
the early nineties of the last century and in the last ten-fifteen
years, a quiet revolution has taken place in the quantum arena. This
revolution has progressively indicated and clarified that aspects
once thought to be problematic, such as quantum nonseparability and
``spooky'' actions at a distance, are actually not only problems but
rather some of the key ingredients that are allowing a deeper
understanding of quantum mechanics, its applications to new and
exciting fields of research (such as quantum information and quantum
computation), and tremendous progress in the development of its
mathematical and conceptual foundations. Among the key elements of
the current re-foundation of quantum mechanics, entanglement
certainly plays a very important role because it is a concept that
can be mathematically qualified and quantified in a way that allows
it to provide new and general characterizations of quantum
properties, operations, and states.

In the context of this special issue on {\em Quantum Information},
we will review the main aspects of entanglement
in continuous variable systems, as they are currently understood.
With the aim to be as self-contained
as possible, we start with a tutorial summary of the structural
properties of continuous variable systems --systems associated to
infinite-dimensional Hilbert spaces whose possible applications in quantum
information and communication tasks are gaining increasing interest and
attention-- focusing on the specially relevant family of Gaussian states.
The reason why this review is mainly focused on the discussion of separability
criteria and entanglement properties of Gaussian states is due to their
great practical relevance in applications to quantum optics and quantum
information, to the very clean framework that they allow for
the study of the structure of nonlocal correlations, and to the obvious
consequence that in the last years most studies and results on continuous-variable
entanglement have been obtained for Gaussian states.

In Sections 2 and 3 we give a self-contained
introduction to phase-space and symplectic methods in the study of Gaussian
states of infinite-dimensional bosonic systems, we discuss the covariance
matrix formalism, and we provide a classification of pure and mixed Gaussian
states according to the various forms that the associated covariance matrices
can take. In section 4 we introduce and describe the separability problem, the
inseparability criteria, and the machinery of bipartite entanglement evaluation
and distillation in Gaussian states. In Section 5 we review some important
results specific on two-mode Gaussian states, including the existence of
extremally entangled states, minimal and maximal, at a given degree of mixedness,
and the different orderings induced on the set of two-mode Gaussian states by
different measures of entanglement such as the Gaussian entanglement of formation
and the logarithmic negativity. In Section 6 we describe the unitary (and therefore
reversible) localization of bipartite multimode entanglement to bipartite two-mode
entanglement in fully symmetric and bisymmetric multimode Gaussian states, and the
scaling of bipartite entanglement with the number of modes in general multimode Gaussian
states. In Section 7 we then discuss recent crucial advances in the
understanding of entanglement sharing in multimode Gaussian states, including the proof
of the monogamy inequality of distributed entanglement for all Gaussian states.
Multipartite entanglement of Gaussian states is reviewed in Section 8 by discussing its
qualification according to different classes of separability, and the main consequences
of the monogamy inequality, such as the quantification of genuine tripartite entanglement
in three-mode Gaussian states via the residual Gaussian tangle, the promiscuous nature of
entanglement
sharing in Gaussian states with symmetry constraints, and the possible coexistence
of unlimited bipartite and multipartite entanglement. The last two properties (promiscuity
and coexistence of unbound bipartite and multipartite entanglement) are elucidated at length
by discussing three- and four-mode Gaussian states endowed with full and partial symmetry
constraints under the exchange of modes. Section 9 concludes this review with a brief discussion
about recent advances in the qualification and quantification of entanglement in non Gaussian states,
a field of research that is to a large extent yet to be fully explored, a summary on various
applications of Gaussian entanglement in quantum information and computation, and on overview
on open problems and current research directions.

\section{Introduction to continuous variable systems.}
\label{secIntroCV}

A continuous variable (CV) system \cite{brareview,eisplenio,COVAQIAL} of $N$ canonical bosonic modes
is described by a Hilbert space $\hh=\bigotimes_{k=1}^{N} \hh_{k}$
resulting from the tensor product structure of infinite dimensional
Fock spaces $\hh_{k}$'s, each of them associated to a single mode.
For instance, one can think of the non interacting quantized
electromagnetic field, whose Hamiltonian describes a system
of an arbitrary number $N$ of harmonic oscillators of different frequencies,
the {\em modes} of the field,
\begin{equation}\label{CV:Ham}
\op{H} = \sum_{k=1}^N \hbar \omega_k \left(\adj{\op{a}}_k\op{a}_k +
\frac12\right)\,.
\end{equation}
Here $\op{a}_k$ and $\adj{\op{a}}_k$ are the  annihilation and
creation operators of a photon in mode $k$ (with frequency
$\omega_k$), which satisfy the bosonic commutation relation
\begin{equation}\label{CV:comm}
\comm{\op{a}_k}{\adj{\op{a}}_{k'}}=\delta_{kk'}\,,\quad
\comm{\op{a}_k^}{\op{a}_{k'}}=\comm{\adj{\op{a}}_k}{\adj{\op{a}}_{k'}}=0\,.
\end{equation}
From now on we will assume for convenience natural units with
$\hbar=2$. The corresponding quadrature phase operators (position
and momentum)  for each mode are defined as
\begin{eqnarray}
  \hat q_{k} &=& (\op a_{k}+\op a^{\dag}_{k})\,, \label{CV:q}\\
  \hat p_{k} &=& (\op a_{k}-\op a^{\dag}_{k})/i \label{CV:p}
\end{eqnarray}
We can group together the canonical operators in the vector
\be\label{CV:R}
\hat{R}=(\hat{q}_1,\hat{p}_1,\ldots,\hat{q}_N,\hat{p}_N)\T\,,\ee
which enables us to write in compact form the  bosonic commutation
relations between the quadrature phase operators, \be
[\hat{R}_k,\hat{R}_l]=2 i\Omega_{kl} \; ,\label{ccr}\ee where
$\Omega$ is the symplectic form \be
\Omega=\bigoplus_{k=1}^{N}\omega\, , \quad \omega=
\left(\begin{array}{cc}
0&1\\
-1&0
\end{array}\right)\, . \label{symform}
\ee

The space $\hh_k$ is spanned by the Fock basis $\{\ket{n}_k\}$ of
eigenstates of the number operator $\hat{n}_k = \hat a_k^{\dag}\hat
a_k$, representing the Hamiltonian of the noninteracting mode via
\eq{CV:Ham}. The Hamiltonian of each mode is bounded from below, thus
ensuring the stability of the system. For each mode  $k$
there exists a different vacuum state $\ket{0}_k\in \hh_k$
such that $\hat a_k\ket{0}_k=0$.
The vacuum state of the global Hilbert space will be denoted by
$\ket{0}=\bigotimes_k \ket{0}_k$. In the single-mode Hilbert space
${\mathcal H}_k$, the eigenstates of $\hat a_k$ constitute the
important set of {\em coherent states} \cite{BWallsMilburn}, which
is overcomplete in $\hh_k$. Coherent states result from applying the
single-mode Weyl displacement operator $\hat D_k$ to the vacuum
$\ket{0}_k$, $\ket{\alpha}_k= \hat D_k(\alpha)\ket{0}_k$, where
\be\label{CV:Weyl}\hat D_k(\alpha)=\,{\rm e}^{\alpha \hat
a_k^{\dag}-\alpha^{*}\hat a_k}\,,\ee and the coherent amplitude
$\alpha\in{\mathbbm C}$ satisfies $\hat
a_k\ket{\alpha}_k=\alpha\ket{\alpha}_k$. In terms of the Fock basis
of mode $k$ a coherent state reads
\be\label{CV:coh}\ket{\alpha}_k=\,{\rm e}^{-\frac12|\alpha|^2}
\sum_{n=1}^{\infty}\frac{\alpha^{n}}{\sqrt{n!}}\ket{n}_k\,.\ee
Tensor products of coherent states for $N$ different modes are obtained
by applying the $N$-mode Weyl operators $\hat D_{\xi}$ to the global
vacuum $\ket{0}$. For future convenience, we define the operators
$D_{\xi}$ in terms of the canonical operators $\hat{R}$, \be \hat
D_{\xi} = \,{\rm e}^{i\hat{R}^{\sf T} \Omega \xi}\, , \quad {\rm
with} \quad\xi\in {\mathbbm R}^{2N} \; . \ee
One then has
$\ket{\xi}=\hat D_{\xi}\ket{0}$.

\subsection{Quantum phase-space picture.}

The states of a CV system are the set of positive trace-class
operators $\{\varrho\}$ on the Hilbert space $\hh=\bigotimes_{k=1}^N
\hh_k$. However, the complete description of any quantum state
$\varrho$ of such an infinite dimensional system can be provided by
one of its $s$-ordered {\em characteristic functions} \cite{barnett}
\be \chi_s (\xi) = \,{\rm Tr}\,[\varrho \hat D_{\xi}] \,{\rm
e}^{s\|\xi\|^2/2} \; , \label{cfs} \ee with $\xi\in\R^{2N}$,
$\|\centerdot\|$ standing for the Euclidean norm of $\R^{2N}$. The
vector $\xi$ belongs to the real $2N$-dimensional space
$\Gamma=({\mathbbm R}^{2N},\Omega)$, which is called {\em phase
space}, in analogy with classical Hamiltonian dynamics. One can see
from the definition of the characteristic functions that in the
phase space picture, the tensor product structure is replaced by a
direct sum structure, so that the $N$-mode phase space $\Gamma =
\bigoplus_k \Gamma_k$, where $\Gamma_k=({\mathbbm R}^{2},\omega)$ is
the local phase space associated with mode $k$.

The family of characteristic functions is in turn related, via
complex Fourier transform, to the {\em quasi-probability
distributions} $W_s$, which constitute another set of complete
descriptions of the quantum states \be
W_s(\xi)=\frac{1}{\pi^2}\int_{{\mathbbm R}^{2N}}\kappa
\chi_s(\kappa) \,{\rm e}^{i\kappa^{\sf T} \Omega \xi}\,{\rm d}^{2N}
\, . \label{qps} \ee There exist states for which the function $W_s$
is not a regular probability distribution for any $s$, because it
can be singular or assume negative values. Note that the value
$s=-1$ corresponds to the Husimi `Q-function' \cite{Husimi40}
$W_{-1}(\xi)=\bra{\xi}\varrho\ket{\xi}/\pi$ and thus always yields a
regular probability distribution. The case $s=0$ corresponds to the
so called Wigner function \cite{Wigner32}, which will be denoted
simply by $W$. Likewise, for the sake of simplicity, $\chi$ will
stand for the symmetrically ordered characteristic function
$\chi_0$. Finally, the case $s=1$ yields the singular
P-representation, which was introduced, independently, by Glauber
\cite{Glauber63} and Sudarshan \cite{Sudarshan63}.

The quasiprobability distributions of integer order $W_{-1}$, $W_0$
and $W_1$ are respectively associated the antinormally ordered,
symmetrically ordered and normally ordered expressions of operators.
More precisely, if the operator $\hat{O}$ can be expressed as
$\hat{O}= f(\hat a_k,\hat a^{\dag}_k)$ for $k=1,\ldots,N$, where $f$
is a, say, symmetrically ordered function of the field operators,
then one has \cite{cahillglau1,cahillglau2}
\[
\,{\rm Tr}[\varrho \hat{O}] =
\int_{\R^{2N}}W_0(\kappa)\bar{f}(\kappa) \,{\rm d}^{2N}\kappa \, ,
\]
where
$\bar{f}(\kappa)=f(\kappa_k+i\kappa_{k+1},\kappa_k-i\kappa_{k+1})$
and $f$ takes the same form as the operatorial function previously
introduced. The same relationship holds between $W_{-1}$ and the
antinormally ordered expressions of the operators and between
$W_{1}$ and the normal ordering. We also recall that the normally
ordered function of a given operator is provided by its Wigner
representation. This entails the following equalities for the trace
\be 1 = \tr\varrho = \int_{\R^{2N}} W(\kappa) \,{\rm d}^{2N}\kappa =
\chi(0) \,\label{wigtr} , \ee and for the purity
\cite{pariserafozzi}
 \be \mu = \tr\varrho^2 = \int_{\R^{2N}}
W^2(\kappa) \,{\rm d}^{2N}\kappa = \int_{\R^{2N}} |\chi(\xi)|^2
\,{\rm d}^{2N}\xi    \,\label{wigpr} , \ee of a state $\varrho$.
These expressions will be useful in the following.

The (symmetric) Wigner function can be written as follows in terms
of the (unnormalized) eigenvectors $\ket{x}$ of the quadrature
operators $\{\hat q_j\}$ (for which $\hat q_j \ket{x}=q_j \ket{x}$,
$x\in\rr^{N}$, for $j=1,\ldots,N$) \cite{Simon00} \be W(x,p) =
\frac{1}{\pi^{N}} \int_{\rr^{N}} \bra{x-x'}\varrho\ket{x+x'}\,{\rm
e}^{ix'\cdot p} \,{\rm d}^{N}x' \: , \quad x,p\in\rr^{N} \label{wig}
\ee From an operational point of view, the Wigner function admits a
clear interpretation in terms of homodyne measurements
\cite{sculzub}: the marginal integral of the Wigner function over
the variables $p_1,\ldots,p_N,x_1,\ldots,x_{N-1}$
\[\int_{\rr^{2N-1}}W(x,p)\,{\rm d}^{N}p\,{\rm d}\,x_1\ldots{\rm
d}\,x_{N-1}\] gives the probability of the results of homodyne
detections on the remaining quadrature $x_{N}$
\cite{francamentemeneinfischio}.

Table \ref{CVtab} summarizes the mapping of properties and tools
between Hilbert and phase spaces. In the next section more properties
and tools of phase space analysis will be introduced that are specially
useful in the study of Gaussian states.

\begin{table}[t!]   \centering{
\begin{tabular}{c||c|c}
& Hilbert space $\hh$ & Phase space $\Gamma$ \\
\hline \hline
dimension & $\infty$ & $2N$ \\ \hline
structure & $\bigotimes$ & $\bigoplus$ \\ \hline
description & $\ro$ & $\chi_s,\,W_s$ \\ \hline
\end{tabular}}
\caption{Schematic comparison between Hilbert-space  and  phase-space \newline  pictures
for $N$-mode continuous variable systems.}
\label{CVtab}
\end{table}

\section{Mathematical description of Gaussian states.}

The set of {\em Gaussian states} is, by definition, the set of
states with Gaussian characteristic functions and quasi-probability
distributions on the multimode quantum phase space.  Gaussian states
include, among others, coherent, squeezed, and thermal states. Therefore
they are of central importance in quantum optics and in quantum information
and quantum communication with CV systems \cite{adebook}.
Their entanglement properties will thus be one
of the main subjects of this review.

\subsection{Covariance matrix formalism.}

From the definition it follows that  a Gaussian state $\varrho$ is
completely characterized by the first and second statistical moments
of the quadrature field operators, which will be denoted,
respectively, by the vector of first moments $\bar R =
\left(\langle\hat R_{1} \rangle,\langle\hat
R_{1}\rangle,\ldots,\langle\hat R_{N}\rangle, \langle\hat
R_{n}\rangle\right)$ and by the covariance matrix (CM) $\sig$ of
elements
\begin{equation}
\sig_{ij} = \frac{1}{2}\langle \hat{R}_i \hat{R}_j + \hat{R}_j
\hat{R}_i \rangle - \langle \hat{R}_i \rangle \langle \hat{R}_j
\rangle \, , \label{covariance}\,.
\end{equation}

First moments can be arbitrarily adjusted by local unitary
operations, namely displacements in phase space, \ie applications of
the single-mode Weyl operator \eq{CV:Weyl} to locally re-center the
reduced Gaussian corresponding to each single mode (recall that the
reduced state obtained from a Gaussian state by partial tracing over
a subset of modes is still Gaussian). Such operations leave any
informationally relevant property, such as entropy and entanglement,
invariant. Therefore, from now on (unless otherwise stated) we
will adjust first moments to $0$ without any loss of generality
for the scopes of our analysis.

With this position, the Wigner function of a Gaussian state can be
written as follows in terms of phase-space quadrature variables
\begin{equation}
W(X)=\frac{\,{\rm
e}^{-\frac{1}{2}X\boldsymbol{\sigma}^{-1}X\T}}{\pi\sqrt{{\rm
Det}\,\boldsymbol{\sigma}}}{\:,}\label{wigner}
\end{equation}
where $R$ stands for the real phase-space vector
$(q_{1},p_{1},\ldots,q_{n},p_{n})\in\Gamma$. Therefore,
despite the infinite dimension of the associated Hilbert space,
the complete description of an arbitrary Gaussian state (up to
local unitary operations) is given by the $2N \times 2N$ CM $\sig$.
In the following $\sig$ will be assumed indifferently to denote the
matrix of second moments of a Gaussian state, or the Gaussian state itself.
In the language of statistical mechanics, the elements of the CM
are the two-point truncated correlation functions between the $2N$ canonical
continuous variables. We notice also that the entries of the CM can
be expressed as energies by multiplying them by the level spacing
$\hbar \omega_k$, where $\omega_k$ is the frequency of each mode $k$.
In this way $\tr{\sig}$ is related to the mean energy of the
state, \ie the average of the noninteracting Hamiltonian \eq{CV:Ham}.

As the real $\sig$ contains the complete locally-invariant
information on a Gaussian state, we can expect some constraints to
exist to be obeyed by any {\em bona fide} CM, reflecting in
particular the requirements of positive-semidefiniteness of the
associated density matrix $\varrho$. Indeed, such condition together
with the canonical commutation relations
 imply
\begin{equation}
\sig+ i\Omega\ge 0 \; , \label{bonfide}
\end{equation}
\ineq{bonfide} is the necessary and sufficient constraint the matrix
$\sig$ has to fulfill to be a CM corresponding to a physical
Gaussian state \cite{simon87,simon94}. More in general, the previous
condition is necessary for the CM of {\em any}, generally non
Gaussian, CV state (characterized in principle by the moments of any
order). We note that such a constraint implies $\sig\ge0$.
\ineq{bonfide} is the expression of the uncertainty principle on the
canonical operators in its strong, Robertson--Schr\"odinger form
\cite{robertson30,schrodinger30,serafozziprl}.

For future convenience, let us define and write down the CM
$\sig_{1\ldots N}$ of an $N$-mode Gaussian state in terms of two by
two submatrices as \be \sig_{1\ldots N} = \left(\begin{array}{cccc}
\sig_{1} & \eps_{1,2}\; & \cdots & \eps_{1,N} \\
\eps_{1,2}^{\sf T}\; & \ddots & \ddots & \vdots \\
\vdots & \ddots & \ddots & \eps_{N-1,N} \\
\eps_{1,N}^{\sf T}& \cdots & \eps_{N-1,N}^{\sf T} & \sig_{N} \\
\end{array}\right) \; . \label{CM}
\ee Each diagonal block $\sig_k$ is respectively the local CM
corresponding to the reduced state of mode $k$, for all
$k=1,\ldots,N$. On the other hand, the off-diagonal matrices
$\eps_{i,j}$ encode the intermodal correlations (quantum and
classical) between subsystems $i$ and $j$. The matrices $\eps_{i,j}$
all vanish for a product state.

In this preliminary overview, let us just mention an important
instance of two-mode Gaussian state, the {\em two-mode squeezed
state} $\ket{\psi^{sq}}_{i,j}=\hat U_{i,j}(r)
\left(\ket{0}_i\!\otimes\ket{0}_j\right)$ with squeezing factor  $r
\in \R$, where  the (phase-free) two-mode squeezing operator is
given by
\begin{equation}\label{tmsU}
\hat U_{i,j}(r) = \exp \left[-\frac{r}{2} (\hat {a}_i^\dag \hat
{a}_j^\dag -\hat {a}_i \hat {a}_j ) \right]\,,
\end{equation}
In the limit of infinite squeezing ($r\to \infty )$, the state
approaches the ideal Einstein-Podolsky-Rosen (EPR) state
\cite{EPR35}, simultaneous eigenstate of total momentum and relative
position of the two subsystems, which thus share {\em infinite
entanglement}. The EPR state is unnormalizable and unphysical.
However, in principle, an EPR state can be approximated with
an arbitrarily high degree of accuracy by two-mode squeezed states
with sufficiently large squeezing. Therefore, two-mode squeezed
states are of key importance as entangled resources for
practical implementations of CV quantum information protocols
\cite{brareview}. They thus play a central role in the subsequent
study of the entanglement properties of Gaussian states. A two-mode
squeezed state with squeezing $r$ (also known in quantum optics as a
{\em twin-beam} state) is described by a CM
\begin{equation}\label{tms}
\sig^{sq}_{i,j}(r)= \left(\begin{array}{cccc}
\cosh(2r)&0&\sinh(2r)&0\\
0&\cosh(2r)&0&-\sinh(2r)\\
\sinh(2r)&0&\cosh(2r)&0\\
0&-\sinh(2r)&0&\cosh(2r)
\end{array}\right)\! .
\end{equation}

\subsection{Symplectic operations.} \label{SecSympl}

An important role in the theoretical and experimental manipulation
of Gaussian states is played by unitary operations which preserve
the Gaussian character of the states on which they act. These
unitary Gaussian operations are all those generated by Hamiltonian
terms at most quadratic in the field operators. As a consequence of
the Stone-Von Neumann theorem, the so-called {\em metaplectic}
representation entails that any such unitary operation at the
Hilbert space level corresponds, in phase space, to a symplectic
transformation, {\ie}to a linear transformation $S$ which preserves
the symplectic form $\Omega$: \be\label{symplectic}S\T\Omega S =
\Omega\,.\ee Symplectic transformations on a $2N$-dimensional phase
space form the (real) symplectic group $\sy{2N}$. Such
transformations act linearly on first moments and by congruence on
covariance matrices, $\sig\mapsto S \sig S\T$. \eq{symplectic}
implies $\det{S}=1$, $\forall\,S\in\sy{2N}$. Ideal beam splitters,
phase shifters and squeezers are all described by some kind of
symplectic transformation (see \eg \cite{francamentemeneinfischio}).
For instance, the two-mode squeezing operator \eq{tmsU} corresponds
to the symplectic transformation
\begin{equation}\label{tmsS}
S_{i,j}(r)=\left(\begin{array}{cccc}
\cosh r&0&\sinh r&0\\
0&\cosh r&0&-\sinh r\\
\sinh r&0&\cosh r&0\\
0&-\sinh r&0&\cosh r
\end{array}\right)\, ,
\end{equation}
where the matrix is understood to act on the pair of modes $i$ and
$j$. In this way, the two-mode squeezed state, \eq{tms}, can be
obtained as $\sig^{sq}_{i,j}(r) = S_{i,j}(r) \id S_{i,j}\T(r)$
exploiting the fact that the CM of the two-mode vacuum state is
the $4 \times 4$ identity matrix.

Another common unitary operation is the ideal (phase-free) {\em
beam splitter}, whose action $\hat{B}_{i,j}$ on a pair of modes $i$
and $j$ is defined as
\begin{equation}\label{bsplit}
\hat{B}_{i,j}(\theta):\left\{
\begin{array}{l}
\hat a_i \rightarrow \hat a_i \cos\theta + \hat a_j\sin\theta \\
\hat a_j \rightarrow \hat a_i \sin\theta - \hat a_j\cos\theta \\
\end{array} \right.\,,
\end{equation}
with $\hat a_l$ being the annihilation operator of mode $k$. A beam
splitter with transmittivity $\tau$ corresponds to a rotation of
$\theta = \arccos\sqrt{\tau}$ in phase space ($\theta=\pi/4$
corresponds to a balanced 50:50 beam splitter, $\tau=1/2$),
described by a symplectic transformation
\begin{equation}\label{bbs}
B_{i,j}(\tau)=\left(
\begin{array}{cccc}
 \sqrt{\tau } & 0 & \sqrt{1 - \tau } & 0 \\
 0 & \sqrt{\tau } & 0 & \sqrt{1 - \tau } \\
 \sqrt{1 - \tau } & 0 & - \sqrt{\tau } & 0 \\
 0 & \sqrt{1 - \tau } & 0 & - \sqrt{\tau }
\end{array}
\right)\,.
\end{equation}

Single-mode symplectic operations are easily introduced
as linear combinations of planar (orthogonal) rotations and of
single-mode squeezings of the form \be\label{sqz} S_j(r)
=\diag(\,{\rm e}^{r},\,{\rm e}^{-r})\,,\ee acting on mode $j$, for
$r>0$.

In general, symplectic transformations in phase space are generated
by the exponentiation of matrices written as $J\Omega$, where $J$ is
antisymmetric \cite{pramana}. Such generators can be symmetric or
antisymmetric. The operations $B_{ij}(\tau)$, \eq{bbs}, generated by
antisymmetric operators are orthogonal and, acting by congruence on
the CM $\sig$, preserve the value of $\tr{\sig}$. Since $\tr{\sig}$
gives the contribution of the second moments to the average of the
Hamiltonian $\bigoplus_k \hat a_k^{\dag}\hat a_k$, these
transformations are said to be {\em passive} (they belong to the
compact subgroup of $\sy{2N}$). Instead, operations $S_{i,j}(r)$,
\eq{tmsS}, generated by symmetric operators, are not orthogonal and
do not preserve $\tr{\sig}$ (they belong to the non compact subgroup
of $\sy{2N}$). This mathematical difference between squeezers and
phase-space rotations accounts, in a quite elegant way, for the
difference between {\em active} (energy non preserving) and passive
(energy preserving) optical transformations \cite{passive}.

Let us remark that {\em local} symplectic operations belong to the
group $\mathrm{Sp}(2,\mathbb{R})^{\oplus N}$. They correspond, at
the Hilbert space level, to tensor products of unitary
transformations, each acting on the state space of a single mode. It is
useful to notice that the determinants of each $2\times2$ submatrix
of a $N$-mode CM, \eq{CM}, are {\em all} invariants under local
symplectic operations $S \in \mathrm{Sp}(2,\mathbb{R})^{\oplus N}$.
This mathematical property reflects the physical requirement that
marginal informational properties and correlations between
individual subsystems cannot be altered by local operations alone.

\subsubsection{Symplectic eigenvalues and
invariants.}\label{SecWilly} A symplectic transformation of major
importance is the one that diagonalizes a Gaussian state in
the basis of normal modes. Through this decomposition, thanks
to Williamson theorem \cite{williamson36}, the CM of a $N$-mode
Gaussian state can always be written in the so-called Williamson normal,
or diagonal form
\begin{equation}
\sig=S\T \gr\nu S \; , \label{willia}
\end{equation}
where $S\in Sp_{(2N,\mathbb{R})}$ and $\gr\nu$ is the CM
\begin{equation}
\gr{\nu}=\bigoplus_{k=1}^{N}\left(\begin{array}{cc}
\nu_k&0\\
0&\nu_k
\end{array}\right) \, , \label{therma}
\end{equation}
corresponding to a tensor product state with a diagonal density
matrix $\varrho^{_\otimes}$ given by \be
\varrho^{_\otimes}=\bigotimes_{k}
\frac{2}{\nu_{k}+1}\sum_{n=0}^{\infty}\left(
\frac{\nu_{k}-1}{\nu_{k}+1}\right)\ket{n}_{k}{}_{k}\bra{n}\; ,
\label{thermas} \ee where $\ket{n}_k$ denotes the number state of
order $n$ in the Fock space $\hh_{k}$. In the Williamson form, each
mode with frequency $\omega_k$ is a Gaussian state in thermal
equilibrium at a temperature $T_k$, characterized by a Bose-Einstein
statistical distribution of the thermal photons $n_k$, with average

\begin{equation}\label{temperature}
\bar n_k = \frac{\nu_k-1}{2} =
\frac{1}{\exp\left(\frac{\hbar\omega_k}{k_B T_k}\right)-1}\,.
\end{equation}

The $N$ quantities $\nu_{k}$'s form the {\em symplectic spectrum} of
the CM $\sig$, and are invariant under the action of global
symplectic transformations on the matrix $\sig$. The symplectic
eigenvalues can be computed as the orthogonal eigenvalues of the
matrix $|i\Omega\sig|$ \cite{SeralePHD} and are thus determined by
$N$ invariants of the characteristic polynomial of such a matrix
\cite{serafozziprl}. One global symplectic invariant is simply the
determinant of the CM (whose invariance is a consequence of the fact
that ${\rm Det}\,S=1$ $\forall S\in \sy{2N}$), which, once computed
in the Williamson diagonal form, reads
\begin{equation}\label{detsigma}
\det\sig = \prod_{k=1}^N \nu_k^2\,.
\end{equation}
Another important invariant under global symplectic operations is
the so-called {\em seralian} $\Delta$ \cite{polacchi}, defined as
the sum of the determinants of all $2 \times 2$ submatrices of a CM
$\sig$, \eq{CM}, which can be readily computed in terms of its
symplectic eigenvalues as
\begin{equation}\label{seralian}
\Delta(\sig) = \sum_{k=1}^N \nu_k^2\,.
\end{equation}
The invariance of $\Delta_{\sig}$ in the multimode case
\cite{serafozziprl} follows from its invariance in the case of
two-mode states, proved in Ref.~\cite{SymplecticInvariants}, and
from the fact that any symplectic transformation can be decomposed
as the product of two-mode transformations \cite{agarwal94}.

\subsubsection{Symplectic representation of the uncertainty
principle.}\label{SecSympHeis}

The symplectic eigenvalues $\nu_{k}$ provide a powerful tool to
access essential information on the properties of the Gaussian state
$\gr{\sigma}$ \cite{SymplecticInvariants}. For instance, let us
consider the uncertainty relation \pref{bonfide}. Since the inverse
of a symplectic operation is itself symplectic, one has from
\eq{symplectic}, ${S^{-1}}\T {\Omega}S^{-1}= {\Omega}$, so that
\ineq{bonfide} is {\em equivalent} to $\gr{\nu}+i{\Omega}\ge 0$. In
terms of the symplectic eigenvalues $\nu_{k}$ the uncertainty
relation then simply reads \be {\nu}_{k}\ge1 \; . \label{sympheis}
\ee Inequality \pref{sympheis} is completely equivalent to the
uncertainty relation \pref{bonfide} provided that the CM $\sig$
satisfies $\sig\ge 0$.

Without loss of generality, one can rearrange the modes of a
$N$-mode state such that the corresponding symplectic eigenvalues
are sorted in ascending order \[ \nu_- \equiv \nu_1 \le \nu_2 \le
\ldots \le \nu_{N-1} \le \nu_N \equiv \nu_+\,.\] With this notation,
the uncertainty relation reduces to $\nu_1 \ge 1$. We remark that
the full saturation of the uncertainty principle can only be
achieved by {\em pure} $N$-mode Gaussian states, for which
\[\nu_i=1\,\,\forall i=1,\ldots, N\,,\] meaning that the Williamson
normal form of any pure Gaussian state is the vacuum $\ket0$ of the
$N$-mode Hilbert space $\hh$. Instead, mixed states such that
$\nu_{i\le k}=1$ and $\nu_{i>k}>1$, with $1\le k\le N$, only
partially saturate the uncertainty principle, with partial
saturation becoming weaker with decreasing $k$. Such states are
minimum uncertainty mixed Gaussian states in the sense that the
phase quadrature operators of the first $k$ modes satisfy the
Robertson-Schr\"odinger minimum uncertainty, while, for the remaining
$N-k$ modes, the state indeed contains some additional thermal
correlations which are responsible for the global mixedness of the state.

The {\em symplectic rank} $\aleph$ of a CM $\sig$ is defined as the
number of its symplectic eigenvalues different from $1$,
corresponding to the number of non-vacua normal modes
\cite{generic}. A Gaussian state is pure if and only if $\aleph=0$.
For mixed $N$-mode states one has $1\le \aleph \le N$ depending on
their degree of partial minimum-uncertainty saturation. This is in
analogy with the standard rank of finite-dimensional (density)
matrices, defined as the number of nonvanishing eigenvalues; in that
case, only pure states $\varrho=\ketbra{\psi}{\psi}$ have rank $1$,
and mixed states have in general higher rank. As we will now show,
{\em all} the informational properties of Gaussian states can be
recast in terms of their symplectic spectra.

A mnemonic summary of the main structural features of Gaussian
states in the phase-space/CM description (including the definition
of purity given in the next subsection) is provided by Table
\ref{CVtabG}.

\begin{table}[t!]   \centering{
\begin{tabular}{c||c|c}
& Hilbert space $\hh$ & Phase space $\Gamma$ \\
\hline\hline
dimension & $\infty$ & $2N$ \\ \hline
structure & $\bigotimes$ & $\bigoplus$ \\ \hline
description & $\ro$ & $\sig$ \\ \hline
{\em bona fide} & $\ro \ge 0$ & $\sig + i \Omega \ge 0$ \\ \hline
operations & $\underset{\ro \mapsto U \ro \adj{U}}{\overset{}{U}: \adj U U = \id}$ &
$\underset{\sig \mapsto S \sig S\T}{\overset{}{S}: S\T \Omega S =
\Omega}$
\\ \hline
spectra&
$\overset{{}}{\underset{0 \le \lambda_k \le 1}{U \ro \adj{U} = {\rm diag}\{\lambda_k\}}}$ &
$\underset{1 \le \nu_k < \infty}{S \sig S\T = {\rm diag}\{\nu_k\}}$
\\ \hline
pure states & $\lambda_i = 1,\,\lambda_{j \neq i}=0$ & $\nu_j = 1,\, \forall j=1\ldots N$
\\ \hline
purity & $\overset{}{\tr{\ro^2} = \sum_k \lambda_k^2}$ &
$\overset{}{1/\sqrt{\det{\sig}}=\prod_k \nu_k^{-1}}$ \\ \hline
\end{tabular}}
\caption{Schematic comparison between Hilbert-space  and  phase-space  pictures
for $N$-mode Gaussian states. The first two rows are taken from Table \ref{CVtab} and apply to
general states of CV systems. The following rows are special to Gaussian states, relying on
the covariance matrix description and the properties of the symplectic group.}
\label{CVtabG}\end{table}

\subsection{Degree of information contained in a Gaussian state.}
\label{SecEntroGG}

\subsubsection{Measures of information.}\label{ParInfo}

The degree of {\em information} contained in a quantum state
corresponds to the amount of knowledge that we possess {\em a priori}
on predicting the outcome of any test performed on the state \cite{BPeres}.

The simplest measure of such information is the {\em purity}
of a quantum state $\ro$:
\begin{equation}\label{QM:purity}
\mu(\ro) = \tr{\ro^2}\,.
\end{equation}
For states belonging to a given Hilbert space $\hh$ with $\dim\hh=D$,
the purity varies in the range
\[
\frac1D \le \mu \le 1\, .
\]
The minimum is reached by the totally random mixture; the upper
bound is saturated by pure states. In the limit of CV systems $(D
\rightarrow \infty)$, the minimum purity tends asymptotically to
zero. Accordingly, the ``impurity'' or degree of {\em mixedness} of
a quantum state $\ro$, which characterizes our ignorance before
performing any quantum test on $\ro$, can be quantified by the
functional
\begin{equation}\label{QM:SL}
S_L(\ro) = \frac{D}{D-1} \left( 1- \mu \right) = \frac{D}{D-1}
\left( 1- \tr{\ro^2} \right)\,.
\end{equation}
The {\em linear entropy} $S_L$ (ranging between $0$ and $1$) defined by
\eq{QM:SL} is a very useful measure of mixedness in quantum mechanics
and quantum information theory due to its direct connection with the purity
and its computational simplicity.

In general, the degree of mixedness of a quantum state $\varrho$ can
be characterized completely by the knowledge of all the associated
Schatten $p$--norms \cite{bathia} \be \label{schatten} \|\varrho\|_p
= (\,{\rm Tr}\,|\varrho|^p)^{\frac1p} =(\,{\rm
Tr}\,\varrho^p)^{\frac1p}\, , \quad \,{\rm with} \: p \ge 1. \ee In
particular, the case $p=2$ is directly related to the purity $\mu$,
\eq{QM:purity}, as it is essentially equivalent (up to a
normalization) to the linear entropy, \eq{QM:SL}. The $p$-norms are
multiplicative on tensor product states and thus determine a family
of non-extensive ``generalized entropies'' $S_{p}$
\cite{bastiaans,tsallis}, defined as \be S_{p}(\ro) = \frac{1-\,{\rm
Tr}\,\varrho^p}{p-1} \; , \quad p > 1. \label{pgen} \ee The
generalized entropies $S_p$'s range from $0$ for pure states to
$1/(p-1)$ for completely mixed states with fully degenerate
eigenspectra.   We also note that, for any given quantum state,
$S_{p}$ is a monotonically decreasing function of $p$. Finally,
another important class of entropic measures includes the R\'{e}nyi
entropies \cite{renyi} \be S_{p}^{R}(\ro) = \frac{\log \, {\rm Tr}
\, \varrho^{p}}{1-p} \; , \quad p > 1. \ee It can be  shown that
\cite{SeralePHD} \be
\lim_{p\rightarrow1+}S_{p}(\ro)=\lim_{p\rightarrow1+}S_{p}^{R}(\ro)=
-\,{\rm Tr}\,(\varrho\log\varrho)\, . \label{genvneu} \ee Therefore,
even the {\em von Neumann entropy}
\begin{equation}\label{QM:SV}
S_V(\ro) = - \Tr{\ro\, \log \ro}
\end{equation}
can be defined, in terms of
$p$-norms, as a suitable limit within the framework
of generalized entropies.

The von Neumann entropy is {\em subadditive} \cite{Wehrl78}.
Consider a bipartite system $\s$
(described by the Hilbert space  $\hh = \hh_1 \otimes
\hh_2$) in the state  $\ro$. Then
\begin{equation}\label{QM:SVsubadd}
S_V(\ro) \le S_V(\ro_1) + S_V(\ro_2)\,,
\end{equation}
where $\ro_{1,2}$ are the reduced density matrices $\ro_{1,2} = \ptr{2,1}{\ro}$
associated to subsystems  $\s_{1,2}$.
For states of the form $\ro^{\otimes} = \ro_1 \otimes
\ro_2$, \eq{QM:SVsubadd} is saturated, yielding that
the von Neumann entropy is
\emph{additive} on tensor product states:
\begin{equation}\label{QM:SVaddprod}
S_V(\ro_1 \otimes \ro_2) = S_V(\ro_1) + S_V(\ro_2)\,.
\end{equation}
The purity, \eq{QM:purity}, is instead \emph{multiplicative} on
product states, as the trace of a product equates the product of
the traces:
\begin{equation}\label{QM:mumultprod}
\mu(\ro_1 \otimes \ro_2) = \mu(\ro_1) \cdot \mu(\ro_2)\,.
\end{equation}

Considering a joint classical probability distribution over the
variables $X$ and $Y$, one has for the Shannon entropy,
\begin{equation}\label{QM:Shanarale}
S(X,Y) \ge S(X),\,S(Y)\,.
\end{equation}
The Shannon entropy of a joint  probability distribution is always
greater than the Shannon entropy of each marginal probability
distribution, meaning that there is less information in a global
classical system than in any of its parts. On the other hand,
consider a bipartite quantum system in a pure state
$\ro=\ketbra{\psi}{\psi}$ . We have then for the von Neumann
entropies: $S_V(\ro)=0$, $S_V(\ro_1) = S_V(\ro_2) \ge 0$. The global
state  $\ro$ has been prepared in a well defined way, but if we
measure local observables on the subsystems, the measurement
outcomes are unavoidably affected by random noise. One cannot
reconstruct the whole information about how the global system was
prepared in the state $\ro$ (apart from the trivial instance of
$\ro$ being a product state $\ro=\ro_1 \otimes \ro_2$), by only
looking separately at the two subsystems. Information is rather
encoded in nonlocal and non factorizable quantum correlations --
{\em entanglement} -- between the two subsystems. This clearly
highlights the fundamental difference between classical and quantum
distributions of information.

\subsubsection{Entropic measures for Gaussian states.}\label{SecEntroG}

We now illustrate how to evaluate the measures of
information (or lack thereof) defined above for
Gaussian states \cite{extremal}.

The generalized purities ${\rm Tr}\,\varrho^p$ defined by
\eq{schatten} are invariant under global unitary operations.
Therefore, for any $N$-mode Gaussian state they are only functions
of the symplectic eigenvalues $\nu_{k}$ of $\gr{\sigma}$. In fact, a
symplectic transformation acting on $\gr{\sigma}$ is embodied by a
unitary (trace preserving) operator acting on $\varrho$, so that
${\rm Tr}\,\varrho^{p}$ can be easily computed on the Williamson
diagonal state $\gr\nu$ of \eq{therma}. One obtains \cite{extremal}
\be {\rm Tr}\,\varrho^{p}=\prod_{i=1}^{N}g_{p}(\nu_i)\; ,
\label{pgau} \ee where
\[
g_{p}(x)=\frac{2^p}{(x+1)^p-(x-1)^p} \, .
\]
A remarkable consequence of \eq{pgau} is that \be
\mu(\varrho)=\frac{1}{\prod_i \nu_{i}}=\frac{1}{\sqrt{{\rm
Det}\,\gr{\sigma}}} \, . \label{purgau} \ee Regardless of the number
of modes, the purity of a Gaussian state is {\em fully determined}
by the global symplectic invariant ${\rm Det}\,\gr{\sigma}$ alone,
\eq{detsigma}. We recall that the purity is related to the linear
entropy $S_L$ via \eq{QM:SL}, which in CV systems simply becomes
$S_L = 1-\mu$. A second consequence of \eq{pgau} is that, together
with Eqs.~\pref{pgen} and \pref{genvneu}, it allows for the
computation of the von Neumann entropy $S_{V}$, \eq{QM:SV}, of a
Gaussian state $\varrho$, yielding \cite{SymplecticInvariants} \be
S_{V}(\varrho)=\sum_{i=1}^{N}f(\nu_{i}) \; , \label{vneugau} \ee
where \be\label{entfunc} f(x) \equiv
\frac{x+1}{2}\log\left(\frac{x+1}{2}\right)-
\frac{x-1}{2}\log\left(\frac{x-1}{2}\right) \, . \ee Such an
expression for the von Neumann entropy of a Gaussian state was first
explicitly given in Ref.~\cite{holevo99}. Notice that $S_V$ diverges
on infinitely mixed CV states, while $S_L$ is normalized to $1$. Let
us remark that, clearly, the symplectic spectrum of single-mode
Gaussian states, which consists of only one eigenvalue $\nu_1$, is
fully determined by the invariant ${\rm
Det}\,\gr{\sigma}=\nu_{1}^2$. Therefore, all the entropies $S_{p}$'s
(and $S_{V}$ as well) are just increasing functions of ${\rm
Det}\,\gr{\sigma}$ (\emph{i.e.}~of $S_L$) and induce \emph{the same}
hierarchy of mixedness on the set of one-mode Gaussian states. This
is no longer true for multi-mode states, even in the
simplest instance of two-mode states \cite{extremal}.

\subsection{Standard forms of special Gaussian states.}
\label{SecSFCM}

The symplectic analysis applied on Gaussian CMs implies that some of
them can be suitably reduced under local operations. Such reductions
go under the name of {\em standard forms}.

\subsubsection{Pure states: Phase-space Schmidt decomposition.}\label{SecSchmidtPS} In general, {\em
pure}
Gaussian states of a bipartite CV system admit a physically
insightful decomposition at the CM level
\cite{holevo01,botero03,giedkeqic03}, which can be regarded as the
direct analogue of the Schmidt decomposition for pure
discrete-variable states \cite{BPreskill}. Let us recall what
happens in the finite-dimensional case. With respect to a
bipartition of a pure state $\ps_{A|B}$ into two subsystems $\s_A$
and $\s_B$, one can diagonalize (via an operation $U_A \otimes U_B$
which is local unitary according to the considered bipartition) the
two reduced density matrices $\ro_{A,B}$, to find that they have the
same rank and exactly the same nonzero eigenvalues $\set{\lambda_k}$
($k=1\ldots\min\{\dim\hh_A,\,\dim\hh_B\}$). The reduced state of the
higher-dimensional subsystem (say $\s_B$) will accomodate
$(\dim\hh_b-\dim\hh_A)$ additional $0$'s in its spectrum. The state
$\ps_{A|B}$ takes thus the Schmidt form $\ps = \sum_{k=1}^d
\lambda_k \ket{u_k,v_k}$.

Looking at the mapping provided by Table \ref{CVtabG} one can deduce
what happens for Gaussian states. Given a Gaussian CM $\sig_{A|B}$
of an arbitrary number $N$ of modes, where subsystem $\s_A$
comprises $N_A$ modes and subsystem $\s_B$ $N_B$ modes (with
$N_A+N_B=N$), then one can perform the Williamson decomposition
\eq{willia} in both reduced CMs (via a local symplectic operation
$S_A \oplus S_B$), to find that they have the same symplectic rank,
and the same non-unit symplectic eigenvalues $\set{\nu_k}$
($k=1\ldots\min\{N_A,\,N_B\}$). The reduced state of the
higher-dimensional subsystem (say $\s_B$) will accomodate
$(N_B-N_A)$ additional $1$'s in its symplectic spectrum. With
respect to an arbitrary $A|B$ bipartition, therefore, the CM
$\sig^p$ of any pure $N$-mode Gaussian state is locally equivalent
to the form $\sig^p_S = (S_A \oplus S_B) \sig^p (S_A \oplus S_B)\T$,
with
\begin{equation}\label{CMschmidt}
\sig^p_S = \tiny{\left(
\begin{array}{c|c}
\overbrace{\left.
\begin{array}{cccc}
\gr C_1 & \diamond  & \diamond  & \diamond  \\
\diamond  & \gr C_2 & \diamond  & \diamond  \\
\diamond  & \diamond  & \ddots & \diamond  \\
\diamond  & \diamond  & \diamond  & \gr C_{N_A}
\end{array}
\right.}^{N_A} & \overbrace{\left.
\begin{array}{ccccccc}
\gr S_1 & \diamond  & \diamond  & \diamond  & \diamond  & \diamond  & \diamond  \\
\diamond  & \gr S_2 & \diamond  & \diamond  & \diamond  & \diamond  & \diamond  \\
\diamond  & \diamond  & \ddots & \diamond  & \diamond  & \diamond  & \diamond  \\
\diamond  & \diamond  & \diamond  & \gr S_{N_A} & \diamond  & \diamond  & \diamond
\end{array}
\right.}^{N_B} \\ \hline
\left.
\begin{array}{cccc}
\gr S_1 & \diamond  & \diamond  & \diamond  \\
\diamond  & \gr S_2 & \diamond  & \diamond  \\
\diamond  & \diamond  & \ddots\,  & \diamond  \\
\diamond  & \diamond  & \diamond  & \gr S_{N_A} \\
\diamond  & \diamond  & \diamond  & \diamond  \\
\diamond  & \diamond  & \diamond  & \diamond  \\
\diamond  & \diamond  & \diamond  & \diamond
\end{array}
\right. & \left.
\begin{array}{ccccccc}
\gr C_1 & \diamond  & \diamond  & \diamond  & \diamond  & \diamond  & \diamond  \\
\diamond  & \gr C_2 & \diamond  & \diamond  & \diamond  & \diamond  & \diamond  \\
\diamond  & \diamond  & \ddots & \diamond  & \diamond  & \diamond  & \diamond  \\
\diamond  & \diamond  & \diamond  & \gr C_{N_A}\!\! & \diamond  & \diamond  & \diamond   \\
\diamond  & \diamond  & \diamond  & \diamond  & \!\!\gr\id\!\! &
\diamond  & \diamond \vspace*{-0.2cm} \\
\diamond  & \diamond  & \diamond  & \diamond  & \diamond  & \!\!\!\ddots\!\!\! & \diamond  \\
\diamond  & \diamond  & \diamond  & \diamond  & \diamond  & \diamond  &
\!\!\gr\id\!\!
\end{array}
\right.
\end{array}
\right).}
\end{equation}
Here each element denotes a $2\times 2$ submatrix, in particular the
diamonds ($\diamond$) correspond to null matrices, $\gr\id$ to the
identity matrix, and \[ \gr C_k = \left(
\begin{array}{cc}
\nu_k & 0 \\
0 & \nu_k \\
\end{array}
\right)\,,\quad \gr S_k = \left(
\begin{array}{cc}
\sqrt {\nu_k^2-1} & 0 \\
0 & -\sqrt {\nu_k^2-1} \\
\end{array}
\right)\,. \]
The matrices $\gr C_k$ contain the symplectic
eigenvalues $\nu_k \neq 1$ of both reduced CMs. By expressing them
in terms of hyperbolic functions, $\nu_k = \cosh(2 r_k)$ and by
comparison with \eq{tms}, one finds that each two-mode CM
\[
\left(
\begin{array}{cc}
\gr C_k & \gr S_k \\
\gr S_k & \gr C_k \\
\end{array}
\right)\,,
\]
encoding correlations between a single mode from $\s_A$ and a single
mode from $\s_B$, is a two-mode squeezed state with squeezing $r_k$.
Therefore, the Schmidt form of a pure $N$-mode Gaussian state with
respect to a ($N_A \times N_B)$-mode bipartition (with $N_B \ge
N_A$) is that of a direct sum \cite{holevo01,giedkeqic03}
\begin{equation}\label{CVschmidt}
\sig^p_S = \bigoplus_{i=1}^{N_A} \sig^{sq}_{i,j}(r_i)
\bigoplus_{k=2N_A+1}^{N} \sig^0_k\,,
\end{equation}
where mode $i \in \s_A$, mode $j\equiv i+N_A \in \s_B$, and
$\sig^0_k = \id_2$ is the CM of the vacuum state of mode $k  \in
\s_B$. This corresponds, on the Hilbert space level, to the product
of two-mode squeezed states, tensor additional uncorrelated vacuum
modes in the higher-dimensional subsystem ($\s_B$ in our notation)
\cite{botero03}. The phase-space Schmidt decomposition is a very
useful tool both for the understanding of the structural features of
Gaussian states in the CM formalism, and for the evaluation of their
entanglement properties. Notice that the validity of such a
decomposition can be extended to mixed states with a fully degenerate
symplectic spectrum, \ie a Williamson normal form proportional to the
identity \cite{botero03,giedkeqic03}. As a straightforward
consequence of \eq{CVschmidt}, any pure two-mode Gaussian state is
equivalent, up to local unitary operations, to a two-mode squeezed
state of the form \eq{tms}.

We will now show that for (generally mixed) Gaussian states with
some local symmetry constraints, a similar phase-space reduction  is
available, such that multimode properties (like entanglement) can be
unitarily reduced to two-mode ones.

\subsubsection{Symmetric and bisymmetric states.} \label{SecSymm}

Very often in quantum information, and in particular in the theory
of entanglement, peculiar roles are played by {\em symmetric
states}, that is, states that are either invariant under a
particular group of transformations --- like Werner states of
qu$d$its \cite{Werner89} --- or under permutation of two or more
parties in a multipartite system, like ground and thermal states of
translationally invariant Hamiltonians ({\eg}of harmonic
lattices) \cite{chain}. Here we will introduce classes of Gaussian
states globally invariant under all permutation of the modes (fully
symmetric states), or locally invariant in  each ofvthe two subsystems
across a global bipartition of the modes (bisymmetric states).
The properties of their symplectic spectrum allow to determine
explicitely a standard form in both cases. Here we will limit
ourself to list and collect the results that will be useful for
the computation and exploitation of entanglement in the
corresponding states. The detailed discussions and all the
rigorous proofs can be found in Ref.~\cite{unitarily}.
Unless explicitly stated, in the following we will be dealing
with generally mixed states.

A multimode Gaussian state $\varrho$ is  ``fully
symmetric'' if it is invariant under the exchange of any two modes.
In the following, we will consider the fully symmetric $M$-mode and
$N$-mode Gaussian states $\varrho_{\alp^M}$ and $\varrho_{\bet^N}$,
with CMs $\sig_{\alp^M}$ and $\sig_{\bet^N}$. Due to symmetry, we
have that
\begin{equation}\label{fscm}
\sig_{\alp^M}={\left(%
\begin{array}{cccc}
\gr\alpha & \gr\varepsilon & \cdots & \gr\varepsilon \\
\gr\varepsilon & \gr\alpha & \gr\varepsilon & \vdots \\
\vdots & \gr\varepsilon & \ddots & \gr\varepsilon \\
\gr\varepsilon & \cdots & \gr\varepsilon & \gr\alpha \\
\end{array}%
\right)}\,, \quad
\sig_{\bet^N}={\left(%
\begin{array}{cccc}
\gr\beta & \gr\zeta & \cdots & \gr\zeta \\
\gr\zeta & \gr\beta & \gr\zeta & \vdots \\
\vdots & \gr\zeta & \ddots & \gr\zeta \\
\gr\zeta & \cdots & \gr\zeta & \gr\beta \\
\end{array}%
\right)}\,,
\end{equation}
where $\gr\alpha$, $\gr\varepsilon$, $\gr\beta$ and $\gr\zeta$ are
$2\times2$ real symmetric submatrices (the symmetry of
$\gr\varepsilon$ and $\gr\zeta$ stems again from the symmetry under
the exchange of any two modes).

A fully symmetric $N$-mode Gaussian states with CM $\sig_{\beta^N}$
admits the following standard form \cite{adescaling,unitarily}. The
$2\times 2$ blocks $\bet$ and $\gr\zeta$ of $\sig_{\beta^N}$,
defined by \eq{fscm}, can be brought by means of local, single-mode
symplectic operations $S\in \sy{2}^{\oplus N}$ into the form
$\bet=\,{\rm diag}\,(b,b)$ and $\gr\zeta=\,{\rm diag}\,(z_1,z_2)$.
Obviously, analogous results hold for the $M$-mode CM
$\sig_{\alpha^M}$ of \eq{fscm}, whose $2\times 2$ submatrices can be
brought to the form $\alp=\,{\rm diag}\,(a,a)$ and
$\gr\varepsilon=\,{\rm diag}\,(e_1,e_2)$.

Let us now generalize this analysis to the $(M+N)$-mode Gaussian
states with CM $\sig$, which results from a correlated combination
of the fully symmetric blocks $\sig_{\alp^M}$ and $\sig_{\bet^N}$:
\be \sig = \left(\begin{array}{cc}
\sig_{\alp^M} & \gr\Gamma\\
\gr\Gamma^{\sf T} & \sig_{\bet^N}
\end{array}\right) \; , \label{fulsim}
\ee where $\gr\Gamma$ is a $2M\times 2N$ real matrix formed by
identical $2\times 2$ blocks $\gr\gamma$. Clearly, $\gr\Gamma$ is
responsible for the correlations existing between the $M$-mode and
the $N$-mode parties. Once again, the identity of the submatrices
$\gr\gamma$ is a consequence of the local invariance under mode
exchange, internal to the $M$-mode and $N$-mode parties. States of
the form of \eq{fulsim} will be henceforth referred to as
``bisymmetric'' \cite{adescaling,unitarily}. By evaluating the
symplectic spectrum of such a bisymmetric state, and by comparing it
with the spectra of the reduced fully symmetric CMs, one can show
the following central result \cite{unitarily}, which applies to all
(generally mixed) bisymmetric Gaussian states, and is  somehow
analogous to --- but independent of --- the phase-space Schmidt
decomposition of pure Gaussian states (and of mixed states with
fully degenerate symplectic spectrum). {\it The bisymmetric
$(M+N)$-mode Gaussian state with CM $\sig$, \eq{fulsim} can be
brought, by means of a local unitary (symplectic) operation with
respect to the $M\times N$ bipartition with reduced CMs
$\sig_{\alpha^M}$ and $\sig_{\beta^N}$, to a tensor product of
single-mode uncorrelated states and of a two-mode Gaussian state
comprised of one mode from the $M$-mode block and one mode from the
$N$-mode block.}

We will explore the consequences of such a unitary localization on
the entanglement properties of bisymmetric states, in the next
sections. Let us just note that fully symmetric Gaussian states,
\eq{fscm}, are special instances of bisymmetric states with respect
to any global bipartition of the modes.

\section{Theory of bipartite entanglement for Gaussian states.}

\subsection{Entanglement and nonlocality.}

When two quantum systems come to interact, some quantum
correlation is established between the two of them.
This correlation persists even when the interaction is
switched off and the two systems are spatially
separated\footnote{Entanglement can be also created without direct
interaction between the subsystems, via the common interaction
with other {\it assisting} parties . This and related mechanisms
define the so-called entanglement swapping \cite{Telep}.}.
If we measure a local observable on the
first system, its state collapses in an eigenstate of that
observable. In the ideal case of no environmental decoherence,
the state of the second system is instantly modified. The
question then arises about what is the mechanism responsible
for this ``spooky action at a distance'' \cite{EPR35}.

Suppose we have a bipartite or multipartite quantum state.
An apparently innocent question like
\begin{quote}
\centering {\emph{Does this state contain quantum correlations?}}
\end{quote}
turns out to be very hard to answer
\cite{heiss,audenplenio,PlenioVirmani}. The first step towards
a solution concerns a
basic understanding of what such a question really means.

One may argue that a system contains quantum correlations if the
observables associated to the different subsystems are correlated,
and their correlations cannot be reproduced with purely classical
means. This implies that some form of inseparability or
non factorizability is necessary to properly take into account those
correlations. For what concerns globally {\em pure states} of the
composite quantum system, it is relatively easy to control if the
correlations are of genuine quantum nature. In particular, it is
enough to check if a Bell-CHSH inequality \cite{Bell64,CHSH69} is
violated \cite{Gisin91}, to conclude that a pure quantum state is
entangled. There are in fact many different criteria to characterize
entanglement, but all of them are practically based on equivalent
forms of nonlocality in pure quantum states.

This scheme fails with {\em mixed states}. Contrary to a pure state,
a statistical mixture can be prepared in (generally infinitely)
many different ways. Being it impossible to reconstruct the
original preparation of the state, one cannot extract all the
information it contains. Accordingly, there is no completely
general and practical criterion to decide whether correlations in a
mixed quantum state are of classical or quantum nature. Moreover,
different manifestations of quantum inseparability are in general
{\em not} equivalent. For instance, one pays more (in units of
singlets $(\ket{00}+\ket{11})/\sqrt2$) to create an entangled mixed
state $\ro$ (entanglement cost \cite{Bennett96pra}) than to
reconvert $\ro$ into a product of singlets
(distillable entanglement \cite{Bennett96pra}) via local operations
and classical communication (LOCC) \cite{Synak}. An important example
has been introduced by Werner \cite{Werner89}, who defined
a parametric family of mixed states (known as Werner
states) which, in some range of the parameters, are entangled
(inseparable) without violating any Bell inequality on local
realistic models.
These states thus admit a description in terms of local hidden
variables. It is indeed an open question in quantum information
theory to prove whether any entangled state violates some kind
of Bell-type inequality \cite{QIProb,Terhal00}.

In fact, entanglement and nonlocality are different resources
from an operational point of view \cite{GisinNJP}.
This can be understood within the general framework
of no-signalling theories which exhibit stronger nonlocal
features than those of quantum mechanics. Let us briefly recall
what is intended by nonlocality according to Bell \cite{BBell}: there
exists in Nature a channel that allows one to distribute
correlations between distant observers, such that  the correlations
are not already established at the source, and  the correlated
random variables can be created in a configuration of space-like
separation, \ie no normal signal (assuming no superluminal
transmission) can be the cause of the correlations
\cite{EPR35,Bell64}. A convenient understanding of the
phenomenon of nonlocality is then given by quantum mechanics, which
describes the channel as a pair of entangled particles. But such
interpretation is not the only possible one. In recent years, there
has been a growing interest in constructing alternative
descriptions of this channel, mainly assuming a form of
communication \cite{tapp}, or the use of
an hypothetical ``nonlocal machine'' \cite{prbox} able to violate
the CHSH inequality \cite{CHSH69} up to its algebraic value of $4$
(while the local realism threshold is $2$ and the maximal violation
admitted by quantum mechanics is $2\sqrt2$, the Cirel'son bound
\cite{cirel}). Looking at these alternative descriptions allows
to quantify how powerful quantum mechanics is by
comparing its performance with that of other resources
\cite{Gisinonlocal}.

\subsection{Entanglement of pure states.}

It is now well understood that entanglement in a pure bipartite
quantum state $\ps$ is equivalent to the degree of mixedness of each
subsystem. Accordingly, it can be properly quantified by the {\em
entropy of entanglement} $E_V(\ps)$, defined as the von Neumann
entropy, \eq{QM:SV}, of the reduced density matrices
\cite{Bennett96},
\begin{equation}\label{E:E}
E_V(\ps) = S_V(\ro_1) = S_V(\ro_2) = -\sum_{k=1}^d \lambda_k^2 \,
\log \lambda_k^2\,.
\end{equation}
The entropy of entanglement is by definition invariant under local
unitary operations
\begin{equation}\label{E:Elocalinv}
E_V\Big((\op{U}_1 \otimes \op{U}_2) \ps\Big) = E_V\Big(\ps\Big)\,,
\end{equation}
and it can be shown \cite{Popescu97} that $E_V(\ps)$ cannot increase
under LOCC performed on the state $\ps$: this is a fundamental
physical requirement as it reflects the fact that entanglement
cannot be created via LOCC only \cite{VedralPRK,Vidal00}. It can be
formalized as follows. Let us suppose, starting with a state $\ps$
of the global system $\s$, to perform local measurements on  $\s_1$
and $\s_2$, and to obtain, after the measurement, the state
$\ket{\ph_1}$ with probability $p_1$, the state $\ket{\ph_2}$ with
probability $p_2$, and so on. Then
\begin{equation}\label{E:EnoLO}
E_V(\ps) \ge \sum_k p_k E_V(\ket{\ph_k})\,.
\end{equation}
Note that entanglement cannot increase \emph{on average}, that is
nothing prevents, for a given $k$, that $E_V(\ket{\ph_k})
> E_V(\ps)$. The concept of {\em entanglement distillation}
is based on this fact \cite{Bennett96prl,Bennett96pra,Gisin96}:
with a probability $p_k$, it is possible to increase entanglement
via LOCC manipulations.

\subsection{Entanglement of mixed states.}

A mixed state $\ro$ can be decomposed as a convex combination of
pure states,
\begin{equation}\label{E:rodec}
\ro = \sum_k p_k \ketbra{\psi_k}{\psi_k}\,.
\end{equation}
\eq{E:rodec} tells us how to create the state described by the
density matrix $\ro$: we have to prepare the state  $\ket{\psi_1}$
with probability  $p_1$, the state $\ket{\psi_2}$ with probability
$p_2$, etc. For instance, we could collect $N$ copies ($N \gg 1$) of
the system, prepare $n_k \simeq N p_k$ of them in the state
$\ket{\psi_k}$, and pick a copy at random.

The difficulty lies in the fact that the decomposition of \eq{E:rodec} is not
unique: apart from pure states, there exist infinitely many
decompositions of a generic $\ro$, meaning that the mixed state can
be prepared in infinitely many different ways. This has important
consequences on the determination of mixed-state entanglement.
Suppose that for a bipartite system in a mixed state we detect,
by local measurements, the presence
of correlations between the two subsystems. Given the ambiguity on
the state preparation, we cannot know {\em a priori} if those
correlations arose from a quantum interaction between the subsystems
(meaning entanglement) or were induced by means of LOCC (meaning
classical correlations). It is thus clear that a mixed state can be
defined separable (classically correlated) if there exists at least
one way of engineering it by LOCC; on the other hand it is entangled
(quantumly correlated) if, among the infinite possible procedures
for its preparation, there is no one which relies on LOCC alone
\cite{Werner89}.

However,deciding separability
according to the above definition would imply checking all the
infinitely many decompositions of a state $\ro$ and looking for at
least one, expressed as a convex combination of product states, to
conclude that the state is not entangled. This is clearly
impractical. For this reason, several {\em operational} criteria
have been developed in order to detect entanglement in mixed quantum
states \cite{CiracPrimer,BrussJMP,BInfo}. Some of them, of special
relevance to Gaussian states of CV systems, are discussed in the
following.

\subsection{Separability and distillability of Gaussian states.}

\subsubsection{PPT criterion.}\label{SecPPTG}

One of the most powerful results to date in the context of
separability criteria is the Peres--Horodecki condition
\cite{Peres96,Horodecki96}. It is based on the operation of {\em
partial transposition} of the density matrix of a bipartite system,
obtained by performing transposition  with respect to the degrees of
freedom of one subsystem only. Peres criterion states that, if a
state $\ro_s$ is separable, then its partial transpose $\ro_s\PT1$
(with respect \eg to subsystem $\s_1$) is a valid density matrix, in
particular positive semidefinite, $\ro_s\PT1 \ge 0.$ Obviously, the
same holds for $\ro_s\PT2$. Positivity of the partial transpose (PPT)
is therefore a necessary condition for separability \cite{Peres96}.
The converse (\ie $\ro\PT1\ge0 \Rightarrow \ro$ separable) is in
general false, but is has been proven true for low-dimensional
systems, specifically bipartite systems with Hilbert state space of
dimensionality $2 \times 2$ and  $2 \times 3$. In these cases the
PPT property is equivalent to separability \cite{Horodecki96}. For higher
dimensional tensor product structures of Hilbert spaces, PPT entangled
states (with  $\ro\PT1 \ge 0$) have been shown to exist \cite{Horodecki97}.
These states are known as \emph{bound entangled} \cite{Horodecki98},
because their entanglement
cannot be distilled to obtain maximally entangled states. The
existence of bound entangled (undistillable) states with negative
partial transposition has been conjectured \cite{Dur00,DiVincenzo00},
but at present there is not yet evidence of this property \cite{QIProb}.

Recently, the PPT criterion has been generalized to continuous
variable systems by Simon  \cite{Simon00}, who showed how the operation
of transposition acquires in infinite-dimensional Hilbert
spaces an elegant geometric interpretation in terms of time
inversion (mirror reflection of the momentum operator). The
PPT criterion is necessary and sufficient for the
separability of all $(1 \times N)$-mode Gaussian states
\cite{Simon00,Duan00,werewolf}. Under
partial transposition, the CM $\sig_{A|B}$, where subsystem $\s_A$
groups $N_A$ modes, and subsystem $\s_B$ is formed by $N_B$ modes,
is transformed into a new matrix
\begin{equation}\label{cmpt}
\tilde{\sig}_{A|B} \equiv \gr\theta_{A|B}\ \sig_{A|B}\
\gr\theta_{A|B}\,,
\end{equation}
with \[\gr\theta_{A|B} = {\rm diag}
\{\underbrace{1,\,-1,\,1,\,-1,\,\ldots,\,1,-1}_{2N_A},\,
\underbrace{1,\,1,\,1,\,1,\,\ldots,\,1,\,1}_{2N_B}\}\,.\]
Referring to the notation of \eq{CM}, the partially transposed
matrix $\tilde{\sig}_{A|B}$ differs from $\sig_{A|B}$ by a sign flip
in the determinants of the intermodal correlation matrices, $\det
\eps_{ij}$, with modes $i \in \s_A$ and modes $j \in s_B$.

The PPT criterion yields that a Gaussian state
$\sig_{A|B}$ (with $N_A=1$ and $N_B$ arbitrary) is {\em separable}
if and only if the partially transposed $\tilde{\sig}_{A|B}$ is a
{\em bona fide} CM, that is it satisfies the uncertainty principle
\eq{bonfide},
\begin{equation}\label{bonfidept}
   \tilde{\sig}_{A|B} + i \Omega \ge 0\,.
\end{equation}
This property in turn
reflects the positivity of the partially transposed density matrix
$\varrho\PT{A}$ associated to the state $\varrho$. For Gaussian
states with $N_A >1$, not endowed with special symmetry
constraints, the PPT condition is only necessary for separability, as
bound entangled Gaussian states, whose entanglement is
undistillable, exist already in the instance
$N_A=N_B=2$ \cite{werewolf}.

We have demonstrated the existence of ``bisymmetric'' $(N_A +
N_B)$-mode Gaussian states for which PPT is again equivalent to
separability  \cite{unitarily}.  In view of the invariance of the
PPT criterion under local unitary transformations,
and considering the results of section
\ref{SecSymm} on the unitary localization of bisymmetric Gaussian
states, it is immediate to verity that the following property holds
\cite{unitarily}. {\it For generic $N_A \times N_B$-mode
bipartitions, the positivity of the partial transpose (PPT) is a
necessary and sufficient condition for the separability of
bisymmetric $(N_A+N_B)$-mode mixed Gaussian states of the form
\eq{fulsim}. In the case of fully symmetric mixed Gaussian states,
\eq{fscm}, of an arbitrary number of bosonic modes, PPT is
equivalent to separability across any global bipartition of the modes.}

This statement generalizes to multimode
bipartitions the equivalence between separability and PPT for
$1 \times N$ bipartite Gaussian states \cite{werewolf}.
In particular, it implies that no bisymmetric bound entangled
Gaussian states can exist \cite{werewolf,giedkeqic01} and all the
$N_A \times N_B$ multimode block entanglement of such states is
distillable. Moreover, it justifies the use of the negativity and
the logarithmic negativity as measures of entanglement for this
class of multimode Gaussian states \ref{ChapUniLoc}.

The PPT criterion has an elegant symplectic representation. The
partially transposed matrix $\tilde{\sig}$ of any $N$-mode Gaussian
$CM$ is still a positive and symmetric matrix. As such, it admits a
Williamson normal-mode decomposition \cite{williamson36},
\eq{willia}, of the form
\begin{equation}
\tilde{\sig}=S\T \tilde{\gr\nu} S \; , \label{williapt}
\end{equation}
where $S\in Sp_{(2N,\mathbb{R})}$ and $\tilde{\gr\nu}$ is the CM
\begin{equation}
\tilde {\gr{\nu}}=\bigoplus_{k=1}^{N}\left(\begin{array}{cc}
\tilde \nu_k&0\\
0&\tilde\nu_k
\end{array}\right) \, , \label{thermapt}
\end{equation}

The $N$ quantities $\tilde\nu_{k}$'s are the symplectic eigenvalues
of the partially transposed CM $\tilde{\sig}$. The symplectic
spectrum $\set{\nu_{k}}$ of $\sig$ encodes the structural and
informational properties of a Gaussian state. The partially
transposed spectrum $\set{\tilde\nu_{k}}$ encodes the
qualitative (and, to some extent, quantitative --- see section
\ref{secbibquantify}) characterization of entanglement in the state.
The PPT condition \pref{bonfidept}, \ie the uncertainty
relation for $\tilde{\sig}$, can be equivalently recast in terms of
the parameters $\tilde\nu_{k}$'s as
 \be
{\tilde\nu}_{k}\ge1 \; . \label{sympheispt} \ee

We can, without loss of generality, rearrange the modes of a
$N$-mode state such that the corresponding symplectic eigenvalues of
the partial transpose $\tilde{\sig}$ are sorted in ascending order
\[ \tilde\nu_-\equiv \tilde\nu_1 \le \tilde\nu_2 \le \ldots \le
\tilde\nu_{N-1} \le \tilde\nu_N \equiv \tilde \nu_+\,,\] in analogy
to what done in section \ref{SecSympHeis} for the spectrum of
$\sig$. With this notation, the PPT criterion across an arbitrary
bipartition reduces to $\tilde\nu_1 \ge 1$ for all separable
Gaussian states. If $\tilde\nu_1 < 1$, the corresponding
Gaussian state $\sig$ is entangled. The symplectic characterization
of physical and PPT Gaussian states is
summarized in Table \ref{CVtabGE}.

\begin{table}[t!]   \centering{
\begin{tabular}{c||c|c}
& Physical & Separable \\
\hline \hline
density matrix & ${\ro \ge 0}$ & $\overset{}{\ro\PT{A}\ge 0}$  \\ \hline
covariance matrix & $\sig + i \Omega \ge 0$ & $\overset{}{\tilde{\sig}
+ i \Omega \ge 0}$ \\ \hline
symplectic spectrum & $\nu_k \ge 1$ & $\overset{}{\tilde\nu_k \ge 1}$  \\
\hline
\end{tabular}}
\caption{Schematic comparison between the conditions of existence and
the conditions of separability for Gaussian states, as expressed in
different representations. The second column
qualifies the PPT condition, which is always implied by
separability, and equivalent to it in $1 \times N$, and in
$M \times N$ bisymmetric Gaussian states.} \label{CVtabGE}
\end{table}

The distillability problem for Gaussian states has been
solved in quite general terms \cite{giedkeqic01}:
the entanglement of any non-PPT
bipartite Gaussian state is distillable by LOCC. However, we remind
that this entanglement can be distilled only resorting to
non Gaussian LOCC \cite{browne}, since distilling Gaussian states
with Gaussian operations is impossible \cite{nogo1,nogo2,nogo3}.

\subsubsection{Additional criteria for separability.}

Let us briefly mention alternative characterizations of separability
for Gaussian states.

For a general Gaussian state of any $N_A\times N_B$ bipartition, a
necessary and sufficient condition states that a CM $\sig$
corresponds to a separable state if and only if there exists a pair
of CMs $\sig_{A}$ and $\sig_{B}$, relative to the subsystems $\s_A$
and $s_B$ respectively, such that the following inequality holds
\cite{werewolf}, $\sig \geq \sig_{A} \oplus \sig_{B}$. This
criterion is not very useful in practice. Alternatively, one can
introduce an operational criterion based on iterative applications
of a nonlinear map, that is independent of (and strictly stronger
than) the PPT condition \cite{Giedke01}, and completely qualifies
separability  for all bipartite Gaussian states.

Another powerful tool establish the separability of quantum states is
given by the so-called {\em entanglement witnesses}. A state $\ro$ is
entangled if and only if there exists a Hermitian operator $\W$ such
that $\Tr{\W\,\ro}<0$ and $\Tr{\W\,{\sigma}}\ge0$ for any state
$\op{\sigma} \in \dd$, where $\dd \subset \hh$ is the convex and
compact subset of separable states \cite{Horodecki96,Terhal00}.
The operator $\W$ is the {\em witness} responsible for detecting
entanglement in the state $\ro$. According to the Hahn-Banach
theorem, given a convex and compact set $\dd$ and given $\ro \not\in
\dd$, there exists an hyperplane which separates $\ro$ from $\dd$.
Optimal entanglement witnesses induce an hyperplane which is tangent
to the set $\dd$ \cite{Lewenstein00}. A sharper detection of
separability can be achieved by means of nonlinear entanglement
witnesses, curved towards the set $\dd$ \cite{NorbertoGuhne}.
A comprehensive characterization of linear and
nonlinear entanglement witnesses is available for CV systems
\cite{illuso}, and can be efficiently applied to the detection of
separability in arbitrary Gaussian states, both in the bipartite and
in the multipartite context.

Finally, several operational criteria have been developed, that are
specially useful in experimental settings. They are based on the
violation of inequalities involving combinations of variances of
canonical operators, and their validity ranges from the two-mode
\cite{Duan00} to the multimode setting \cite{vloock03}.

\subsection{Quantification of bipartite entanglement in Gaussian states.}
\label{secbibquantify}

The question of the quantification of bipartite entanglement for
general (pure {\em and} mixed) cannot
be considered completely solved yet. We have witnessed a proliferation
of entanglement measures, each one motivated by specific contexts in which
quantum correlations play a central role, and accounting for different,
and in some cases inequivalent, operational characterizations
and orderings of entangled states. Detailed expositions on the
subject can be found \eg
in Refs.~\cite{BrussJMP,QIC01,PlenioVirmani,ChristandlPHD}.

The first natural generalization of the quantification of entanglement
to mixed states is certainly the {\em
entanglement of formation} $E_F(\ro)$ \cite{Bennett96pra}, defined
as the convex-roof extension \cite{OsborneCROOF} of the entropy of
entanglement \eq{E:E}, \ie the weighted average of pure-state
entanglement,
\begin{equation}\label{E:EF}
E_F(\ro) = \min_{\{p_k,\,\ket{\psi_k}\}} \sum_k
p_k\,E_V(\ket{\psi_k})\,,
\end{equation}
minimized over all decompositions of the mixed state $\ro=\sum_k p_k
\ketbra{\psi_k}{\psi_k}$. This is clearly an optimization problem of
formidable difficulty, and an explicit solution is known only
for the mixed states of two qubits \cite{Wootters98}, and
for highly symmetric states like Werner states and isotropic states
in arbitrary dimension \cite{Terhal01,VollWerner01}. In CV systems,
an explicit expression for the entanglement of formation is
available only for symmetric, two-mode Gaussian states \cite{giedke03}.
To date, the additivity of the entanglement of
formation remains an open problem \cite{QIProb}.

\subsubsection{Negativities.}\label{secnega} An important class of entanglement
monotones is defined by the negativities, which quantify the
violation of the PPT criterion for separability (see section
\ref{SecPPTG}), \ie how much the partial transposition of $\ro$
fails to be positive. The {\em negativity} $\N(\ro)$
\cite{Zyczkowski98,EisertPHD} is defined as
\begin{equation}\label{E:N}
\N(\ro) = \frac{\norm{\ro\PT{i}}_1-1}{2}\,,
\end{equation}
where
\begin{equation}\label{E:tracenorm}
\|\op{O}\|_1 = \tr{\sqrt{\adj{\op{O}}\op{O}}}
\end{equation}
is the trace norm of the operator $\op O$. The negativity has
the advantage of being a computable measure of entanglement, being
\begin{equation}\label{E:Nsum}
\N(\ro)=\max\set{0,-{\sum_k \lambda_k^-}}\,,
\end{equation}
where the $\set{\lambda_k^-}$'s are the negative eigenvalues of the
partial transpose.

The negativity can be defined for CV systems as well
\cite{VidalWerner02}, even though a related measure is more often
used, the \emph{logarithmic negativity} $E_\N(\ro)$
\cite{VidalWerner02,EisertPHD},
\begin{equation}\label{E:EN}
E_{\N}(\ro)= \log\|\ro\PT{i}\|_{1} = \log \left[1+2\N(\ro)\right]\,.
\end{equation}
The logarithmic negativity is additive and, despite not being
convex, is a full entanglement monotone under LOCC
\cite{Plenio05}; it is an upper bound for the distillable
entanglement \cite{EisertPHD}, $E_\N(\ro) \ge E_D(\ro)$, and coincides
with the entanglement cost under operations preserving the
positivity of the partial transpose \cite{Eisert03}. Both
the negativity and the logarithmic negativity fail to be continuous
in trace norm on infinite-dimensional Hilbert spaces; however, this
problem can be circumvented by restricting to physical states of
finite mean energy \cite{eisert02}.

The great advantage of the negativities is that they are easily {\em
computable} for general Gaussian states; they provide a proper
quantification of entanglement in particular for arbitrary $1 \times
N$ and bisymmetric $M \times N$ Gaussian states, directly
quantifying the degree of violation of the necessary and sufficient
PPT criterion for separability, \eq{sympheispt}. Following
\cite{VidalWerner02,SeralePHD,extremal,adescaling}, the negativity
of a Gaussian state with CM $\sig$ is given by \be
\N(\sig)=\left\{\begin{array}{l} \frac12\left(\prod_k
{\tilde{\nu}_k}^{-1}-1\right) , \quad {\rm for}\;
k: \tilde{\nu}_k<1   \; . \\
\\
0 \quad \,{\rm if}\; \tilde{\nu}_i\ge 1 \; \forall\, i \; .
\end{array}\right.\label{negagau}
\ee Here the set $\{\tilde{\nu}_k\}$ is constituted by the
symplectic eigenvalues of the partially transposed CM
$\tilde{\gr{\sigma}}$. Accordingly, the logarithmic negativity reads
\be E_\N(\sig)=\left\{\begin{array}{l} -\sum_k {\log{\tilde{\nu}_k}}
, \quad {\rm for}\;
k : \tilde{\nu}_k<1   \; . \\
\\
0 \quad \,{\rm if}\; \tilde{\nu}_i\ge 1 \; \forall\, i \; .
\end{array}\right.\label{lognegau}
\ee

The following lemma is quite useful
for the interpretation and the computation of the negativities in
Gaussian states \cite{serafozziprl}. In a  $(N_A +N_B)$-mode
Gaussian state with CM
$\sig_{A|B}$, at most
\begin{equation}\label{littlelemm}
N_{\min} \equiv \min\{N_A,\,N_B\}
\end{equation}
symplectic eigenvalues $\tilde\nu_k$ of the partial transpose $\tilde{\sig}_{A|B}$ can
violate the PPT inequality~\pref{sympheispt}  with respect to a $N_A
\times N_B$ bipartition. Thanks to this result, in all $1 \times N$
Gaussian states and in all bisymmetric $M \times N$ Gaussian states
(whose symplectic spectra exhibit degeneracy, see section
\ref{SecSymm}), the negativities are quantified in terms of
the smallest symplectic eigenvalue $\tilde\nu_-$ of the partially
transposed CM alone. For $\tilde \nu_- \ge 1$ the state is
separable, otherwise it is entangled; the smaller $\tilde \nu_-$,
the more entangled is the corresponding Gaussian
state. In the limit of a vanishing partially transposed symplectic
eigenvalue, $\tilde \nu_- \rightarrow 0$, the negativities grow
unboundedly. In the special instance of two-mode Gaussian states,
such a result had been originally derived in \cite{prl,extremal}.

\subsubsection{Gaussian convex-roof extended measures.}\label{SecGEMS}

It is possible to define a family of entanglement measures
exclusively defined for Gaussian states of CV systems
\cite{ordering}. The formalism of {\em Gaussian entanglement
measures} (Gaussian EMs) has been introduced in Ref.~\cite{GEOF}
where the ``Gaussian entanglement of formation'' has been defined
and analyzed. The framework developed in
Ref.~\cite{GEOF} is general and allows to define generic Gaussian
EMs of bipartite entanglement by applying the Gaussian convex roof,
that is, the convex roof over pure Gaussian decompositions only, to
any {\em bona fide} measure of bipartite entanglement defined for
pure Gaussian states.

The original motivation for the introduction of Gaussian EMs stems
from the fact that the entanglement of formation
\cite{Bennett96pra}, defined by \eq{E:EF}, involves a nontrivial
minimization of the pure-state entropy of entanglement over convex
decompositions of bipartite mixed Gaussian states in ensemble of
pure states. These pure states may be, in principle, non Gaussian
states of CV systems, thus rendering the analytical solution of the
optimization problem in \eq{E:EF} extremely difficult even in the
simplest instance of one mode per party. Nevertheless, in the special
subset of two-mode symmetric mixed Gaussian states, the optimal
convex decomposition of \eq{E:EF} has been exactly determined, and
is realized in terms of pure Gaussian states \cite{giedke03}. Apart
from that case (which will be discussed in section
\ref{SecEOFGauss}), the determination of the entanglement of formation
of nonsymmetric two-mode Gaussian states (and more general Gaussian states)
is an open problem in the theory of entanglement \cite{QIProb}.
However, inspired by the results achieved on two-mode symmetric states,
one can at first try to restrict the problem only to decompositions into
pure Gaussian states. The resulting measure, named as Gaussian entanglement
of formation in Ref.~\cite{GEOF}, is an upper bound to the true
entanglement of formation and coincides with it for symmetric
two-mode Gaussian states.

In general, we can define a Gaussian EM $G_E$ as follows. For any
pure Gaussian state $\psi$ with CM $\sig^p$, one has
\begin{equation}\label{Gaussian EMp}
G_E (\sig^p) \equiv E(\psi)\,,
\end{equation}
where $E$ can be {\em any} proper measure of entanglement of pure
states, defined as a monotonically increasing function of the
entropy of entanglement ({\ie}the von Neumann entropy of the reduced
density matrix of one party).

For any mixed Gaussian state $\varrho$ with CM $\sig$, one has
\cite{GEOF}
\begin{equation}\label{Gaussian EMm}
G_E (\sig) \equiv \inf_{\sig^p \le \sig} G_E(\sig^p)\,.
\end{equation}
If the function $E$ is taken to be exactly the entropy of
entanglement, \eq{E:E}, then the corresponding Gaussian EM is known
as {\em Gaussian entanglement of formation} \cite{GEOF}.
Any Gaussian EM is an entanglement monotone under
Gaussian LOCC. The proof given in Sec. IV of Ref.~\cite{GEOF} for
the Gaussian entanglement of formation automatically
extends to every Gaussian EM constructed via the Gaussian convex
roof of any proper measure $E$ of pure-state entanglement.

In general, the definition \eq{Gaussian EMm} involves an
optimization over all pure Gaussian states with CM $\sig^p$ smaller
than the CM $\sig$ of the mixed state whose entanglement one wishes
to compute. This corresponds to taking the Gaussian convex roof.
Despite being a simpler optimization problem than that appearing in
the definition \eq{E:EF} of the true entanglement of formation, the
Gaussian EMs cannot be expressed in a simple closed form, not even
in the instance of (nonsymmetric) two-mode Gaussian states.
Some recent results on this issue have been obtained in \cite{ordering},
and will be discussed in section \ref{secorder}, in relation with
the problem of the ordering of quantum states according to different
measures of entanglement.

\section{Entanglement of two-mode Gaussian states.} \label{Chap2M}

We now discuss the characterization of the prototypical entangled states
of CV systems, \ie the two-mode Gaussian states. This includes the
explicit determination of the negativities and their relationship
with global and marginal entropic measures \cite{prl,extremal,polacchi},
and the evaluation of the Gaussian measures of entanglement
\cite{ordering}. We will then compare the two families of measures
according to the ordering that they establish on entangled states.

\subsection{Symplectic parametrization of two-mode Gaussian states.}
\label{SecSympParam}

To study entanglement and informational properties (like global and
marginal entropies) of two-mode Gaussian states, we can consider
without loss of generality states whose CM $\sig$ is in the
$\sy{2}\oplus\sy{2}$-invariant standard form derived in
Refs.~\cite{Simon00,Duan00}. Let us recall it here for the sake of
clarity,
\begin{equation}\label{stform2}
 {\sig}= \left(\begin{array}{cc}
{\alp}&{\gr\gamma}\\
{\gr\gamma}^{\sf T}&{\bet}
\end{array}\right) = \left(\begin{array}{cccc}
a&0&c_{+}&0\\
0&a&0&c_{-}\\
c_{+}&0&b&0\\
0&c_{-}&0&b
\end{array}\right)\,.
\end{equation}

For two-mode states, the uncertainty principle \ineq{bonfide} can be
recast as a constraint on the $Sp_{(4,{\mathbb R})}$ invariants
(invariants under global, two-mode symplectic operations) ${\rm
Det}\sig$ and $\Delta(\sig)={\rm Det}{\alp}+\,{\rm Det}{\bet}+2
\,{\rm Det}{\gr\gamma}$ \cite{SymplecticInvariants},
\begin{equation}
\Delta(\sig)\le1+\,{\rm Det}\sig \label{sepcomp}\; .
\end{equation}

The symplectic eigenvalues of a two-mode Gaussian state will be
denoted as $\nu_{-}$ and $\nu_{+}$, with $\nu_{-}\le \nu_{+}$, with
the uncertainty relation \pref{sympheis} reducing to \be
\label{symptwo} \nu_{-}\ge 1 \; . \ee A simple expression for the
$\nu_{\mp}$ can be found in terms of the two $Sp_{(4,\mathbb{R})}$
invariants \cite{VidalWerner02,SymplecticInvariants}
\cite{prl,extremal}
\begin{equation}
2{\nu}_{\mp}^2=\Delta(\sig)\mp\sqrt{\Delta^2(\sig) -4\,{\rm
Det}\,\sig} \, . \label{sympeig}
\end{equation}

According to \eq{stform2}, two-mode Gaussian states can be
classified in terms of  their four standard form covariances $a$,
$b$, $c_{+}$, and $c_{-}$. It is relevant to provide a
reparametrization of standard form states in terms of symplectic
invariants which admit a direct interpretation for generic Gaussian
states \cite{prl,extremal,polacchi}. Namely, the parameters of
\eq{stform2} can be determined in terms of the two local symplectic
invariants
\begin{equation}\label{mu12}
\mu_1 = (\det\gr\alpha)^{-1/2} = 1/a\,,\quad \mu_2 =
(\det\gr\beta)^{-1/2} = 1/b\,,
\end{equation}
which are the marginal purities of the reduced single-mode states,
and of the two global symplectic invariants
\begin{equation}\label{globinv}
\mu = (\det\gr\sigma)^{-1/2} =
[(ab-c_{+}^2)(ab-c_{-}^2)]^{-1/2}\,,\quad \Delta =
a^2+b^2+2c_+c_-\,,
\end{equation}
which are, respectively, the global purity \eq{purgau} and the seralian
\eq{seralian}. Eqs.~{\rm(\ref{mu12}--\ref{globinv})}
can be inverted to give a physical parametrization of two-mode
states in terms of the four independent parameters
$\mu_1,\,\mu_2,\,\mu$, and $\Delta$. This parametrization is
particularly useful for the evaluation of entanglement \cite{prl,extremal}.


\subsection{Qualifying and quantifying two-mode entanglement.}
\subsubsection{Partial transposition and negativities.}
\label{SecNega2M}

The PPT condition for separability, \eq{bonfidept} has obviously a
very simple form for two-mode Gaussian states. In terms of
symplectic invariants, partial transposition corresponds to flipping
the sign of ${\rm Det}\,\gr{\gamma}$,
\begin{equation}\label{GS:sig2PT}
\sig =\left(%
\begin{array}{cc}
\gr\alpha & \gr\gamma \\
\gr\gamma\T & \gr\beta \\
\end{array}%
\right)\quad \overset{\ro\,\rightarrow\,\ro\PT{i}}
{\overrightarrow{\quad\quad\quad}} \quad
\tilde{\sig} =\left(%
\begin{array}{cc}
\gr\alpha & \tilde{\gr\gamma} \\
\tilde{\gr\gamma}\T & \gr\beta \\
\end{array}%
\right)\,,
\end{equation}
with $\det\tilde{\gr\gamma} = - \det\gr\gamma$. For a standard form
CM \eq{stform2}, this simply means $c_+ \rightarrow c_+$, $c_-
\rightarrow - c_-$. Accordingly, the seralian
$\Delta=\det\alp+\det\bet+2\,\det\gr\gamma$, \eq{seralian}, is
mapped, under partial transposition, into
\begin{eqnarray}
\tilde{\Delta}&=&\det\alp+\det\bet+2\,\det\tilde{\gr\gamma}
=\det\alp+\det\bet-2\,\det\gr\gamma \nonumber \\ &=&\Delta-4\,{\rm
Det}\,\gr{\gamma} = -\Delta + 2/\mu_1^2 + 2/\mu_2^2\,.
\label{deltatilde}
\end{eqnarray}
From \eq{sympeig}, the symplectic eigenvalues of the partial
transpose $\tilde{\gr{\sigma}}$ of a two-mode CM $\sig$ are promptly
determined in terms of symplectic invariants,
\cite{SymplecticInvariants}
\begin{equation}
2\tilde{\nu}_{\mp}^2 = \tilde{\Delta}\mp\sqrt{\tilde{\Delta}^2
-{\frac{4}{\mu^2}}}\,. \label{n1}
\end{equation}
The PPT criterion is then reexpressed by the following inequality
\begin{equation}
\tilde\Delta\le1+1/\mu^2 \label{sepcomppt}\,,
\end{equation}
equivalent to separability. The state
$\gr{\sigma}$ is separable if and only if
$\tilde{\nu}_{-}\ge 1$.
Accordingly, the logarithmic negativity \eq{lognegau} is a
decreasing function of $\tilde{\nu}_{-}$ alone,
\begin{equation}\label{en}
E_{\N}=\max\{0,-\log\,\tilde{\nu}_{-}\}\,,
\end{equation}
as for the biggest symplectic eigenvalue of the partial transpose
one has $\tilde{\nu}_{+} > 1$ for all two-mode Gaussian states
\cite{prl,extremal}.

Note that from
Eqs.~{\rm(\ref{stform2},\ref{sepcomp},\ref{deltatilde},\ref{sepcomppt})}
the following necessary condition for two-mode entanglement follows
\cite{Simon00},
\begin{equation}\label{detgammanegative}
\sig\hbox{ entangled}\quad \Rightarrow \quad \det{\gr\gamma} <0\,.
\end{equation}

\subsubsection{Entanglement of formation for symmetric states.}
\label{SecEOFGauss}

The optimal convex decomposition involved in the definition
\eq{E:EF} of the entanglement of formation \cite{Bennett96pra}
has been remarkably solved in the special instance of
two-mode symmetric mixed Gaussian states [\ie with
$\det\gr\alpha=\det\gr\beta$ in \eq{stform2}], and turns out to be
Gaussian. Namely, the absolute minimum is realized within the set of
pure two-mode Gaussian states \cite{giedke03}, yielding \be E_F =
\max\left[ 0,h(\tilde{\nu}_{-}) \right] \; , \label{eofgau} \ee with
\begin{equation}\label{hentro}
h(x)=\frac{(1+x)^2}{4x}\log \left[\frac{(1+x)^2}{4x}\right]-
\frac{(1-x)^2}{4x}\log \left[\frac{(1-x)^2}{4x}\right].
\end{equation}
Such a quantity is, again, a monotonically decreasing function of
the smallest symplectic eigenvalue $\tilde{\nu}_{-}$ of the partial
transpose $\tilde{\sig}$ of a two-mode symmetric Gaussian CM $\sig$,
thus providing a quantification of the entanglement of symmetric
states {\em equivalent} to the one provided by the negativities.
Lower bounds on the entanglement of formation have been derived
for nonsymmetric two-mode Gaussian states \cite{rigesc04}.

As a consequence of the equivalence between negativities and Gaussian
measures of entanglement in symmetric two-mode Gaussian states,
it is tempting to conjecture
that there exists a unique quantification of entanglement for all
two-mode Gaussian states, embodied by the smallest symplectic
eigenvalue $\tilde \nu_-$ of the partially transposed CM, and that
the different measures simply provide trivial rescalings of the same
unique quantification. In particular, the {\em ordering} induced on
the set of entangled Gaussian state is uniquely defined for the
subset of symmetric two-mode states, and it is independent of the
chosen measure of entanglement. However, in Sec. \ref{secorder} we
will indeed show, within the general framework of Gaussian measures
of entanglement (see section \ref{SecGEMS}), that different
families of entanglement monotones induce, in general, different
orderings on the set of nonsymmetric Gaussian states, as
demonstrated in \cite{ordering}.

\subsubsection{Gaussian measures of entanglement: geometric framework.}

The problem of evaluating Gaussian measures of entanglement (Gaussian
EMs) for a generic two-mode Gaussian state has been solved in
Ref.~\cite{GEOF}. However, the explicit result contains rather
cumbersome expressions, involving the solutions of a
fourth-order algebraic equation. As such, they were judged of no
particular insight to be reported explicitly by the authors
of Ref.~\cite{GEOF}.

We recall here the computation procedure \cite{ordering}.
For any two-mode Gaussian state with CM $\sig
\equiv \sig_{sf}$ in standard form \eq{stform2}, a generic Gaussian
EM $G_E$ is given by the entanglement $E$ of the least entangled
pure state with CM $\sig^p \le \sig$, see \eq{Gaussian EMm}.
Denoting by $\gamma_q$ (respectively $\gamma_p$) the $2 \times 2$
submatrix obtained from $\sig$ by canceling the even (resp. odd)
rows and columns, we have
\begin{equation}\label{cpcq}
\gamma_q = \left(
\begin{array}{ll}
a & c_+ \\
c_+ & b \\
\end{array}
\right)\,,\quad \gamma_p = \left(
\begin{array}{ll}
a & c_- \\
c_- & b \\
\end{array}
\right)\,.
\end{equation}
All the covariances relative to the ``position'' operators of the
two modes are grouped in $\gamma_q$, and analogously for the
``momentum'' operators in $\gamma_p$. The total CM can then be
written as a direct sum $\sig = \gamma_q \oplus \gamma_p$.
Similarly, the CM of a generic pure two-mode Gaussian state in
block-diagonal form  (it has been proven that the CM of the optimal
pure state has to be with all diagonal $2 \times 2$ submatrices as
well \cite{GEOF}) can be written as $\sig^p = \gamma_q^p \oplus
\gamma_p^p$, where the global purity of the state imposes
$(\gamma_p^p)^{-1} = \gamma_q^p \equiv \Gamma$. The pure states
involved in the definition of the Gaussian EM must thus fulfill the
condition
\begin{equation}\label{rim}
\gamma_p^{-1} \le \Gamma \le \gamma_q\,.
\end{equation}

This problem admits an interesting geometric formulation
\cite{GEOF}. Writing the matrix $\Gamma$ in the basis constituted by
the identity matrix and the three Pauli matrices,
\begin{equation}\label{Gamma} \Gamma = \left(
\begin{array}{cc}
x_0 + x_3 & x_1 \\
x_1 & x_0 - x_3 \\
\end{array}
\right)\,,
\end{equation}
the expansion coefficients $(x_0,x_1,x_3)$ play the role of
coordinates in a three-dimensional Minkowski space. In
this picture, for example, the rightmost inequality in \eq{rim} is
satisfied by matrices $\Gamma$ lying on a cone, which is equivalent
to the (backwards) light cone of $C_q$ in the Minkowski space; and
similarly for the leftmost inequality. Indeed, one can show that,
for the optimal pure state $\sig^p_{opt}$ realizing the minimum in
\eq{Gaussian EMm}, the two inequalities in \eq{rim} have to be
simultaneously saturated \cite{GEOF}. From a geometrical point of
view, the optimal $\Gamma$ has then to be found on the rim of the
intersection of the forward and the backward cones of
$\gamma_p^{-1}$ and $\gamma_q$, respectively. This is an ellipse,
and one is left with the task of minimizing the entanglement $E$ of
$\sig^p = \Gamma \oplus \Gamma^{-1}$ (see \eq{Gaussian EMp}) for
$\Gamma$ lying on this ellipse.\footnote{The geometric picture
describing the optimal two-mode state which enters in the
determination of the Gaussian EMs is introduced in \cite{GEOF}. A
more detailed discussion, including the explicit expression of the
Lorentz boost needed to move into the plane of the ellipse, can be
found in \cite{noteole}.}

We recall that any pure two-mode
Gaussian state $\sig^p$ is locally equivalent to a two-mode squeezed
state with squeezing parameter $r$, described by the CM of \eq{tms}.
The following statements are then equivalent: (i) $E$ is a
monotonically increasing function of the entropy of entanglement;
(ii) $E$ is a monotonically increasing function of the single-mode
determinant $m^2\equiv\det\gr\alpha \equiv \det\gr\beta$ (see
\eq{stform2}); (iii) $E$ is a monotonically decreasing function of
the local purity $\mu_i\equiv\mu_1\equiv\mu_2$ (see \eq{purgau});
(iv) $E$ is a monotonically decreasing function of the smallest
symplectic eigenvalue $\tilde\nu_-^p$ of the partially transposed CM
$\tilde{\sig}^p$; (v) $E$ is a monotonically increasing function of
the squeezing parameter $r$. This chain of equivalences is
immediately proven by simply recalling that a pure state is
completely specified by its single-mode marginals, and that for a
single-mode Gaussian state there is a unique symplectic invariant
(the determinant), so that all conceivable entropic quantities are
monotonically increasing functions of this invariant, as shown in
section \ref{SecEntroG} \cite{extremal}. In particular, statement
(ii) allows us to minimize directly the single-mode determinant over
the ellipse:
\begin{equation}\label{mdef}
m^2 = 1 + \frac{x_1}{\det \Gamma}\,,
\end{equation}
with $\Gamma$ given by \eq{Gamma}.

To simplify the calculations, one can move to the plane of the
ellipse with a Lorentz boost which preserves the relations between
all the cones; one can then choose the transformation so that the
ellipse degenerates into a circle (with fixed radius), and introduce
polar coordinates on this circle. The calculation of the Gaussian EM
for any two-mode Gaussian state is thus finally reduced to the
minimization of $m^2$ from \eq{mdef}, at given standard-form
covariances of $\sig$, as a function of the polar angle $\theta$ on
the circle \cite{noteole}.  After some tedious but straightforward
algebra, one finds \cite{ordering}
\begin{eqnarray}\label{mfunc}
&\!\!\!&\!\!\!\!\!\! \hspace*{-2cm} m^2_\theta (a,b,c_+, c_-)\ = \nonumber \\
&\!\!\!&\!\!\!\!\!\! \hspace*{-1.5cm}1 +
\left\{\left[c_+(ab-c_-^2)-c_-+\cos \theta \sqrt{\left[a -
b(ab-c_-^2)\right]\left[b-a(ab-c_-^2)\right]}\right]^2\right\}
\nonumber \\
&\!\!\!&\!\!\!\!\!\! \hspace*{-1.3cm}\times  \Bigg\{
2\left(ab-c_-^2\right)\left(a^2+b^2+2c_+c_- \right)  \nonumber \\
&\!\!\!&\!\!\!\!\!\! \hspace*{-1.3cm}-\ \frac{\cos
\theta\left[2abc_-^3+\left(a^2+
b^2\right)c_+c_-^2+\left(\left(1-2b^2\right)a^2+
b^2\right)c_--ab\left(a^2+b^2- 2\right)c_+\right]}{\sqrt{\left[a -
b(ab-c_-^2)\right]\left[b-a(ab-c_-^2)\right]}} \nonumber \\
&\!\!\!&\!\!\!\!\!\! \hspace*{-1.3cm}+\ \sin \theta\left(a^2-
b^2\right)\sqrt{1-\frac{\left[c_+(ab-c_-^2)+c_-\right]^2}{\left[a -
b(ab-c_-^2)\right]\left[b-a(ab-c_-^2)\right]}} \, \Bigg\}^{-1}\,,
\end{eqnarray}
where we have assumed $c_+ \ge |c_-|$ without loss of
generality. Therefore, for any entangled state, $c_+ > 0$
and $c_- < 0$, see \eq{detgammanegative}. The Gaussian EM (defined
in terms of the function $E$ on pure states, see \eq{Gaussian EMp})
of a generic two-mode Gaussian state coincides then with the
entanglement $E$ computed on the pure state with $m^2=m^2_{opt}$,
with $m^2_{opt} \equiv \min_\theta (m^2_\theta)$. Accordingly, the
symplectic eigenvalue $\tilde \nu_-$ of the partial transpose of the
corresponding optimal pure-state CM $\sig^{p}_{opt}$, realizing the
infimum in \eq{Gaussian EMm}, would read (see \eq{n1})
\begin{equation}\label{nutopt}
\tilde\nu_{- opt}^{p} \equiv \tilde \nu_-(\sig^{p}_{opt}) = m_{opt}
- \sqrt{m^2_{opt}-1}\,.
\end{equation}
As an example, for the Gaussian entanglement of formation
\cite{GEOF} one has \be \label{geofm} G_{E_F}(\sig) =
h\left[\tilde\nu_{- opt}^{p}(m^2_{opt})\right]\,,\ee with $h(x)$
defined by \eq{hentro}.

Finding the minimum of \eq{mfunc} analytically for a generic state
is an involved task, but numerical optimizations may be quite
accurate.

\subsection{Extremal entanglement versus information.}\label{secEntvsMix}

Here we review the characterization of entanglement of
two-mode Gaussian states with respect to its
relationship with the degrees of information associated with the
global state of the system, and with the reduced states of each of
the two subsystems.

As extensively discussed in the previous sections, the concepts of
entanglement and information encoded in a quantum state are
closely related. Specifically, for pure states bipartite
entanglement is equivalent to the lack of information (mixedness) of
the reduced state of each subsystem. For mixed states, each
subsystem has its own level of impurity, and moreover the global
state is itself characterized by a nonzero mixedness. Each of these
properties can be interpreted as information on the preparation of
the respective (marginal and global) states, as clarified in section
\ref{ParInfo}. Therefore, by properly accessing these degrees of
information it should be possible to deduce, at least to some extent,
the status of the correlations between the subsystems.

Indeed, the original studies revealed that,
at fixed global and marginal degrees of purity (or of
generalized entropies), the negativities of arbitrary (mixed)
two-mode Gaussian states are analytically constrained by rigorous
upper and lower bounds \cite{prl,extremal,polacchi}.
This follows by reparametrizing, as already
anticipated, the standard form CM \eq{stform2} in terms of the
invariants $\mu_1,\ \mu_2,\ \mu,\ \Delta$ and by observing that, at
fixed purities, the negativities are monotonically decreasing
function of $\Delta$. Further constraints
imposed on $\Delta$ by the uncertainty principle and by the
existence condition of the radicals involved in the
reparametrization, \be
\frac{2}{\mu} + \frac{(\mu_1 - \mu_2)^2}{\mu_1^2 \mu_2^2}
\le \Delta  \,\,\le\,\,  \min \left\{ \frac{(\mu_1 + \mu_2)^2}{\mu_1^2 \mu_2^2}
- \frac{2}{\mu} \; , \; 1+\frac{1}{\mu^2}\right\}  \, , \label{deltabnd}
\ee immediately lead to the definition of {\em extremally} --
maximally and minimally -- entangled Gaussian states at fixed global
and local purities. They are known, respectively, as
``GMEMS'' (saturating the leftmost inequality in \eq{deltabnd}),
alias nonsymmetric thermal squeezed states, and ``GLEMS''
(saturating the rightmost inequality in \eq{deltabnd}), alias mixed
states of partial minimum uncertainty \cite{prl,extremal}.
Nonsymmetric thermal squeezed states have also been proven to be
maximally entangled Gaussian mixed states at fixed global purity and
mean energy \cite{ziman}.

Summarizing, {\em the entanglement, quantified by the negativities,
of two-mode (mixed) Gaussian states is strictly bound from above and
from below by the amounts of global and marginal purities, with only
one remaining degree of freedom related to the symplectic invariant
$\Delta$.}

\begin{figure}[t!]
\centering
\includegraphics[width=8.5cm]{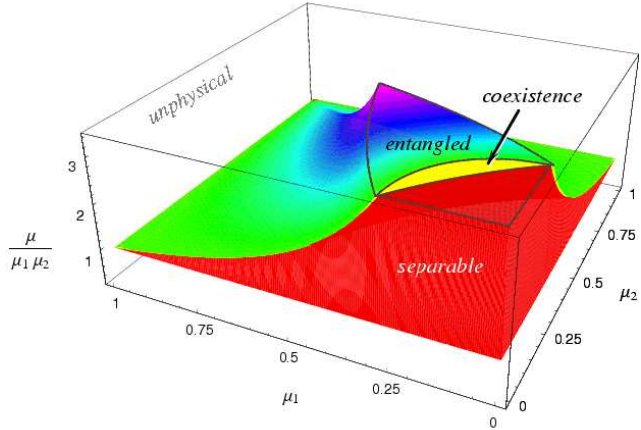}\\
\caption{
Summary of entanglement properties of two-mode (nonsymmetric)
Gaussian states in the space of marginal purities $\mu_{1,2}$
and global purity $\mu$ (we plot the normalized ratio
$\mu/\mu_1\mu_2$ to gain a better graphical
clarity). In this space, separable states (red zone) and entangled
states (green to magenta zone, according to the average entanglement)
are well separated except for a narrow region of coexistence (depicted
in yellow).
The mathematical relations defining the boundaries between the three regions
are collected in Table~\ref{table1}.
The three-dimensional envelope is cut at $z=3.5$.}
\label{fig2D}
\end{figure}

The existence of GMEMS and GLEMS has two consequences.
First, it allows for a classification of the properties of
separability of all two-mode Gaussian states according to their
degree of global and marginal purities, as summarized in
Table~\ref{table1}. Namely, from the separability properties of the
extremally entangled states, necessary and/or sufficient conditions
for entanglement --- which constitute the strongest entropic
criteria  for separability \cite{NielsKempe01} to date in the case
of Gaussian states --- are straightforwardly derived, which allow
one to decide almost unambiguosly if a given two-mode Gaussian state
is entangled or not based on its degree of global and local
purities. There is only a narrow region where, for given purities,
both separable and entangled states can coexist, as pictorially
shown in Fig.~\ref{fig2D}.

\begin{table}[b!]\centering{
\begin{tabular*}{0.8\textwidth}{@{\extracolsep{\fill}} c c }
\hline
\hline
\vspace*{-.2cm}\\
{\rm  Degrees of purity} &
{\rm  Entanglement properties} \vspace*{0.1cm} \\  \hline \vspace*{-.2cm}\\
$\mu<\mu_1 \mu_2$ & {\rm unphysical region} \vspace*{0.3cm}\\
$\mu_1 \mu_2 \; \le \; \mu \; \le \; \frac{\mu_1 \mu_2}{\mu_1 + \mu_2 - \mu_1 \mu_2}$ &
{\rm \emph{separable} states}
\vspace*{0.1cm} \\
$\frac{\mu_1 \mu_2}{\mu_1 + \mu_2 - \mu_1 \mu_2} < \mu \le
\frac{\mu_1 \mu_2}{\sqrt{\mu_1^2 + \mu_2^2 - \mu_1^2 \mu_2^2}}$ &
{\rm \emph{coexistence} region}
\vspace*{0.1cm}\\
$\frac{\mu_1 \mu_2}{\sqrt{\mu_1^2 + \mu_2^2 - \mu_1^2 \mu_2^2}} <
\mu \le \frac{\mu_1 \mu_2}{\mu_1 \mu_2 + \abs{\mu_1-\mu_2}}$ &
{\rm \emph{entangled} states}
\vspace*{0.1cm}\\
$\mu > \frac{\mu_1 \mu_2}{\mu_1 \mu_2 + \abs{\mu_1-\mu_2}}$ &
{\rm unphysical region}
\vspace*{0.2cm}\\ \hline
\end{tabular*}
\caption{Classification of two-mode Gaussian states and of their
properties of separability according to their degrees of global
purity $\mu$ and of marginal purities $\mu_1$ and $\mu_2$
\cite{prl,extremal}.} \label{table1}}
\end{table}

The second consequence is of a quantitative nature. The quantitative
analysis of the maximal and the minimal entanglement allowed for a Gaussian
state with given purities, reveals that they
are very close to each other, their
difference narrowing exponentially with increasing entanglement.
One can then define the {\em average logarithmic negativity} (mean value of
the entanglements of the GMEMS and the GLEMS corresponding to a
given triplet of purities) as a reliable estimator of bipartite
entanglement in two-mode Gaussian states and its
accurate quantification by knowledge of the
global and marginal purities alone \cite{prl}. The purities are
nonlinear functionals of the state, but, assuming some prior knowledge
about it (essentially, its Gaussian character), be measured through
direct methods, in particular by means of single-photon detection schemes
\cite{fiurasek04} (of which preliminary experimental verifications
are available \cite{wenger04}) or by interferometric quantum-network
architectures \cite{Ekert02,Filip02,Oi-Aberg}. Very recently, a
scheme to measure locally all symplectic invariants (and hence the
entanglement) of two-mode Gaussian states has been proposed
\cite{rigolinew}. Notice that no
complete homodyne reconstruction \cite{homotomo} of the covariance
matrix is needed in all these schemes.

The average estimate of the logarithmic negativity becomes
indeed an {\em exact} quantification in the two important instances of
GMEMS (nonsymmetric thermal squeezed states) and GLEMS (mixed states
of partial minimum uncertainty), whose logarithmic negativity is
completely determined as a function of the three purities alone
\cite{prl,extremal}.

\subsection{Ordering Gaussian states with measures of entanglement.}
\label{secorder}

The role of GMEMS and GLEMS in the characterization of entanglement
of two-mode Gaussian states is not confined to the choice of the
negativities. In fact, the Gaussian EMs
for these two families of states can be obtained in closed form, as
the optimization involved in \eq{mfunc} admits a rather simple
analytical solutions in this two instances, as we have shown in
Ref.~\cite{ordering}. One might therefore raise the question whether
such extremally entangled states (with respect to the negativities)
conserve their role once the entanglement is measured by the Gaussian
EMs. The answer, somehow surprisingly, is a negative one: actually,
there is a large region in the space of purities, where GMEMS are
strictly less entangled than the corresponding GMEMS, when their
entanglement is measured according to the Gaussian EMs.

The Gaussian EMs and the negativities are thus {\em not}
equivalent for the quantification of entanglement in mixed,
nonsymmetric two-mode Gaussian states \cite{ordering}.
The interpretation of this
result needs to be discussed with some care. On the one hand, one
could think that the ordering induced by the negativities is a
natural one, due to the fact that such measures of entanglement are
directly inspired by the necessary and sufficient PPT criterion for
separability. Thus, one would expect that the ordering induced by
the negativities should be preserved by any {\em bona fide} measure
of entanglement, especially if one considers that the extremal
states, GLEMS and GMEMS, have a clear physical interpretation.
Therefore, as the Gaussian entanglement of formation is an upper
bound to the true entanglement of formation, one could be tempted to
take this result as an evidence that the latter is globally
minimized on non Gaussian decomposition, at least for GLEMS.
However, this is only a qualitative/speculative argument: proving or
disproving that the Gaussian entanglement of formation is the true
one for any two-mode Gaussian state remains an open question
\cite{QIProb}.

On the other hand, one could take the simplest discrete-variable
instance, constituted by a two-qubit system, as a test-case for
comparison. There, although for pure states the negativity coincides
with the concurrence, an entanglement monotone equivalent to the
entanglement of formation for all states of two qubits
\cite{Wootters97}, the two measures cease to be equivalent for
mixed states, and the orderings they induce on the set of entangled
states can be different \cite{eisert99,VerstraeteJPA}. This analogy
seems to support the stand that, in the arena of mixed states, the
definition of a unique measure of entanglement cannot really
be pursued, due to the different operational meanings and physical
processes (in the cases when it has been possible to identify them)
that are associated to each definition: one could think, for
instance, of the operational difference existing between the
definitions of distillable entanglement and entanglement cost
\cite{Bennett96pra}. In other words, from this point of view, each
inequivalent measure of entanglement introduced for mixed states
should capture physically distinct aspects and forms of quantum correlations
existing in these states. Gaussian EMs might then still be considered
as proper measures of CV entanglement, adapted to a different context than
negativities. This view seems particularly appropriate when
constructing Gaussian EMs to investigate entanglement sharing in
multipartite Gaussian states \cite{contangle,hiroshima}, as
discussed in section \ref{seccontangle}.

\begin{figure}[t!]
\centering
\includegraphics[width=8.5cm]{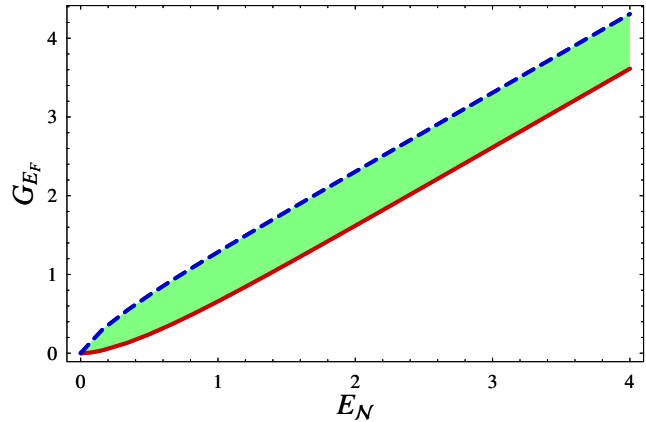}\\
\caption{Comparison between the Gaussian entanglement of formation
$G_{E_F}$ and the logarithmic negativity $E_\N$ for two--mode
Gaussian states. Symmetric states accommodate on the lower boundary
(solid line). States of maximal negativities at fixed (infinite)
 average local mixedness, lie on the dashed line. All
GMEMS and GLEMS lie below the dashed line. The latter is
conjectured, with strong numerical support, to be the upper boundary
for the Gaussian entanglement of formation of all two-mode Gaussian
states, at fixed negativity.  Cfr. Ref.~\cite{ordering} for the
complete discussion and the detailed mathematical proofs.}
\label{disordine}
\end{figure}

The inequivalence between the two families of CV entanglement
measures is somehow tempered.
Namely, we
have rigorously proven that, at fixed negativities, the Gaussian
measures of entanglement are bounded from below (the states which
saturate this bound are simply symmetric two-mode states); moreover,
we provided some strong evidence suggesting that they are as well
bounded from above \cite{ordering}. A direct comparison between the
logarithmic negativity and the Gaussian entanglement of formation
for all two-mode Gaussian states is shown in Fig.~\ref{disordine}.

\section{Multimode bipartite entanglement: localization and scaling.}
\label{ChapUniLoc}

We shall now discuss the properties of entanglement in
multimode Gaussian states endowed with particular symmetry
constraints under permutations of the modes \cite{adescaling,unitarily}.
The usefulness of these states arise in contexts like quantum error correction
\cite{BraunsteinERR}, where some redundancy is required for a
fault-tolerant encoding of information. Bisymmetric and, as a
special case, fully symmetric Gaussian states have been introduced
in section \ref{SecSymm}. The study of the symplectic spectra of
$(M + N)$-mode Gaussian states reveals that, with respect to
the bipartition across which they exhibit the local permutation
invariance (any bipartition is valid for fully symmetric states),
local symplectic diagonalizations of the $M$-mode and the $N$-mode
blocks result in a complete reduction of the multimode state to an
equivalent two-mode state, tensored with $M+N-2$ uncorrelated thermal
single-mode states. The equivalent two-mode state possesses all the
information of the original bisymmetric multimode state for what
concerns entropy and entanglement. As a consequence, the validity of
the PPT criterion as a necessary and sufficient condition for
separability has been extended to bisymmetric Gaussian states in
section \ref{SecPPTG}.

\subsection{Unitarily localizable entanglement of Gaussian states.}

Here, equipped with the tools introduced for the analysis of two-mode
entanglement in Gaussian states, we discuss multimode entanglement
in symmetric and bisymmetric Gaussian states. In particular, we will
investigate how the block entanglement scales with the number of
modes at fixed squeezing. The form of the scaling hints at the presence
of genuine multipartite entanglement that progressively arises
among all the modes as their total number increases.

The central observation of the present section is contained in the
following result \cite{adescaling,unitarily}, straightforwardly
deducible from the discussions in section \ref{SecSymm} and section
\ref{SecPPTG}. {\it The bipartite entanglement of bisymmetric $(M +
N)$-mode Gaussian states under $M \times N$ partitions is
``unitarily localizable'', namely, through local unitary
(reversible) operations, it can be completely concentrated onto a
single pair of modes, each of them belonging respectively to the
$M$-mode and to the $N$-mode blocks.} Hence the multimode block
entanglement ({\ie} the bipartite entanglement between two
blocks of modes) of
bisymmetric (generally mixed) Gaussian states can be determined as
an equivalent
two-mode entanglement. The entanglement will be quantified by the
logarithmic negativity in the general instance because the PPT
criterion holds, but we will also show some explicit nontrivial
cases in which the entanglement of formation, \eq{E:EF},
between $M$-mode and $N$-mode
parties can be exactly computed.

We remark that the notion of ``unitarily localizable entanglement'' is
different from that introduced by Verstraete, Popp, and Cirac for
spin systems \cite{localiz}. There, it was defined as the maximal
entanglement that can be concentrated on two chosen spins through
local {\em measurements} on all the other spins. Here, the local operations
that concentrate all the multimode entanglement on two modes are
{\em unitary} and involve the two chosen modes as well, as parts of
the respective blocks. The localizable entanglement in the sense of
\cite{localiz} can be computed as well for (mixed) symmetric
Gaussian states of an arbitrary number of modes, and is in direct
quantitative connection with the optimal fidelity of multiparty
teleportation networks \cite{telepoppate}.

It is important to observe that the unitarily localizable
entanglement (when computable) is always stronger than the
localizable entanglement in the sense of \cite{localiz}. In fact, if
we consider a generic bisymmetric multimode state of a $M \times N$
bipartition, with each of the two target modes owned respectively by
one of the two parties (blocks), then the ensemble of optimal local
measurements on the remaining (``assisting'') $M+N-2$ modes belongs
to the set of  LOCC with respect to the considered bipartition. By
definition the entanglement cannot increase under LOCC, which
implies that the localized entanglement (in the sense of
\cite{localiz}) is always less or equal than the original $M\times
N$ block entanglement. On the contrary, {\em all} of the same
$M\times N$ original bipartite entanglement can be unitarily
localized onto the two target modes.

This is a key point, as such local unitary transformations are {\em
reversible} by definition.
Therefore,  by only
using passive and active linear optics elements such as beam
splitters, phase shifters and squeezers, one can in principle
implement a reversible machine ({\em entanglement switch}) that,
from mixed, bisymmetric multimode states with strong quantum
correlations between all the modes (and consequently between the
$M$-mode and the $N$-mode partial blocks) but weak couplewise
entanglement, is able to extract a highly pure, highly entangled
two-mode state (with no entanglement loss, as all the $M \times N$
entanglement can be localized). If needed, the same machine would be
able, starting from a two-mode squeezed state and a collection of
uncorrelated thermal or squeezed states, to distribute the two-mode
entanglement between all modes, converting the two-mode into
multimode, multipartite quantum correlations, again with no loss of
entanglement. The bipartite or multipartite entanglement can then be
used on demand, the first for instance in a CV quantum teleportation
protocol \cite{Braunstein98}, the latter to secure quantum key
distribution or to perform multimode entanglement swapping
\cite{Multientswap}. We remark, once more, that such an entanglement
switch is endowed with maximum ($100 \%$) efficiency, as no
entanglement is lost in the conversions.

\subsection{Quantification and scaling of entanglement in fully
symmetric states.} \label{SecScal}

Here we will review the explicit evaluation of
the bipartite block entanglement for some
instances of multimode Gaussian states. We will discuss its scaling
behavior as a function of the number of modes and explore in some
detail the localizability of the multimode entanglement. We focus
our attention on fully symmetric $L$-mode Gaussian states (the
number of modes is denoted by $L$ in general to avoid confusion),
endowed with complete permutation invariance under mode exchange,
and described by a $2L \times 2L$ CM $\sig_{\beta^{L}}$ given by
\eq{fscm}. These states are trivially bisymmetric under any
bipartition of the modes, so that their block entanglement is always
localizable by means of local symplectic operations. Let us recall
that concerning the covariances in normal forms of fully symmetric
states (see section \ref{SecSymm}), {\em pure} $L$-mode states are
characterized by
\begin{eqnarray}\label{fspure}
z_1 &=& \frac{(L - 2) (b^2 -1)  + \sqrt{\left(b^2 -
1\right) \left[L \left(\left(b^2 - 1\right) L + 4\right) -
4\right]}}{2 b (L - 1)}\,, \nonumber \\ & & \\
z_2 &=& \frac{(L - 2) (b^2 -1)
- \sqrt{\left(b^2 -
1\right) \left[L \left(\left(b^2 - 1\right) L + 4\right) -
4\right]} }{2 b (L - 1)}\,. \nonumber
\end{eqnarray}
Pure, fully symmetric Gaussian states are generated as the outputs of
the application of a sequence of $L-1$ beam splitters to $L$
single-mode squeezed inputs. \cite{network,vloock03}.
The CM $\sig^{p}_{\beta^{L}}$ of this class of pure states, for a
given number of modes, depends only on the parameter $b\equiv
1/\mu_\beta \ge 1$, which is an increasing function of the
single-mode squeezing needed to prepare the state. Correlations
between the modes are induced according to the above expression for
the covariances $z_{i}$. Their multipartite entanglement sharing has
been studied in Ref.~\cite{contangle} (see next section), and their
use in teleportation networks have been investigated in
Refs.~\cite{network,telepoppate}.

In general, one can compute the
entanglement between a block of $K$ modes and the remaining $L-K$
modes for pure states (in this case the block entanglement is
simply equivalent to the von Neumann entropy of each of the reduced
blocks) and for mixed, fully symmetric states under
any bipartition of the modes.

\subsubsection{$1 \times N$ entanglement.} \label{Sec1N}

Based on Ref.~\cite{adescaling}, we begin by assigning a single mode
to subsystem $\s_A$, and an arbitrary number $N$ of modes to
subsystem $\s_B$, forming a CV system globally prepared in a fully
symmetric $(1+N)$-mode Gaussian state of  modes.

\begin{figure}[t!]\centering{
\includegraphics[width=8cm]{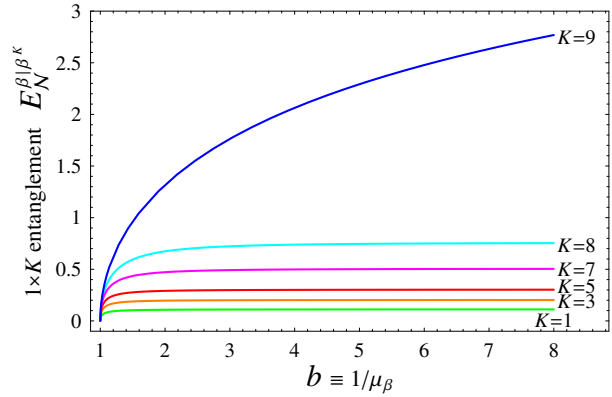}
\caption{Entanglement hierarchy for $(1+N)$-mode fully symmetric
pure Gaussian states ($N=9$).} \label{fighz}}
\end{figure}

We consider pure, fully symmetric states with CM
$\sig^p_{\beta^{1+N}}$, obtained by inserting \eq{fspure} with $L
\equiv (1+N)$. Exploiting our previous analysis, we can compute the
entanglement between a single mode with reduced CM $\sig^\beta$ and
any $K$-mode partition of the remaining modes ($1 \le K \le N$), by
determining the equivalent two-mode CM
$\sig_{eq}^{\beta\vert\beta^K}$.
We remark that, for every $K$,
the $1 \times K$ entanglement is always equivalent to a $1 \times 1$
entanglement, so that the quantum correlations between the different
partitions of $\sig$ can be directly compared to each other: it is
thus possible to establish a multimode entanglement hierarchy
without any problem of ordering.

\begin{figure}[t!]\centering{
\includegraphics[width=7cm]{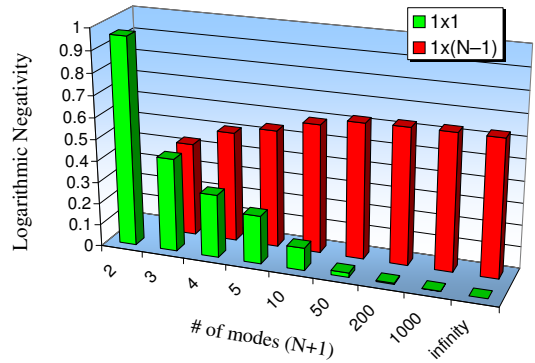}
\caption{Scaling as a function of $N$ of the $1 \times 1$ entanglement (green bars)
and of the $1 \times (N-1)$ entanglement (red bars)
for a $(1+N)$-mode pure fully symmetric Gaussian
state, at fixed squeezing ($b=1.5$).}
\label{figscal}}
\end{figure}

The $1\times K$ entanglement quantified by the logarithmic
negativity $E_\N^{\beta\vert\beta^K}$ is determined by the smallest
symplectic eigenvalue $\tilde{\nu}_-^{(K,N)}$ of the partially
transposed CM $\tilde{\sig}_{eq}^{\beta\vert\beta^K}$. For any
nonzero squeezing ({\em i.e.} $b>1$) one has that
$\tilde{\nu}_-^{(K,N)}<1$, meaning that the state exhibits genuine
multipartite entanglement: each mode is entangled with any other
$K$-mode block, as first remarked in Ref.~\cite{vloock03}. Further,
the genuine multipartite nature of the entanglement can be precisely
quantified by observing that \[E_\N^{\beta\vert\beta^K} \ge
E_\N^{\beta\vert\beta^{K-1}}\,,\] as shown in Fig.~\ref{fighz}.

The $1\times 1$ entanglement between two modes is weaker than the
$1\times 2$ one between a mode and other two modes, which is in turn
weaker than the $1\times K$ one, and so on with increasing $K$ in
this typical cascade structure. From an operational point of view, a
signature of {\em genuine multipartite entanglement} is revealed by
the fact that performing {\em e.g.}~a local measurement on a single
mode will affect {\em all} the other $N$ modes. This means that the
quantum correlations contained in the state with CM
$\sig^{p}_{\beta^{1+N}}$ can be fully recovered only when
considering the $1 \times N$ partition.

In particular, the pure-state $1 \times N$ logarithmic negativity
is, as expected, independent of $N$, being a simple monotonic
function of the entropy of entanglement $E_V$, \eq{E:E} (defined as
the von Neumann entropy of the reduced single-mode state with CM
$\sig_\beta$). It is worth noting that, in the limit of infinite
squeezing ($b \rightarrow \infty$), only the $1\times N$
entanglement diverges while all the other $1\times K$ quantum
correlations remain finite (see Fig.~\ref{fighz}). Namely, \be
\label{logneg1k}
E_\N^{\beta\vert\beta^K}\!\!
\big(\!\sig^{p}_{\beta^{1+N}}\!\big)
\overset{b\rightarrow\infty}{\longrightarrow}
-\frac12 \log\left[\frac{1- 4K}{N(K+1)-K(K-3)}\right]\,, \ee
which cannot exceed
$\log\sqrt5 \simeq 0.8$ for any $N$ and for any $K<N$.

At fixed squeezing (\ie fixed local properties, $b \equiv
1/\mu_\beta$), the {\em scaling} with $N$ of the $1 \times (N-1)$
entanglement compared to the $1\times 1$ entanglement is shown in
Fig.~\ref{figscal} (we recall that the $1\times N$ entanglement is
independent on $N$). Notice how, with increasing number of modes,
the multimode entanglement increases to the detriment of the
two-mode one. The latter is indeed being {\em distributed} among all
the modes: this feature will be properly quantified within the
framework of CV entanglement sharing in the next section
\cite{contangle}.

We remark that such a scaling feature occurs both in
fully symmetric and bisymmetric states (think, for instance, to a
single-mode squeezed state coupled with a $N$-mode symmetric thermal
squeezed state), pure or mixed.  The simplest example of a mixed
state in which our analysis reveals the presence of genuine
multipartite entanglement is obtained from $\sig^{p}_{\beta^{1+N}}$
by tracing out some of the modes. Fig.~\ref{figscal} can then also
be seen as a demonstration of the scaling in such a $N$-mode mixed
state, where the $1 \times (N-1)$ entanglement is the strongest one.
Thus, with increasing $N$, the global mixedness can limit but not
destroy the distribution of entanglement in multiparty form among
all the modes.

\subsubsection{$M \times N$ entanglement.}

Based on the results of Ref.~\cite{unitarily}, one can
consider a generic $2N$-mode fully symmetric mixed state with CM
$\sig_{\beta^{2N}}^{p\backslash Q}$, see \eq{fscm}, obtained from a
pure fully symmetric $(2N+Q)$-mode state by tracing out $Q$ modes.

\begin{figure}[t!]\centering{
\includegraphics[width=8cm]{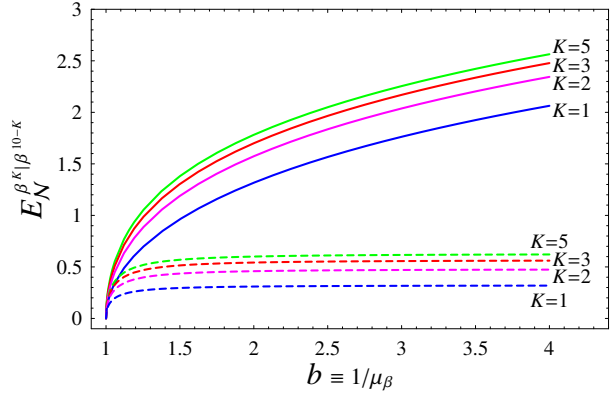}
\caption{Hierarchy of block entanglements of fully symmetric
$2N$-mode Gaussian states of $K \times (2N-K)$ bipartitions
($2N=10$) as a function of the single-mode squeezing $b$. The block
entanglements are depicted both for pure states (solid lines) and
for mixed states obtained from fully symmetric $(2N+4)$-mode pure
Gaussian states by tracing out $4$ modes (dashed lines).}
\label{fiscalb}}
\end{figure}

For any $Q$, for any dimension $K$ of the block ($K \leq N$), and
for any non zero squeezing ({\em i.e.~}for $b>1$) one has that
$\tilde{\nu}_K<1$, meaning that the state exhibits genuine
multipartite entanglement, generalizing the $1 \times N$ case
described before: each $K$-mode party is entangled with the
remaining $(2N-K)$-mode block. The genuine multipartite
nature of the entanglement can be determined by observing
that, again, $E_\N^{\beta^K\vert\beta^{2N-K}}$ is an increasing
function of the integer $K \le N$, as shown in Fig.~\ref{fiscalb}.
The multimode entanglement of mixed states
remains finite also in the limit of infinite squeezing, while the
multimode entanglement of pure states diverges with respect to any
bipartition, as shown in Fig.~\ref{fiscalb}.

In fully symmetric Gaussian states, the block entanglement is
unitarily localizable with respect to any $K \times (2N-K)$
bipartition. Since in this instance {\em all} the entanglement can
be concentrated on a single pair of modes, after the partition has
been decided, no strategy could grant a better yield than the local
symplectic operations bringing the reduced CMs in Williamson form
(because of the monotonicity of the entanglement under general
LOCC). However, the amount of block entanglement, which is the
amount of concentrated two--mode entanglement after unitary
localization has taken place, actually depends on the choice of a
particular $K \times (2N-K)$ bipartition, giving rise to a hierarchy
of localizable entanglements.

Let us imagine that for a given Gaussian multimode state (say, for
simplicity, a fully symmetric state) its
entanglement is meant to serve as a resource for a given protocol.
Let us next suppose that the protocol is optimally implemented if
the entanglement is concentrated between only two modes of the
global systems, as it is the case, {\em e.g.}, in a CV teleportation
protocol between two single-mode parties \cite{Braunstein98}. Which
choice of the bipartition between the modes allows for the best
entanglement concentration by a succession of local unitary
operations? In this framework, it turns out that assigning $K=1$
mode at one party and all the remaining modes to the other, as
discussed in section \ref{Sec1N}, constitutes the {\em worst}
localization strategy \cite{unitarily}. Conversely, for an even
number of modes the best option for localization is an equal $K=N$
splitting of the $2N$ modes between the two parties. The logarithmic
negativity $E_\N^{\beta^N\vert\beta^{N}}$, concentrated into two
modes by local operations, represents the optimal localizable
entanglement (OLE) of the state $\sig_{\beta^{2N}}$, where
``optimal'' refers to the choice of the bipartition. Clearly, the
OLE of a state with $2N+1$ modes is given by
$E_\N^{\beta^{N+1}\vert\beta^{N}}$. These results may be applied to
arbitrary, pure or mixed, fully symmetric Gaussian states.

\begin{figure}[t!]\centering{
\subfigure[] {\includegraphics[width=7cm]{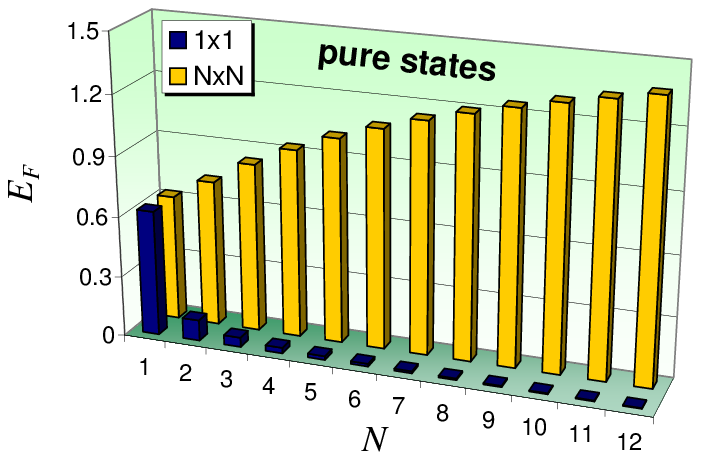}}\\ 
\subfigure[] {\includegraphics[width=7cm]{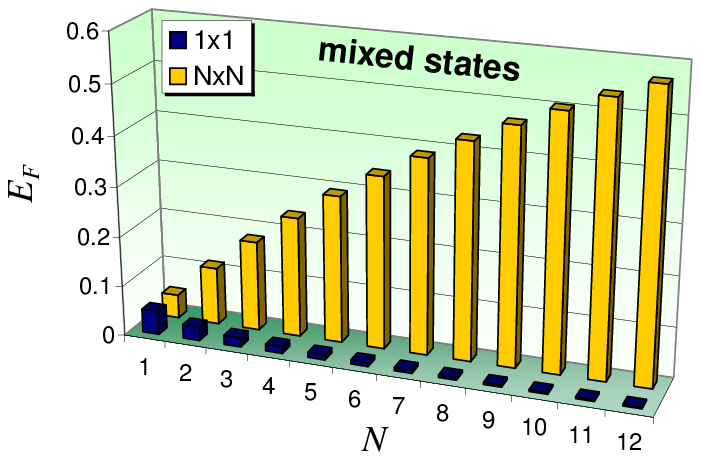}}
\caption{Scaling, with half the number of modes, of the entanglement
of formation in two families of fully symmetric $2N$-mode Gaussian
states. Plot {\rm(a)} depicts pure states, while mixed states
{\rm(b)} are obtained from $(2N+4)$-mode pure states by tracing out
$4$ modes. For each class of states, two sets of data are plotted,
one referring to $N \times N$ entanglement (yellow bars), and the
other to $1 \times 1$ entanglement (blue bars). Notice how the $N
\times N$ entanglement, equal to the optimal localizable
entanglement (OLE) and estimator of genuine multipartite quantum
correlations among all the $2N$ modes, increases at the detriment of
the bipartite $1\times1$ entanglement between any pair of modes. The
single-mode squeezing parameter is fixed at $b=1.5$.}
\label{fisnef}}
\end{figure}

We now turn to the study of the scaling behavior of the OLE
of $2N$-mode Gaussian states when the number of modes is increased,
to understand how the number of local
cooperating parties can improve the maximal entanglement that can be
shared between two parties. For generic (mixed) fully symmetric
$2N$-mode states of $N \times N$ bipartitions, the OLE can be
quantified also by the entanglement of formation $E_F$, \eq{eofgau},
as the equivalent two-mode state is symmetric \cite{unitarily}. It
is then useful to compare, as a function of $N$, the $1 \times 1$
entanglement of formation between a pair of modes (all pairs are
equivalent due to the global symmetry of the state) before the
localization, and the $N \times N$ entanglement of formation, which
is equal to the optimal entanglement concentrated in a specific pair
of modes after performing the local unitary operations. The results
of this study are shown in Fig.~\ref{fisnef}. The two quantities are
plotted at fixed squeezing $b$ as a function of $N$ both for a pure
$2N$-mode state with CM $\sig_{\beta^{2N}}^p$ and a mixed $2n$-mode
state with CM $\sig_{\beta^{2N}}^{p\backslash4}$. As the number of
modes increases, any pair of single modes becomes steadily less
entangled, but the total multimode entanglement of the state grows
and, as a consequence, the OLE increases with $N$. In the limit $N
\rightarrow \infty$, the $N \times N$ entanglement diverges while
the $1\times1$ one vanishes. This exactly holds both for pure {\em
and} mixed states, although the global degree of mixedness produces
the typical behavior that tends to reduce the total entanglement of
the state.

\subsection{Towards multipartite entanglement.}

We have shown that bisymmetric multimode Gaussian states (pure or
mixed) can be reduced, by local symplectic operations, to the tensor
product of a correlated two-mode Gaussian state and of uncorrelated
thermal states (the latter being obviously irrelevant as far as the
correlation properties of the multimode Gaussian state are
concerned). As a consequence, {\em all} the entanglement of
bisymmetric multimode Gaussian states of arbitrary $M \times N$
bipartitions is {\em unitarily localizable} in a single (arbitrary)
pair of modes shared by the two parties. Such a useful reduction to
two-mode Gaussian states is somehow similar to the one holding for
states with fully degenerate symplectic spectra
\cite{botero03,giedkeqic03}, encompassing the relevant instance of
pure states, for which all the symplectic eigenvalues are equal to
$1$. The present result allows to extend the PPT criterion as a
necessary and sufficient condition for separability for all
bisymmetric multimode Gaussian states of arbitrary $M \times N$
bipartitions (as shown in section \ref{SecPPTG}), and to quantify
their entanglement \cite{adescaling,unitarily}.

Notice that, in the general bisymmetric instance addressed in this
section, the possibility of performing a two-mode reduction is
crucially partition-dependent. However, as we have explicitly shown,
in the case of fully symmetric states all the possible bipartitions
can be analyzed and compared, yielding remarkable insight into the
structure of the multimode block entanglement of Gaussian states.
This leads finally to the determination of the maximum, or optimal
localizable entanglement that can be concentrated on a single pair
of modes.

The multipartite entanglement in the
considered class of multimode Gaussian states can be produced, detected
\cite{network,vloock03}, and, by virtue of the present
analysis, reversibly localized by all-optical means. Moreover, the
multipartite entanglement allows for a reliable ({\em i.e.}~with
fidelity ${\cal F}>{\cal F}_{cl}$, where ${\cal F}_{cl}=1/2$ is the
classical threshold \cite{bfkjmo,hammerer}) quantum teleportation
between any two parties with the assistance of the remaining others
\cite{network}. The connection between entanglement in the symmetric
Gaussian resource states and optimal teleportation-network fidelity
has been clarified in \cite{telepoppate}, as briefly recalled in
section \ref{secappl}.

The present section is intended as a bridge
between the two main parts of this review, the one dealing
with bipartite entanglement and the one dealing with
multipartite entanglement. We have characterized
entanglement in multimode Gaussian states by reducing it to a
two-mode problem. By comparing the equivalent two-mode entanglements
in the different bipartitions one is able to detect unambiguously
the presence of genuine multipartite entanglement.
We now analyze in more detail the sharing
phenomenon responsible for the distribution of entanglement from a
bipartite, two-mode form, to a genuine multipartite manifestation in
$N$-mode Gaussian states, both with and without symmetry constraints.

\section{Distributed entanglement and monogamy inequality for all
Gaussian states.}
\label{seccontangle}

We will now review some recent work on multipartite
entanglement sharing in CV systems, that has lead to
the definition of a mathematically and physically {\em bona
fide} measure of genuine tripartite entanglement for arbitrary
three-mode Gaussian states \cite{contangle,3mpra}, the proof of the
monogamy inequality on distributed entanglement for all Gaussian
states \cite{hiroshima}, and the demonstration of the {\em
promiscuous} sharing structure of multipartite entanglement in
Gaussian states \cite{contangle}, which can be unlimited in states
of more than three modes \cite{unlim}.

We begin by introducing some new entanglement monotones, the {\em
contangle}, the {\em Gaussian contangle}, and the {\em Gaussian tangle}
\cite{contangle,hiroshima}, apt to quantify distributed Gaussian
entanglement, thus generalizing to the CV setting the {\em tangle}
\cite{CKW} defined for systems of qubits.

Motivated by the analysis of the block entanglement hierarchy and
its scaling structure in fully symmetric Gaussian states (see
previous section) we will review the central result that CV
entanglement, once properly quantified, is monogamous for {\em all}
Gaussian states, in the sense that it obeys a proper generalization
to CV systems of the Coffman-Kundu-Wootters monogamy inequality
\cite{hiroshima}. In the next section we will review
the results concerning the simplest instance of
tripartite CV entangled states, the three-mode Gaussian
states \cite{3mpra}. For these states, thanks to the monogamy
inequality, it is possible to contruct a measure of genuine tripartite
entanglement, the {\em residual contangle}, that has been shown to
be a full entanglement monotone under Gaussian
LOCC \cite{contangle}. Equipped with such a powerful tool to
quantify tripartite entanglement, we will proceed to review the
entanglement sharing structure in three-mode Gaussian states,
and the related, peculiar property of {\em promiscuity} for CV entanglement.
This property essentially consists in the fact that bipartite and
multipartite entanglement in multimode Gaussian states can be
enhanced and simultaneously maximized without violating the monogamy
inequality on entanglement
sharing, and can even grow unboundedly
in Gaussian states of more than three modes \cite{unlim}.

\subsection{The need for a new continuous-variable entanglement monotone.}

Our primary aim, as in Ref.~\cite{contangle}, is to analyze the
distribution of entanglement between different (partitions of) modes
in Gaussian states of CV systems.

In Ref.~\cite{CKW} Coffman, Kundu
and Wootters (CKW) proved for system of three qubits, and
conjectured for $N$ qubits (this conjecture has now been proven by
Osborne and Verstraete \cite{osborne}), that the bipartite
entanglement $E$ (properly quantified) between, say, qubit A and the
remaining two-qubits partition (BC) is never smaller than the sum of
the A$|$B and A$|$C bipartite entanglements in the reduced states:
\begin{equation}
\label{CKWater} E^{A|(BC)} \ge E^{A|B} + E^{A|C} \; .
\end{equation}
This statement quantifies the so-called {\em monogamy} of quantum
entanglement \cite{monogamy}, in opposition to the classical
correlations, which are not constrained and can be freely shared
\cite{pisa}.

One would expect a similar inequality to hold for three-mode
Gaussian states, namely
\begin{equation}\label{CKWine}
E^{i|(jk)}- E^{i|j} - E^{i|k} \ge 0 \; ,
\end{equation}
where $E$ is a proper measure of bipartite CV entanglement and the
indexes $\{i,j,k\}$ label the three modes. However, the
demonstration of such a property is plagued by subtle difficulties.

Let us consider the simplest conceivable instance of a
pure three-mode Gaussian state completely invariant under mode
permutations. These pure Gaussian states are named fully symmetric
(see section \ref{SecSymm}), and their standard form CM (obtained by
inserting \eq{fspure} with $L=3$ into \eq{fscm}) is only
parametrized by the local mixedness $b=(1/\mu_{\beta}) \ge 1$, an
increasing function of the single-mode squeezing $r_{loc}$, with $b
\rightarrow 1^{+}$ when $r_{loc} \rightarrow 0^{+}$. For these
states, it is not difficult to show that the inequality
\pref{CKWine} can be violated for small values of the local
squeezing factor, using either the logarithmic negativity $E_\N$ or
the entanglement of formation $E_F$ (which is computable in this
case via \eq{eofgau}, because the two-mode reduced mixed states of a
pure symmetric three-mode Gaussian states are again symmetric) to
quantify the bipartite entanglement. This fact implies that none of
these two measures is the proper candidate for approaching the task
of quantifying entanglement sharing in CV systems. This situation is
reminiscent of the case of qubit systems, for which the CKW
inequality holds using the tangle $\tau$, defined as the square of
the concurrence \cite{Wootters97}, but can fail if one chooses
equivalent measures of bipartite entanglement such as the
concurrence itself or the entanglement of formation \cite{CKW}.

It is then necessary to define a proper measure of CV
entanglement that specifically quantifies entanglement sharing
according to a monogamy inequality of the form \pref{CKWine}
\cite{contangle}. A first important hint toward this goal comes by
observing that, when dealing with $1\times N$ partitions of fully
symmetric multimode pure Gaussian states together with their $1
\times 1$ reduced partitions, the desired measure should be a
monotonically decreasing function $f$ of the smallest symplectic
eigenvalue $\tilde \nu_-$ of the corresponding partially transposed
CM $\tilde{\sig}$. This requirement stems from the fact that $\tilde
\nu_-$ is the only eigenvalue that can be smaller than $1$, as shown
in section \ref{SecPPTG}, violating the PPT criterion with respect
to the selected bipartition. Moreover, for a pure symmetric
three-mode Gaussian state, it is necessary to require that the
bipartite entanglements $E^{i|(jk)}$ and $E^{i|j}=E^{i|k}$ be
respectively functions $f(\tilde n_-^{i|(jk)})$ and $f(\tilde
\nu_-^{i|j})$ of the associated smallest symplectic eigenvalues
$\tilde \nu_-^{i|(jk)}$ and $\tilde \nu_-^{i|j}$, in such a way that
they become infinitesimal of the same order in the limit of
vanishing local squeezing, together with their first derivatives:
\begin{equation}
\frac{f(\tilde \nu_-^{i|(jk)})}{2f(\tilde \nu_-^{i|j})} \cong
\frac{f'(\tilde \nu_-^{i|(jk)})}{2f'(\tilde \nu_-^{i|j})}
\rightarrow 1 \; \; \; \; \; \mbox{for} \; \; \; b \rightarrow 1^{+}
\; , \label{requir}
\end{equation}
where the prime denotes differentiation with respect to the
single-mode mixedness $b$. The violation of the sharing inequality
\pref{CKWine} exhibited by the logarithmic negativity can be in fact
traced back to the divergence of its first derivative in the limit
of vanishing squeezing. The above condition formalizes the physical
requirement that in a {\em symmetric} state the quantum correlations
should grow smoothly and be distributed uniformly among all the
three modes. One can then see that the unknown function $f$
exhibiting the desired property is simply the squared logarithmic
negativity\footnote{Notice that an infinite number of functions
satisfying \eq{requir} can be obtained by expanding $f(\tilde
\nu_-)$ around $\tilde \nu_- = 1$ at any even order. However, they
are all monotonic convex functions of $f$. If the inequality
\pref{CKWine} holds for $f$, it will hold as well for any
monotonically increasing, convex function of $f$, such as the
logarithmic negativity raised to any even power $k \ge 2$, but not
for $k=1$. We will exploit this ``gauge freedom'' in the following,
to define an equivalent entanglement monotone in terms of squared
negativity \cite{hiroshima}.}
\begin{equation}\label{funcsq}
f(\tilde \nu_-)=[-\log \tilde \nu_-]^2\,.
\end{equation}
We remind again that for (fully symmetric) $(1+N)$-mode pure
Gaussian states, the partially transposed CM with respect to any $1
\times N$ bipartition, or with respect to any reduced $1 \times 1$
bipartition, has only one symplectic eigenvalue that can drop below
$1$; hence the simple form of the logarithmic negativity (and,
equivalently, of its square) in \eq{funcsq}.

\subsection{Squared negativities as continuous-variable tangles.}

Equipped with this finding,
one can give a formal definition of a bipartite entanglement
monotone $E_\tau$ that, as we will soon show, can be regarded as a
CV analogue of the tangle. Note that the context here is completely
general and we are not assuming that we are dealing with Gaussian
states only. For a generic pure state $\ket{\psi}$ of a $(1+N)$-mode
CV system, we define the square of the logarithmic negativity (the
latter defined by \eq{E:EN}):
\begin{equation}
\label{etaupure} E_\tau (\psi) \equiv \log^2 \| \tilde \ro \|_1 \; ,
\quad \ro = \ketbra\psi\psi \; .
\end{equation}
This is a proper measure of bipartite entanglement, being a convex,
increasing function of the logarithmic negativity $E_\N$, which is
equivalent to the entropy of entanglement \eq{E:E} for arbitrary
pure states in any dimension.  Def.~(\ref{etaupure}) is naturally
extended to generic mixed states $\rho$ of $(N+1)$-mode CV systems
through the convex-roof formalism. Namely, we can introduce the
quantity
\begin{equation}\label{etaumix}
E_\tau(\rho) \equiv \inf_{\{p_i,\psi_i\}} \sum_i p_i
E_\tau(\psi_i)\; ,
\end{equation}
where the infimum is taken over all convex decompositions of $\rho$
in terms of pure states $\{\ket{\psi_i}\}$, and if the index $i$ is
continuous, the sum in \eq{etaumix} is replaced by an integral, and
the probabilities $\{p_i\}$ by a probability distribution
$\pi(\psi)$. Let us now recall that, for two qubits, the tangle can
be defined as the convex roof of the squared negativity
\cite{crnega} (the latter being equal to the concurrence
\cite{Wootters97} for pure two-qubit states
\cite{VerstraeteJPA,crnega}. Here, \eq{etaumix} states that the
convex roof of the squared logarithmic negativity properly defines
the continuous-variable tangle, or, in short, the {\em contangle}
$E_\tau(\rho)$, in which the logarithm takes into account for the
infinite dimensionality of the underlying Hilbert space.

On the other hand, by recalling the equivalence of negativity and
concurrence for pure states of qubits, the {\em tangle} itself can
be defined for CV systems as the convex-roof extension of the
squared negativity. Let us recall that the negativity $\N$, \eq{E:N}
of a quantum state $\ro$ is a convex function of the logarithmic
negativity $E_\N$, \eq{E:EN},
\begin{equation}\label{NvsEN}
\N(\ro) = \frac{\exp[E_\N(\ro)-1]}{2}\,.
\end{equation}

\subsubsection{Gaussian contangle and Gaussian
tangle.}\label{SecTau}

From now on, we will restrict our attention to Gaussian states.

\smallskip

\noindent {\rm {\sc Gaussian contangle}.}--- For any pure multimode
Gaussian state $\ket\psi$, with CM $\sig^p$, of $N+1$ modes assigned
in a generic $1 \times N$ bipartition, explicit evaluation gives
immediately that $E_\tau (\psi) \equiv E_\tau (\sig^{p})$ takes the
form
\begin{equation}
\label{piupurezzapertutti} E_\tau (\sig^{p}) = \log^2 \left(1/\mu_1
- \sqrt{1/\mu_1^2-1}\right) \; ,
\end{equation}
where $\mu_1 = 1/\sqrt{\det\sig_1}$ is the local purity of the
reduced state of mode $1$ with CM $\sig_1$.

For any multimode, mixed Gaussian states with CM $\sig$, we will
then denote the contangle by $E_\tau(\sig)$, in analogy with the
notation used for the contangle $E_\tau(\sig^{p})$ of pure Gaussian
states in \eq{piupurezzapertutti}. Any multimode mixed Gaussian
state with CM $\sig$, admits at least one decomposition in terms of
pure Gaussian states $\sig^p$ only. The infimum of the average
contangle, taken over all pure Gaussian state decompositions,
defines then the {\em Gaussian contangle} $G_\tau$:
\begin{equation}
G_\tau(\sig) \equiv \inf_{\{\pi(d\sig^p ), \sig^{p} \}} \int \pi
(d\sig^p) E_\tau (\sig^p) \; . \label{GaCoRo}
\end{equation}
It follows from the convex roof construction that the Gaussian
contangle $G_\tau(\sig)$ is an upper bound to the true contangle
$E_\tau(\sig)$ (as the latter can be in principle minimized over a
non Gaussian decomposition),
\begin{equation}
E_\tau(\sig) \leq G_\tau(\sig) \; , \label{UpperCut}
\end{equation}
It can be shown that $G_\tau(\sig)$ is a bipartite entanglement
monotone under Gaussian LOCC: in fact, the Gaussian contangle
belongs to the general family of Gaussian entanglement measures,
whose properties as studied in Ref.~\cite{ordering} have been
presented in section \ref{SecGEMS}. Therefore, for Gaussian states,
the Gaussian contangle, similarly to the Gaussian entanglement of
formation \cite{GEOF}, takes the simple form
\begin{equation}
G_\tau (\sig) = \inf_{\sig^p \le \sig} E_\tau(\sig^p) \; ,
\label{simple}
\end{equation}
where the infimum runs over all pure Gaussian states with CM $\sig^p
\le \sig$. Let us remark that, if $\sig$ denotes a mixed symmetric
two-mode Gaussian state, then the Gaussian decomposition is the
optimal one \cite{giedke03} (see section \ref{SecEOFGauss}), and the
optimal pure-state CM $\sig^p$ minimizing $G_\tau(\sig)$ is
characterized by having $\tilde \nu_-(\tilde {\sig}^p) = \tilde
\nu_-(\tilde{\sig})$ \cite{GEOF}. The fact that the smallest
symplectic eigenvalue is the same for both partially transposed CMs
entails that \begin{equation} \label{etausym2} E_\tau(\sig) =
G_\tau(\sig) = [\max\{0,-\log \tilde
\nu_-(\sig)\}]^2\,.\end{equation} We thus consistently retrieve for
the Gaussian contangle (or, equivalently, the contangle, as they
coincide in this specific case),  the expression previously found
for the mixed symmetric reductions of fully symmetric three-mode
pure states, \eq{funcsq}.

To our aims, it is useful here to provide a compact, operative
definition of the Gaussian contangle for $1 \times N$ bipartite
Gaussian states, based on the evaluation of Gaussian entanglement
measures in section \ref{SecGEMS}. If $\sig_{i\vert j} $ is the CM
of a (generally mixed) bipartite Gaussian state where subsystem
$\s_i$ comprises one mode only, then the Gaussian contangle
${G_\tau} $ can be computed as
\begin{equation}
\label{tau} {G_\tau} (\sig_{i\vert j} )\equiv {G_\tau} (\sig_{i\vert
j}^{opt} )=g[m_{i\vert j}^2 ],\;\;\;g[x]={\rm arcsinh}^2[\sqrt
{x-1}].
\end{equation}
Here $\sig_{i\vert j}^{opt} $ corresponds to a pure Gaussian state,
and $m_{i\vert j} \equiv m(\sig _{i\vert j}^{opt} )=\sqrt {\det
\sig_i^{opt} } =\sqrt {\det \sig_j^{opt}}$, with $\sig_{i(j)}^{opt}$
being the reduced CM of subsystem $\s_i$ $(\s_j)$ obtained by
tracing over the degrees of freedom of subsystem $\s_j$ ($\s_i$).
The CM $\sig_{i\vert j}^{opt} $ denotes the pure bipartite Gaussian
state which minimizes $m(\sig_{i\vert j}^p )$ among all pure-state
CMs $\sig_{i\vert j}^p $ such that $\sig_{i\vert j}^p \le
\sig_{i\vert j}$. If $\sig_{i\vert j}$ is a pure state, then
$\sig_{i\vert j}^{opt} =\sig_{i\vert j}$, while for a mixed Gaussian
state \eq{tau} is mathematically equivalent to constructing the
Gaussian convex roof. For a separable state $m(\sig_{i\vert
j}^{opt})=1$. Here we have implicitly used the fact that the
smallest symplectic eigenvalue $\tilde\nu_-$ of the partial
transpose of a pure $1 \times N$ Gaussian state
 $\sig_{i\vert j}^p$ is equal to $\tilde\nu_-=\sqrt{\det\sig_i}
-\sqrt{\det\sig_i-1}$, as follows by recalling that the $1 \times N$
entanglement is equivalent to a $1 \times 1$ entanglement by virtue
of the phase-space Schmidt decomposition (see section
\ref{SecSchmidtPS}) and by exploiting \eq{n1} with $\Delta=2$,
$\mu=1$ and $\mu_1=\mu_2 \equiv 1/\sqrt{\det\sig_i}$.

The Gaussian contangle ${G_\tau}$, like all members of the family
of measures of entanglement (see section \ref{SecGEMS}) is
completely equivalent to the Gaussian entanglement of formation
\cite{GEOF}, which quantifies the cost of creating a given mixed
Gaussian state out of an ensemble of pure, entangled Gaussian
states.

\smallskip

\noindent {\rm {\sc Gaussian tangle}.}--- Analogously, for a $1
\times N$ bipartition associated to a pure Gaussian state
$\ro_{A|B}^{p}$ with $\s_A=\s_{1}$ (a subsystem of a single mode)
and $\s_B=\s_{2}\ldots \s_{N}$, we define the following quantity
\begin{equation} \label{eq:G-tangle_pure}
\tau _{G}(\ro _{A|B}^{p})=\mathcal{N}^{2}(\ro _{A|B}^{p}).
\end{equation}
Here, $\mathcal{N}(\ro)$ is the negativity, \eq{E:N}, of the
Gaussian state $\ro$. The functional $\tau _{G}$, like the
negativity $\mathcal{N}$, vanishes on separable states and does not
increase under LOCC, \ie, it is a proper measure of pure-state
bipartite entanglement. It can be naturally extended to mixed
Gaussian states $\rho_{A|B}$ via the convex roof construction
\begin{equation} \label{eq:G-tangle_mixed}
\tau _{G}(\ro _{A|B})= \inf_{\{p_{i},\ro
_{i}^{(p)}\}}\sum_{i}p_{i}\tau _{G}(\ro _{i}^{p}),
\end{equation}
where the infimum is taken over all convex decompositions of
$\ro_{A|B}$ in terms of pure {\em Gaussian} states $\ro _{i}^{p}$:
$\rho _{A|B}=\sum_{i}p_{i}\ro_{i}^{p}$. By virtue of the Gaussian
convex roof construction,  the Gaussian entanglement measure
$\tau_{G}$ \eq{eq:G-tangle_mixed} is an entanglement monotone under
Gaussian LOCC  (see section \ref{SecGEMS}). Henceforth, given an
arbitrary $N$-mode Gaussian state $\ro _{\s_{1}|\s_{2}\ldots
\s_{N}}$, we refer to $\tau _{G}$, \eq{eq:G-tangle_mixed}  as the
{\em Gaussian tangle} \cite{hiroshima}. Obviously, in terms of CMs,
the analogous of the definition \pref{simple} is valid for the
Gaussian tangle as well, yielding it computable like the contangle
in \eq{tau}. Namely, exploiting \eq{negagau}, one finds
\begin{equation}
\label{Gtau} {\tau_G} (\sig_{i\vert j} )\equiv {\tau_G}
(\sig_{i\vert j}^{opt} )=w[m_{i\vert j}^2 ],\;\;\;w[x]=\frac{1}{4}
\left(\sqrt{x - 1} + \sqrt{x} - 1\right)^2 \, ,
\end{equation}
where we refer to the discussion immediately after \eq{tau} for the
definition of the quantities involved in \eq{Gtau}.

\medskip

We will now proceed to review the structure of entanglement sharing
in Gaussian states and the expressions of the monogamy constraints on its
distribution. We remark that, being the (squared) negativity a
monotonic and convex function of the (squared) logarithmic
negativity, see \eq{NvsEN}, the validity of any monogamy constraint
on distributed Gaussian entanglement using as a measure of entanglement
the ``Gaussian tangle'', is {\em implied} by the proof of
the corresponding monogamy inequality obtained using the ``Gaussian
contangle''. For this reason, when possible, we will always employ
as a preferred choice the primitive entanglement monotone,
represented by the (Gaussian) contangle \cite{contangle} (which
could be generally referred to as a `logarithmic' tangle in quantum
systems of arbitrary dimension).

\subsection{Monogamy inequality for all Gaussian states.}

We are now in the position to recall a collection of recent results
concerning the monogamy of distributed Gaussian entanglement in
multimode Gaussian states.

In the broadest setting we want to investigate whether a monogamy
inequality like \ineq{CKWine} holds in the general case of Gaussian
states with an arbitrary number $N$ of modes. Considering a Gaussian
state distributed among $N$ parties (each owning a single mode), the
monogamy constraint on distributed entanglement can be written as
\begin{equation}
\label{ckwine} E^{\s_i \vert (\s_1 \ldots \s_{i-1} \s_{i+1} \ldots
\s_N )} \ge \sum\limits_{j\ne i}^N {E^{\s_i \vert \s_j } }
\end{equation}
where the global system is multipartitioned in subsystems $\s_k$
($k=1,{\ldots},N$), each owned by a respective party, and $E$ is a
proper measure of bipartite entanglement. The corresponding general
monogamy inequality is known to hold for qubit systems
\cite{osborne}.

The left-hand side of inequality \pref{ckwine} quantifies the
bipartite entanglement between a probe subsystem $\s_i $ and the
remaining subsystems taken as a whole. The right-hand side
quantifies the total bipartite entanglement between $\s_i$ and each
one of the other subsystems $\s_{j\ne i}$ in the respective reduced
states. The non negative difference between these two entanglements,
minimized over all choices of the probe subsystem, is referred to as
the \emph{residual multipartite entanglement}. It quantifies the
purely quantum correlations that are not encoded in pairwise form,
so it includes all manifestations of genuine $K$-partite
entanglement, involving $K$ subsystems (modes) at a time, with
$2<K\le N$.  The study of entanglement sharing and monogamy
constraints thus offers a natural framework to interpret and
quantify entanglement in multipartite quantum systems \cite{pisa}.

With these premises, we have proven that the (Gaussian) contangle
(and the Gaussian tangle, as an implication) is monogamous in fully
symmetric Gaussian states of $N$ modes \cite{contangle}. In general,
we have proven the Gaussian tangle to satisfy inequality
\pref{ckwine} in {\em all} $N$-mode Gaussian states
\cite{hiroshima}. A full analytical proof of the monogamy inequality
for the contangle in all Gaussian states beyond the symmetry, is
currently lacking; however, numerical evidence obtained for randomly
generated non symmetric $4$-mode Gaussian states strongly supports
the conjecture that the monogamy inequality be true for all
multimode Gaussian states, using also the (Gaussian) contangle as a
measure of bipartite entanglement \cite{contangle}. Remarkably, for
all (generally nonsymmetric) three-mode Gaussian states the
(Gaussian) contangle has been proven to be monogamous, leading in
particular to a proper measure of tripartite entanglement in terms
of residual contangle: the analysis of distributed entanglement in
the special instance of three-mode Gaussian states, with all the
resulting implications, is postponed to the next section.

Let us restate again the main result of this section:
{\it The Gaussian tangle $\tau_G$,  an entanglement monotone under Gaussian LOCC, is monogamous for all, pure
and mixed, $N$-mode Gaussian states distributed among $N$ parties, each owning a single mode} \cite{hiroshima}.

\subsection{Discussion.}

The monogamy constraints on entanglement sharing are essential for
the security of CV quantum cryptographic schemes \cite{Cry1,Cry2},
because they limit the information that might be extracted from the
secret key by a malicious eavesdropper. Monogamy is useful as well
in investigating the range of correlations in Gaussian
matrix-product states of harmonic rings \cite{gvbs,generic}, and in
understanding the entanglement frustration occurring in ground
states of many-body harmonic lattice systems \cite{frusta}, which
may be now extended to arbitrary states beyond symmetry constraints.

At a fundamental level, having established the monogamy property for
all Gaussian states paves the way to a proper quantification of
genuine multipartite entanglement in CV systems in terms of the
residual distributed entanglement. In this respect, the intriguing
question arises  whether a {\em stronger} monogamy constraint exists
on the distribution of entanglement in many-body systems, which
imposes a physical trade-off on the sharing of both bipartite and
genuine multipartite quantum correlations.

It would be important to understand whether the inequality
\pref{ckwine} holds as well for discrete-variable qu$d$its ($2 < d <
\infty$), interpolating between qubits and CV systems \cite{pisa}.
If this were the case, the (convex-roof extended) squared
negativity, which coincides with the tangle for arbitrary states of
qubits and with the Gaussian tangle for Gaussian states of CV
systems, would qualify as a universal {\em bona fide},
dimension-independent quantifier of entanglement sharing in all
multipartite quantum systems. At present, this must be considered
as a completely open problem.

\section{Multipartite Gaussian entanglement: quantification and promiscuous
sharing structure.}

This section is mainly devoted to the characterization of
entanglement in the simplest multipartite CV setting, namely that of
three-mode Gaussian states.

To begin with, let us set the notation and review the known results
about three-mode Gaussian states of CV systems. We will refer to the
three modes under exam as mode $1$, $2$ and $3$. The $2 \times 2$
submatrices that form the CM $\sig \equiv \sig_{123}$ of a
three-mode Gaussian state are defined according to \eq{CM}, whereas
the $4 \times 4$ CMs of the reduced two-mode Gaussian states of
modes $i$ and $j$ will be denoted by $\sig_{ij}$. Likewise, the
local (two-mode) seralian invariants $\Delta_{ij}$, \eq{seralian},
will be specified by the labels $i$ and $j$ of the modes they refer
to, while, to avoid any confusion, the three-mode (global) seralian
symplectic invariant will be denoted by $\Delta\equiv\Delta_{123}$.
Let us recall the uncertainty relation \eq{sepcomp} for two-mode
Gaussian states, \be \Delta_{ij} - \det{\sig_{ij}} \le 1 \; .
\label{uncedue} \ee

\subsection{Separability classes for three-mode Gaussian
states.}\label{secbarbie}

As it is clear from the discussion of section \ref{SecPPTG},    a
complete {\em qualitative} characterization of the entanglement of
three-mode Gaussian state is possible because the PPT criterion is
necessary and sufficient for their separability under {\em any},
partial or global (\ie $1\times 1$ or $1\times 2$), bipartition of
the modes. This has lead to an exhaustive classification of
three-mode Gaussian states in five distinct separability classes
\cite{kraus}. These classes take into account the fact that the
modes $1$, $2$ and $3$ allow for three distinct global bipartitions:
\begin{itemize}
\item{{\it Class 1}: states not separable under all the three possible
bipartitions $i \times (jk)$  of the modes (fully inseparable
states, possessing genuine multipartite entanglement).}
\item{{\it Class 2}: states separable under only one of the three possible
bipartitions (one-mode biseparable states).}
\item{{\it Class 3}: states separable under only two of the three possible
bipartitions (two-mode biseparable states).}
\item{{\it Class 4}: states separable under all the three possible bipartitions,
but impossible to write as a convex sum of tripartite products of
pure one-mode states (three-mode biseparable states).}
\item{{\it Class 5}: states that are separable under all the three possible bipartitions,
and can be written as a convex sum of tripartite products of pure
one-mode states (fully separable states).}
\end{itemize}
Notice that Classes 4 and 5 cannot be distinguished by partial
transposition of any of the three modes (which is positive for both
classes). States in Class 4 stand therefore as nontrivial examples
of tripartite entangled states of CV systems with positive partial
transpose \cite{kraus}. It is well known that entangled states with
positive partial transpose possess {\em bound entanglement}, that
is, entanglement that cannot be distilled by means of LOCC.

\subsection{Residual contangle as genuine tripartite entanglement monotone.}

We have proven in Ref.~\cite{contangle}, that all three-mode
Gaussian states satisfy the CKW monogamy inequality \pref{CKWine},
using the (Gaussian) contangle \eq{tau} to quantify bipartite
entanglement.

The sharing constraint leads naturally to the definition of the {\em
residual contangle} as a quantifier of genuine tripartite
entanglement in three-mode Gaussian states, much in the same way as
in systems of three qubits \cite{CKW}. However, at variance with the
three-qubit case (where the residual tangle of pure states is
invariant under qubit permutations), here the residual contangle is
partition-dependent according to the choice of the probe mode, with
the obvious exception of the fully symmetric states. A {\em bona
fide} quantification of tripartite entanglement is then provided by
the {\em minimum} residual contangle \cite{contangle}
\begin{equation}
\label{etaumin} E_\tau^{i|j|k}\equiv\min_{(i,j,k)} \left[
E_\tau^{i|(jk)}-E_\tau^{i|j}-E_\tau^{i|k}\right] \; ,
\end{equation}
where the symbol $(i,j,k)$ denotes all the permutations of the three
mode indexes. This definition ensures that $E_\tau^{i|j|k}$ is
invariant under all permutations of the modes and is thus a genuine
three-way property of any three-mode Gaussian state. We can adopt an
analogous definition for the minimum residual Gaussian contangle
$G_\tau^{res}$, sometimes referred to as {\em arravogliament}
\cite{contangle,3mpra,3mj}:
\begin{equation}
\label{gtaures} G_\tau^{res} \equiv
G_\tau^{i|j|k}\equiv\min_{(i,j,k)} \left[
G_\tau^{i|(jk)}-G_\tau^{i|j}-G_\tau^{i|k}\right] \; .
\end{equation}

One can verify that
\begin{equation}
\label{refsat} (G_\tau^{i|(jk)} \, - \, G_\tau^{i|k}) \, - \,
(G_\tau^{j|(ik)} \, - \, G_\tau^{j|k}) \, \ge \, 0
\end{equation}
if and only if $a_i \ge a_j$, and therefore the absolute minimum in
\eq{etaumin} is attained by the decomposition realized with respect
to the reference mode $l$ of smallest local mixedness $a_l$, i.e.
for the single-mode reduced state with CM of smallest determinant
(corresponding to the largest local purity  $\mu_{l}$).

A crucial requirement for the residual (Gaussian) contangle,
\eq{gtaures}, to be a proper measure of tripartite entanglement is
that it be nonincreasing under (Gaussian) LOCC.  The monotonicity of
the residual tangle was proven for three-qubit pure states in
Ref.~\cite{wstates}. In the CV setting we have proven that for pure
three-mode Gaussian states $G_\tau^{res}$ is an entanglement
monotone under tripartite Gaussian LOCC, and that it is
nonincreasing even under probabilistic operations, which is a
stronger property than being only monotone on average
\cite{contangle}. Therefore, the residual Gaussian contangle
$G_\tau^{res}$ is a proper and computable measure of genuine
multipartite (specifically, tripartite) entanglement in three-mode
Gaussian states.

\subsection{Standard form and tripartite entanglement of pure three-mode
Gaussian states.}
\label{secpuri}

Here we apply the above defined measure of tripartite entanglement
to the relevant instance of {\em pure} three-mode Gaussian states.
We begin by recalling their structural properties.

\subsubsection{Symplectic properties of three-mode covariance matrices.}

The CM $\sig$ of a pure three-mode Gaussian state is characterized
by \be \det{\sig} = 1 \; , \quad \Delta=3 \, . \label{purinv} \ee
The purity constraint requires the local entropic measures of any
$1\times 2$-mode bipartitions to be equal: \be
\det{\sig_{ij}}=\det{\sig_{k}} \; , \label{pur} \ee with $i$, $j$
and $k$ different from each other. This general, well known property
of the bipartitions of pure states may be easily proven resorting to
the Schmidt decomposition (see section \ref{SecSchmidtPS}).

A first consequence of Eqs.~\pref{purinv} and \pref{pur} is rather
remarkable. Combining such equations one easily obtains
\[
(\Delta_{12}-\det{\sig_{12}}) + (\Delta_{13}-\det{\sig_{13}}) +
(\Delta_{23}-\det{\sig_{23}}) = 3 \; ,\] which, together with
Inequality \pref{uncedue}, implies \be \Delta_{ij} = \det{\sig_{ij}}
+ 1 \; , \quad \forall \, i,j: \; i\neq j \, . \label{glems3m} \ee
The last equation shows that any reduced two-mode state of a pure
three-mode Gaussian state saturates the partial uncertainty relation
\eq{uncedue}. The states endowed with such a partial minimal
uncertainty (namely, with their smallest symplectic eigenvalue equal
to $1$) are states of minimal negativity for given global and local
purities, alias GLEMS (Gaussian least entangled mixed states)
\cite{prl,extremal}, introduced in section \ref{secEntvsMix}.

This is relevant considering that the standard form CM of Gaussian
states is completely determined by their global and local
invariants. Therefore, because of \eq{pur}, the entanglement between
any pair of modes embedded in a three-mode pure Gaussian state is
fully determined by the local invariants $\det{\sig_{l}}$, for
$l=1,2,3$, whatever proper measure we choose to quantify it.
Furthermore, the entanglement of a $\sig_i|\sig_{jk}$ bipartition of
a pure three-mode state is determined by the entropy of one of the
reduced states that is, once again, by the quantity
$\det{\sig_{i}}$. Thus, {\em the three local symplectic invariants
$\det{\sig_{1}}$, $\det{\sig_{2}}$ and $\det{\sig_{3}}$ fully
determine the entanglement of any bipartition of a pure three-mode
Gaussian state}. We will show that they suffice to determine as well
the genuine tripartite entanglement encoded in the state
\cite{3mpra}.

For ease of notation, in the following we will denote by $a_l$ the
local single-mode symplectic eigenvalues associated to mode $l$ with
CM $\sig_l$: \be \label{al} a_l\equiv \sqrt{\det{\sig_l}} \; . \ee
\eq{purgau} shows that the quantities $a_l$ are simply related to
the purities of the reduced single-mode states, the local purities
$\mu_l$, by the relation \be \mu_l = \frac{1}{a_{l}} \; . \ee Since
the set $\{a_l\}$ fully determines the entanglement of any of the
$1\times2$ and $1\times1$ bipartitions of the state, it is important
to determine the range of the allowed values for such quantities.
This will provide a complete quantitative characterization of the
entanglement of three-mode pure Gaussian states.

Such an analysis has been performed in Ref.~\cite{3mpra}. Defining
the parameters
\begin{equation}
\label{aprimi} a'_l \equiv a_l-1 \; ,
\end{equation}
one has that the following permutation-invariant triangular
inequality \be \label{triangleprim} |a'_i-a'_j| \, \le \, a'_k \,
\le \, a'_i+a'_j \; . \ee holds as the only condition,  together
with the positivity of each $a'_l$, which fully characterizes the
local symplectic eigenvalues of the CM of three-mode pure Gaussian
states.  All standard forms of pure three-mode Gaussian states and
in particular, remarkably, all the possible values of the
negativities (section \ref{secnega}) and/or of the Gaussian
entanglement measures (section \ref{SecGEMS}) between {\em any} pair
of subsystems, can be determined by letting $a'_1$, $a'_2$ and
$a'_3$ vary in their range of allowed values. Let us remark that
\eq{triangleprim} qualifies itself as an entropic inequality, as the
quantities $\{a'_j\}$ are closely related to the purities and to the
von Neumann entropies of the single-mode reduced states. In
particular the von Neumann entropies $S_{Vj}$ of the reduced states
are given by $S_{Vj}=f(a'_j+1)=f(a_j)$, where the increasing convex
entropic function $f(x)$ has been defined in \eq{entfunc}. Now,
Inequality \pref{triangleprim} is strikingly analogous to the well
known triangle (Araki-Lieb) and subadditivity inequalities for the
von Neumann entropy (holding for general systems
\cite{ArakiLieb70,Wehrl78}), which in our case read \be
|f(a_i)-f(a_j)| \le f(a_k) \le f(a_i) + f(a_j) \; .
\label{arakilieb} \ee However,  the condition imposed by
\eq{triangleprim} is strictly {\em stronger} than the generally
holding inequalities \pref{arakilieb} for the von Neumann entropy
applied to pure quantum states.

We recall that the form of the CM of any Gaussian state can be
simplified through local (unitary) symplectic operations, that
therefore do not affect the entanglement or mixedness properties of
the state, belonging to $Sp_{2,\R}^{\oplus N}$
\cite{generic,sformato,GEOF}. Such reductions of the CMs are called
``standard forms'', as introduced in section \ref{SecSFCM}. For the
sake of clarity, let us write the explicit standard form CM of a
generic {\em pure} three-mode Gaussian state \cite{3mpra},
\begin{equation}
\label{cm3tutta} \sig^p_{sf}=\left(
\begin{array}{cccccc}
a_1 & 0 & e_{12}^+ & 0 & e_{13}^+ & 0 \\
0 & a_1 & 0 & e_{12}^- & 0 & e_{13}^- \\
e_{12}^+ & 0 & a_2 & 0 & e_{23}^+ & 0 \\
0 & e_{12}^- & 0 & a_2 & 0 & e_{23}^- \\
e_{13}^+ & 0 & e_{23}^+ & 0 & a_3 & 0 \\
0 & e_{13}^- & 0 & e_{23}^- & 0 & a_3
\end{array}
\right)\; ,
\end{equation}
with
\begin{eqnarray}\label{eij}\hspace*{-1cm}
e_{ij}^{\pm} &\equiv& \frac{1}{{4 \sqrt{a_i a_j}}} \Bigg\{
\sqrt{\left[\left(a_i-a_j\right)^2-\left(a_k-1\right)^2\right]
\left[\left(a_i-a_j\right)^2-\left(a_k+1\right)^2\right]}
\\ \hspace*{-1cm}
&\pm&
\sqrt{\left[\left(a_i+a_j\right)^2-\left(a_k-1\right)^2\right]
\left[\left(a_i+a_j\right)^2-\left(a_k+1\right)^2\right]}\Bigg\} \,
.\nonumber
\end{eqnarray}

Let us stress that, although useful in actual
calculations, the use of CMs in standard form does not entail any
loss of generality, because all the results concerning the
characterization of bipartite and tripartite entanglement do not
depend on the choice of the specific form of the CMs, but only on
invariant quantities, such as the global and local symplectic
invariants.

\subsubsection{Residual contangle of fully inseparable three-mode Gaussian states.}

We can now analyze separability and entanglement in more detail.
From our study it turns out that, regarding the classification of
section \ref{secbarbie} \cite{kraus}, pure three-mode Gaussian
states may belong either to Class 5, in which case they reduce to
the global three-mode vacuum, or to Class 2, reducing to the
uncorrelated product of a single-mode vacuum and of a two-mode
squeezed state, or to Class 1 (fully inseparable state). No two-mode
or three-mode biseparable pure three-mode Gaussian states are
allowed.

Let us now describe the complete procedure to determine the genuine
tripartite entanglement in a {\em pure} three-mode Gaussian state
with a completely general (not necessarily in standard form) CM
$\sig^p$ belonging to Class 1, as presented in Ref.~\cite{3mpra}.

\begin{description}

\item[{\rm (i)} \it Determine the local purities] The
state is globally pure ($\det\sig^p = 1$). The only quantities
needed for the computation of the tripartite entanglement are
therefore the three local mixednesses $a_l$, defined by \eq{al}, of
the single-mode reduced states $\sig_l,\,l=1,2,3$ (see \eq{CM}).
Notice that the global CM $\sig^p$ needs not to be in the standard
form of \eq{cm3tutta}, as the single-mode determinants are local
symplectic invariants. From an experimental point of view, the
parameters $a_l$ can be extracted from the CM using the homodyne
tomographic reconstruction of the state \cite{homotomo}; or they can
be directly measured with the aid of single photon detectors
\cite{fiurasek04,wenger04}.

\item[{\rm (ii)} \it Find the minimum] From \eq{refsat}, the minimum in the definition
\pref{gtaures} of the residual Gaussian contangle $G_\tau^{res}$ is
attained in the partition where the bipartite entanglements are
decomposed choosing as probe mode $l$ the one in the single-mode
reduced state of smallest local mixedness $a_l \equiv a_{min}$.

\item[{\rm (iii)} \it Check range and compute] Given the mode
with smallest local mixedness $a_{min}$ (say, for instance, mode
$1$) and the parameters $s$ and $d$ defined by
\begin{eqnarray}
s &=& \frac{a_2+a_3}2 \; , \label{s1} \\
d &=& \frac{a_2-a_3}2 \; , \label{d1}
\end{eqnarray}
if $a_{min}=1$ then mode $1$ is uncorrelated from the others:
$G_\tau^{res}=0$. If, instead, $a_{min}>1$ then
\begin{equation}
\label{gtaurespur} G_\tau^{res} (\sig^p) =
\arcsinh^2\!\Big[\sqrt{a_{min}^2-1}\Big] - Q(a_{min},s,d) \; ,
\end{equation}
with $Q \equiv G_\tau^{1|2} + G_\tau^{1|3}$ defined by
\begin{eqnarray}
\label{Qglems} Q(a,s,d) &=& \arcsinh^2 \Big[\sqrt{m^2(a,s,d)-1}\Big]
\\ &+&\arcsinh^2 \Big[\sqrt{m^2(a,s,-d)-1}\Big] \; , \nonumber
\end{eqnarray}
where $m = m_-$ if $D \le 0$, and $m = m_+$ otherwise (one has
$m_+=m_-$ for $D=0$). Here:
\begin{eqnarray*}
\label{unsacco}
m_- & = & \frac{|k_-|}{(s-d)^2-1} \; , \\
m_+ & = & \frac{\sqrt{2\,\left[2 a^2 (1+2 s^2 + 2 d^2) - (4 s^2 -
1)(4 d^2 - 1) -a^4 -
\sqrt{\delta}\right]}}{4(s-d)}\; , \\
D & = & 2 (s - d) - \sqrt{2\left[k_-^2 + 2 k_++|k_-| (k_-^2 + 8
k_+)^{1/2}\right]/k_+} \; , \\
k_\pm & = & a^2 \pm (s+d)^2 \; , \\
\delta & = & (a - 2 d - 1) (a - 2 d + 1) (a + 2 d - 1) (a + 2 d + 1)
\\ & & (a - 2 s - 1) (a - 2 s + 1) (a + 2 s - 1) (a + 2 s + 1) \; .
\end{eqnarray*}
Note (we omitted the explicit dependence for brevity) that
each quantity in \eq{unsacco} is a function of $(a,s,d)$. Therefore,
to evaluate the second term in \eq{Qglems} each $d$ in \eq{unsacco}
must be replaced by $-d$. Note also that if $d<-(a_{min}^2-1)/4s$
then $G_\tau^{1|2}=0$. Instead, if $d>(a_{min}^2-1)/4s$ then
$G_\tau^{1|3}=0$. Otherwise, all terms in
\eq{gtaures} are nonvanishing.
\end{description}

\smallskip

\begin{figure}[t!]
\centering{
\includegraphics[width=8.5cm]{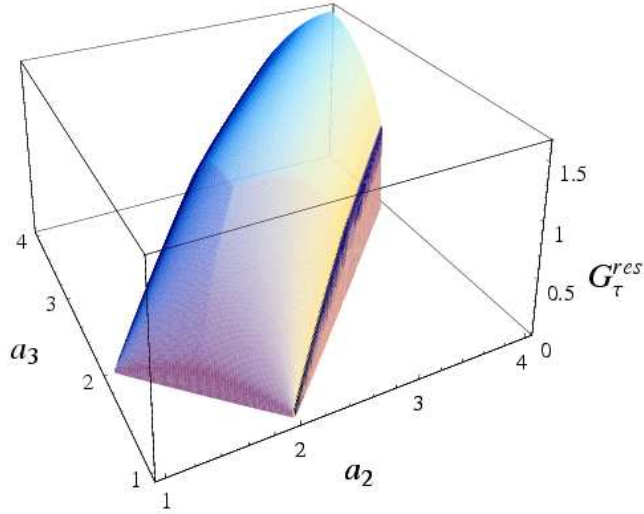}
\caption{Three-dimensional plot of the residual Gaussian contangle
$G_\tau^{res}(\sig^p)$ in pure three-mode Gaussian states $\sig^p$,
determined by the three local mixednesses $a_l$, $l=1,2,3$. One of
the local mixednesses is kept fixed ($a_1=2$). The remaining ones
vary constrained by the triangle inequality \pref{triangleprim}. The
explicit expression of $G_\tau^{res}$ is given by \eq{gtaurespur}.
See text for further details.}
\label{figsupposta}}
\end{figure}

The residual Gaussian contangle \eq{gtaures} in generic pure
three-mode Gaussian states is plotted in Fig.~\ref{figsupposta} as a
function of $a_2$ and $a_3$, at constant $a_1=2$. For fixed $a_1$,
it is interesting to notice that $G_\tau^{res}$ is maximal for
$a_2=a_3$, {\ie}for bisymmetric states. Notice also how the residual
Gaussian contangle of these bisymmetric pure states has a cusp for
$a_1=a_2=a_3$. In fact, from \eq{refsat}, for $a_2=a_3 < a_1$ the
minimum in \eq{gtaures} is attained decomposing with respect to one
of the two modes $2$ or $3$ (the result is the same by symmetry),
while for $a_2=a_3 > a_1$ mode $1$ becomes the probe mode.

\subsubsection{Residual contangle and distillability of mixed states.}

For generic {\em mixed} three-mode  Gaussian states, a quite
cumbersome analytical expression for the $1|2$ and $1|3$ Gaussian
contangles may be written, which explicitly solves the minimization
over the angle $\theta$ in \eq{mfunc}. On the other hand, the
optimization appearing in the computation of the $1|(23)$ bipartite
Gaussian contangle (see \eq{tau}) has to be solved only numerically.
However, exploiting techniques like the unitary localization of
entanglement described in section \ref{ChapUniLoc}, and results like
that of \eq{etausym2}, closed expressions for the residual Gaussian
contangle can be found as well in relevant classes of mixed
three-mode Gaussian states endowed with some symmetry constraints.
Interesting examples of these states and the investigation of their
physical properties are discussed in Refs.~\cite{3mpra,3mj}.

As an additional remark, let us recall that, although the
entanglement of Gaussian states is always distillable with respect
to $1\times N$ bipartitions \cite{werewolf} (see section
\ref{SecPPTG}), they can exhibit bound entanglement in $1 \times 1
\times  1$ tripartitions \cite{kraus}. In this case, the residual
Gaussian contangle cannot detect
tripartite PPT entangled states. For example, the
residual Gaussian contangle in three-mode biseparable Gaussian
states (Class $4$ of Ref.~\cite{kraus}) is always zero, because
those bound entangled states are separable with respect to all
$1\times 2$ bipartitions of the modes. In this sense we can
correctly regard the residual Gaussian contangle as an estimator of
{\em distillable} tripartite entanglement, being strictly nonzero
only  on fully inseparable three-mode Gaussian states (Class 1 in
the classification of section \ref{secbarbie}).

\subsection{Sharing structure of multipartite entanglement:
promiscuous Gaussian states.}\label{secstructex}

We are now in the position to review the sharing structure of CV
entanglement in three-mode Gaussian states by taking the residual
Gaussian contangle as a measure of tripartite entanglement, in
analogy with the study done for three qubits \cite{wstates} using
the residual tangle \cite{CKW}.

The first task we face is that of identifying the three-mode
analogues of the two inequivalent classes of fully inseparable
three-qubit states, the GHZ state \cite{GHZ}
\begin{equation}
\label{qghz} \ket{\psi_{\rm GHZ}} \, = \, \frac{1}{\sqrt2}
\left(\ket{000} + \ket{111}\right) \; ,
\end{equation}
and the $W$ state \cite{wstates}
\begin{equation}
\label{qghz2} \ket{\psi_{W}} \, = \, \frac{1}{\sqrt3}
\left(\ket{001} + \ket{010} + \ket{100}\right) \; .
\end{equation}
These states are both pure and fully symmetric, {\ie}invariant under
the exchange of any two qubits. On the one hand, the GHZ state
possesses maximal tripartite entanglement, quantified by the
residual tangle \cite{CKW,wstates}, with zero couplewise
entanglement in any reduced state of two qubits reductions.
Therefore its entanglement is very fragile against the loss of one
or more subsystems. On the other hand, the $W$ state contains the
maximal two-party entanglement in any reduced state of two qubits
\cite{wstates} and is thus maximally robust against decoherence,
while its tripartite residual tangle vanishes\footnote{The full
inseparability of the $W$ state can be however detected by the
`Schmidt measure' \cite{eisertbriegel}  or by introducing also for
qubits a monogamy inequality in terms of negativity
\cite{cinesino}.}.

\subsubsection{CV finite-squeezing GHZ/{\em W} states.}\label{secghzw}

To define the CV counterparts of the three-qubit states
$\ket{\psi_{\rm GHZ}}$ and $\ket{\psi_{W}}$, one must start from the
fully symmetric (generally mixed) three-mode CM $\sig_s$ of the form
$\sig_{\alp^3}$, \eq{fscm}. Surprisingly enough, in symmetric
three-mode Gaussian states, if one aims at maximizing, at given
single-mode mixedness $a\equiv\sqrt{\det\gr\alpha}$, either the
bipartite entanglement $G_\tau^{i|j}$ in any two-mode reduced state
({\em i.e.}~aiming at the CV $W$-like state), or the genuine
tripartite entanglement $G_\tau^{res}$ ({\em i.e.}~aiming at the CV
GHZ-like state), one finds the same, unique family of states. They
are exactly the {\em pure}, fully symmetric three-mode Gaussian
states (three-mode squeezed states) with CM $\sig^{p}_{s}$  of the
form $\sig_{\alp^3}$, \eq{fscm}, with $\gr\alpha=a \id_2$,
$\gr\varepsilon={\rm diag}\{e^+,\,e^-\}$ and
\begin{equation}
\label{epmfulsym} e^\pm = \frac{a^2-1 \pm \sqrt{\left(a^2 - 1\right)
\left(9 a^2 - 1\right)}}{4a} \; ,
\end{equation}
where we have used \eq{fspure} ensuring the global purity of the
state. In general, we have studied the entanglement scaling in fully
symmetric (pure) Gaussian states by means of the unitary
localization in section \ref{SecScal}. It is in order to mention
that these states were previously known in the literature as CV
``GHZ-type'' states \cite{network,vloock03}, as in the limit of
infinite squeezing ($a \rightarrow \infty$), they approach the
proper (unnormalizable) continuous-variable GHZ state $\int dx
\ket{x,x,x}$, a simultaneous eigenstate of total momentum
$\hat{p}_1+\hat{p}_2+\hat{p}_3$ and of all relative positions
$\hat{q}_i - \hat{q}_j$ ($i,j=1,2,3$), with zero eigenvalues
\cite{cvghz}.

For any finite squeezing (equivalently, any finite local mixedness
$a$), however, the above entanglement sharing study leads ourselves
to re-baptize these states as ``CV GHZ/$W$ states''
\cite{contangle,3mpra,3mj}, and denote their CM by $\sig_{s}^{_{{\rm
GHZ}/W}}$.

The residual Gaussian contangle of GHZ/$W$ states of finite
squeezing takes the simple form \cite{contangle}
\begin{eqnarray}
\label{gresghzw}
G_\tau^{res}(\sig_{s}^{_{{\rm GHZ}/W}})&=& \arcsinh^2\!\left[\sqrt{a^2 -1}\right] \\
&-& \frac{1}{2} \log^2\!\left[\frac{3 a^2 - 1 -\sqrt{9 a^4 - 10 a^2
+ 1}}{2}\right]\, . \nonumber \end{eqnarray} It is straightforward
to see that $G_\tau^{res}(\sig_{s}^{_{{\rm GHZ}/W}})$ is
nonvanishing as soon as $a>1$. Therefore, the GHZ/$W$ states belong
to the class of fully inseparable three-mode states
\cite{kraus,network,vanlokfortshit,vloock03} (Class 1, see section
\ref{secbarbie}). We finally recall that in a GHZ/$W$ state the
residual Gaussian contangle $G_\tau^{res}$ \eq{gtaures} coincides
with the true residual contangle $E_\tau^{1|2|3}$ \eq{etaumin}. This
property clearly holds because the Gaussian pure-state decomposition
is the optimal one in every bipartition, due to the fact that the
global three-mode state is pure and the reduced two-mode states are
symmetric (see section \ref{SecEOFGauss}).

The peculiar nature of entanglement sharing in  CV GHZ/$W$ states is
further confirmed by the following observation. If one requires
maximization of the $1 \times 2$ bipartite Gaussian contangle
$G_\tau^{i|(jk)}$ under the constraint of separability of all the
reduced two-mode states, one finds a class of symmetric mixed states
($T$ states \cite{contangle,3mpra,3mj}) whose residual contangle is
strictly smaller than the one of the GHZ/$W$ states \eq{gresghzw}
for any fixed value of the local mixedness $a$, that is, for any
fixed value of the only parameter (operationally related to the
squeezing of each single mode) that completely determines the CMs of
both families of states up to local unitary operations.

\subsubsection{Promiscuous continuous-variable entanglement sharing.}
\label{secpromis}

The above results lead to the conclusion that in symmetric
three-mode Gaussian states, when there is no bipartite entanglement
in the two-mode reduced states (like in $T$ states) the genuine
tripartite entanglement is not enhanced, but frustrated. More than
that, if there are maximal quantum correlations in a three-party
relation, like in GHZ/$W$ states, then the two-mode reduced states
of any pair of modes are maximally entangled mixed states.

These findings establish the {\em promiscuous} nature of CV entanglement
sharing in symmetric Gaussian states \cite{contangle}. Being associated with
degrees of freedom with continuous spectra, states of CV systems
need not saturate the CKW inequality to achieve maximum couplewise
correlations (as it was instead the case for $W$ states of qubits,
\eq{qghz2}). In fact, without violating the monogamy constraint
\ineq{CKWine}, pure symmetric three-mode Gaussian states are
maximally three-way entangled and, at the same time, possess the
maximum possible entanglement between any pair of modes in the
corresponding two-mode reduced states. The notion of ``promiscuity''
basically means that bipartite and genuine multipartite (in this
case tripartite) entanglement are increasing functions of each
other, while typically in low-dimensional systems like qubits only
the opposite behavior is compatible with monogamy \cite{pisa}. The
promiscuity of entanglement in three-mode GHZ/$W$ states is,
however, {\em partial}. Namely they exhibit, with increasing
squeezing, unlimited tripartite entanglement (diverging in the limit
$a \rightarrow \infty$) and nonzero, accordingly increasing
bipartite entanglement between any two modes, which nevertheless
stays finite even for infinite squeezing. Precisely, from
\eq{gresghzw}, it saturates to the value
\begin{equation}\label{gredmaxghzw}
G_\tau^{i|j}(\sig_s^{_{{\rm GHZ}/W}},\,a \rightarrow \infty) =
\frac{\log^2{3}}{4} \approx 0.3\,.
\end{equation}
In Gaussian states of CV systems with more than three modes,
entanglement can indeed be distributed in an {\em infinitely}
promiscuous way \cite{unlim}, as we will briefly discuss in the
following.

More remarks are in order concerning the tripartite case. The
structure of entanglement in GHZ/$W$ states is such that, while
being maximally three-party entangled, they are also maximally
robust against the loss of one of the modes. This preselects GHZ/$W$
states also as optimal candidates for carrying quantum information
through a lossy channel, being intrinsically less sensitive to
decoherence effects. They have been in fact proven to be maximally
robust against environmental noise among all three-mode Gaussian
states  \cite{3mpra}.

It is natural to question whether {\em all} three-mode Gaussian
states  are expected to exhibit a promiscuous entanglement sharing.
Such a question is addressed in \cite{3mj}, by investigating the
persistency of promiscuity against the lack of each of the two
defining properties of GHZ/$W$ states: full symmetry, and global
purity. One specifically finds that entanglement promiscuity
survives under a quite strong amount of mixedness (up to an impurity
of $1-\mu \approx 0.8$, but is in general lost if the complete
permutation-invariance is relaxed. Pure three-mode Gaussian states
which are only bisymmetric and not fully symmetric (known as {\em
basset hound states}) offer indeed a traditional, not promiscuous
entanglement sharing, with bipartite and tripartite entanglement
being competitors \cite{3mj}. Therefore, in the tripartite Gaussian
setting, {\em `promiscuity'} is a peculiar consequence not of the
global purity, but of the complete symmetry under modes-exchange.
Beside frustrating the maximal entanglement between pairs of modes
\cite{frusta}, symmetry also constrains the multipartite sharing of
quantum correlations. In basset hound states, the separability of
the reduced state of modes $2$ and $3$ prevents the three modes from
having a strong genuine tripartite entanglement among them all,
despite the heavy quantum correlations shared by the two couples of
modes $1|2$ and $1|3$.

\subsubsection{Unlimited promiscuity of entanglement in four-mode Gaussian states.}

The above argument on the origin of promiscuity does not hold
in the case of Gaussian states with four and more modes,
where relaxing the symmetry constraints may allow for an enhancement
of the distributed entanglement promiscuity to an unlimited extent.
In Ref.~\cite{unlim}, we have introduced a class of pure four-mode
Gaussian states which are not fully symmetric, but invariant under
the double exchange of modes $1\leftrightarrow 4$ and
$2\leftrightarrow 3$. They are defined as follows.

One starts with an uncorrelated state of four modes, each one
initially in the vacuum of the respective Fock space, whose
corresponding CM is the identity. One  applies a two-mode squeezing
transformation $S_{2,3}(s)$, \eq{tmsS}, with squeezing $s$ to modes
2 and 3, then two further two-mode squeezing transformations
$S_{1,2}(a)$ and $S_{3,4}(a)$, with squeezing $a$, to the pairs of
modes $\{1,2\}$ and $\{3,4\}$. The two last transformations serve
the purpose of redistributing the original bipartite entanglement,
created between modes 2 and 3 by the first two-mode squeezing
operations, among all the four modes. For any value of the
parameters $s$ and $a$, the output is a pure four-mode Gaussian
state  with CM $\gr\sigma$,
\begin{equation}\label{s4}
\gr\sigma =S_{3,4}(a)S_{1,2}(a)S_{2,3}(s)S_{2,3}\T (s)S_{1,2}\T
(a)S_{3,4}\T (a)\,.
\end{equation}
Explicitly, $\sig$ is of the form \eq{CM} where
\begin{eqnarray*}
\sig_1 = \sig_4 &=& [\cosh ^2(a) + \cosh (2 s) \sinh ^2(a)] \id_2 \,,\\
\sig_2 = \sig_3 &=& [\cosh (2 s) \cosh ^2(a) + \sinh ^2(a)] \id_2 \,,\\
\eps_{1,2} = \eps_{3,4} &=& [\cosh ^2(s) \sinh (2 a)] Z_2 \,,\\
\eps_{1,3} = \eps_{2,4} &=& [\cosh (a) \sinh (a) \sinh (2 s)]
\id_2\,,\\
\eps_{1,4} &=& [\sinh ^2(a) \sinh (2 s)] Z_2\,,\\
\eps_{2,3} &=& [\cosh ^2(a) \sinh (2 s)] Z_2\,,
\end{eqnarray*}
with $Z_2={{1\ \ \ 0}\choose {0 \ -1}}$.

By a proper investigation of the properties of distributed
entanglement in the parametric state \eq{s4}, it can be  shown that
the entanglement between modes 1 and 2 is an unboundedly increasing
function of the squeezing $a$; the same holds for the pair of modes
3 and 4. In addition, the block of modes (1,2) is arbitrarily
entangled with the block of modes (3,4) as a function of $s$. On the
other hand, one can demonstrate that in the same state an unlimited
genuine four-partite entanglement is present, increasing arbitrarily
with increasing $a$, and thus coexisting with (and being mutually
enhanced by) the bipartite entanglement in two pairs of modes
\cite{unlim}.

Such a simple example demonstrates that, when the quantum
correlations arise among degrees of freedom spanning an
infinite-dimensional space of states (characterized by unbounded
mean energy), an accordingly infinite freedom is allowed for
different forms of bipartite and multipartite entanglement.
This phenomenon happens with no violation
of the fundamental monogamy constraint that retains its general
validity in quantum mechanics. In the CV instance
the only effect of monogamy is to bound the divergence rates of the
individual entanglement contributions as the squeezing parameters
are increased. Within the restricted Hilbert space of four or more
qubits, instead, an analogous entanglement structure between the
single qubits is strictly forbidden.

This result opens interesting perspectives for the understanding and
characterization of entanglement in multiparticle systems. Gaussian
states with finite squeezing (finite mean energy) are somehow
analogous to discrete systems with an effective dimension related to
the amount of squeezing \cite{brareview}. As the promiscuous
entanglement sharing arises in Gaussian states by asymptotically
increasing the squeezing to infinity, it is natural to expect that
dimension-dependent families of states will exhibit an entanglement
structure that becomes gradually more promiscuous with increasing
dimension of the Hilbert space towards the CV limit.
A proper investigation on systems of qu$d$its ($2 < d < \infty$)
is therefore a necessary step in order to develop a complete picture
of entanglement sharing in many-body systems \cite{pisa}. This program
has been initiated by establishing a sharp discrepancy between the
two {\it extrema} in the ladder of Hilbert space dimensions: in the
case of CV systems in the limit of infinite squeezing
(infinite mean energy) entanglement has been proven infinitely more
shareable than that of individual qubits \cite{unlim}.
This fact could prelude to implementations of quantum
information protocols with CV systems that {\em cannot} be achieved
and not even devised with qubit resources.

\section{Conclusions and outlook.}
\subsection{Entanglement in non Gaussian states.}

The infinite-dimensional quantum world is obviously not confined to
Gaussian states. In fact, some recent results demonstrate that
the current trends in the theoretical
understanding and experimental control of CV entanglement are
strongly pushing towards the boundaries of the territory of Gaussian
states and Gaussian operations.

The entanglement of
Gaussian states cannot be increased (distilled) by Gaussian
operations \cite{nogo1,nogo2,nogo3}. Similarly, for universal
one-way quantum computation using Gaussian cluster states, a
single-mode non Gaussian measurement is required \cite{menicucci}.
Moreover, a fundamental motivation for investigating entanglement in
non Gaussian states stems from the property of {\em extremality} of
Gaussian states: it has been recently proved that they are the least
entangled states among all states of CV systems with given second
moments \cite{extra}.
Experimentally, it has been recently demonstrated \cite{nongaussexp}
that a two-mode squeezed Gaussian state can be ``degaussified'' by
coherent subtraction of a single photon, resulting in a mixed
non Gaussian state whose nonlocal properties and entanglement degree
are enhanced (enabling a better efficiency for teleporting coherent
states \cite{kitagawa}). Theoretically, even the characterization of
bipartite entanglement (let alone multipartite) in non Gaussian
states stands as a formidable task.

One immediate observation is that any two-mode state with a CM
corresponding to an entangled Gaussian state is itself entangled
\cite{vanlokfortshit}. Therefore, most of the results reviewed in
this paper may serve to detect entanglement in a broader class of
states of infinite-dimensional Hilbert spaces. They are, however,
all sufficient conditions on entanglement based only on the second
moments of the canonical operators. As such, for arbitrary
non Gaussian states, they are in general very inefficient (meaning
that most entangled non Gaussian states fail to be detected by these
criteria). The description of non Gaussian states requires indeed to
consider high order statistical moments. It could then be expected
that inseparability criteria for these states should involve high order
correlations. Recently, some separability criteria based on
hierarchies of conditions involving higher moments of the canonical
operators have been introduced to provide a sharper detection of
inseparability in generic non Gaussian states.

A first step in this direction has been taken by Agarwal and Biswas,
who have applied the method of partial transposition to the
uncertainty relations for the Schwinger realizations of the $SU(2)$
and $SU(1, 1)$ algebras \cite{agarwalbiswas}. This approach can be
successfully used to detect entanglement in the two-mode entangled
non Gaussian state described by the wave function $\psi(x_1, x_2) =
(2/\pi)^{1/2}(\gamma_1 x_1 + \gamma_2 x_2) exp{-(x^2_1 + x^2_2)/2}$,
with $|\gamma_1|^2 + |\gamma_2|^2 = 1$, and in the $SU(2)$
minimum-uncertainty states \cite{agarwalbiswas,nhakim}. The
demonstration of criteria consisting in hierarchies of arbitrary
order moments has been achieved preliminarily by Hillery and Zubairy
\cite{hilzub}, and definitively by Shchukin and Vogel
\cite{shukvog,pianicomment,shukvogreply}. In particular, Shchukin
and Vogel have introduced an elegant and unifying approach, based on
the PPT requirement, that allows to derive, in the form of an
infinite series of inequalities, a necessary and sufficient
condition for the negativity of the partial transposition
$\tilde{\ro}$ of a bipartite quantum state $\ro$. The Shchukin-Vogel
(SV) criterion includes as special cases all the above mentioned
conditions (including the ones on second moments
\cite{Duan00,Simon00} qualifying entanglement in Gaussian states),
thus demonstrating the important role of PPT in building a strong
criterion for the detection of entanglement.

Here we briefly review the main features of the SV criterion first
introduced by Shchukin and Vogel \cite{shukvog}, and later analyzed
in mathematical detail by Miranowicz and Piani
\cite{pianicomment,shukvogreply} (see also \cite{mirapiani}).
Consider two bosonic modes $A_1$ and $A_2$ with the associated
annihilation and creation operators, respectively $ \hat{a}_1,
\hat{a}_1^{\dagger}$ and $\hat{a}_2, \hat{a}_2{^\dagger}$. Shchukin
and Vogel showed that every Hermitian operator $\hat{X}$ that acts
on the Hilbert space of the two modes is nonnegative if and only if
for any operator $\hat{f}$ whose normally-ordered form exists, i.e.,
for any operator $\hat{f}$ that can be written in the form
\begin{equation}
\hat{f} = \sum_{n,m,k,l} c_{nmkl} \hat{a}_1^{\dag n} \hat{a}_1^{m}
\hat{a}_2^{\dag k} \hat{a}_2^{l} \, ,
\end{equation}
it holds that
\begin{equation}
\tr{\hat{f}^\dagger \hat{f} \hat{X}} \geq 0 \, .
\end{equation}

Applying the above result to the partially transposed matrix
$\tilde{\rho}$ of a quantum state density matrix $\rho$, one
finds that $\tilde{\rho}$ is nonpositive if and only if
\begin{equation}
{\rm Tr}[\tilde{\ro} \hat{f}^{\dag} \hat{f}] < 0 \, .
\label{NPTcriterion}
\end{equation}

\noindent \eq{NPTcriterion} is thus a necessary and sufficient
condition for the negativity of the partially transposed density
matrix (NPT). As such, the NPT criterion is in principle able to
detect the bipartite entanglement of all CV two-mode states, pure or
mixed, Gaussian or non Gaussian, with the exception of the PPT bound
entangled states.

Some specific applications to selected classes of non Gaussian states
have been recently discussed in the two-mode setting \cite{Hilleryapplications},
and preliminary extensions of the SV NPT criterion to multimode and
multipartite cases have been introduced and applied to distinguish
between different classes of separability \cite{shukvogmulti}.
To this aim, entanglement witnesses are useful
as well \cite{illuso}. Another, inequivalent condition based on
matrices of moments ({\em realignment criterion}) has also been
recently introduced for the qualification of bipartite entanglement
in non Gaussian states (and in general mixed states of quantum
systems in arbitrary dimension) \cite{mirapiani}.
We should mention a further interesting approach to
non Gaussian entanglement analyzed by McHugh {\it et al.}
\cite{mchugh}, who showed that the entanglement of multiphoton squeezed
states is completely characterized by observing that with respect to
a new set of modes, those non Gaussian states actually assume
a Gaussian character.

The efficiency of some of the above-mentioned inseparability criteria
in detecting the entanglement of non Gaussian, squeezed number states of
two-mode radiation fields has been recently evaluated \cite{fabiotest}.
Detailed studies of the entanglement properties of non Gaussian states
associated to $SU(1,1)$ active and $SU(2)$ passive optical transformations,
and the efficiency of the SV NPT criterion to reveal them are
currently under way, together with the analysis of their dynamical
properties in the presence of noise and decoherence \cite{albanopreparazione}.

\subsection{Applications, open problems, and current perspectives.}
\label{secappl}

The centrality of Gaussian states in CV quantum information is
motivated not only by their peculiar structural properties which
make their description amenable of an analytical analysis, but also
by the ability to produce, manipulate and detect such states with
remarkable accuracy in realistic, experimental settings.

The scope of this review has been almost entirely theoretical. For
reasons of space and time, we cannot discuss in sufficient detail
all the proposals and experimental demonstrations concerning on one
hand the state engineering of two-, three- and in general $N$-mode
Gaussian states, and on the other hand the use of such
states as resources for the realization of quantum information protocols.
Excellent review papers are already available for what concerns both
the optical state engineering of Gaussian and non Gaussian quantum states
of CV systems \cite{fabio}, and
the implementations of quantum information and communication with
continuous variables \cite{vanlokfortshit,brareview}. We just
mention that, concerning the state engineering, an efficient scheme
to produce generic pure $N$-mode Gaussian states in a standard form
not encoding direct correlations between position and momentum
operators (and so encompassing all the instances of multimode
Gaussian states introduced in the previous sections) has been recently
proposed \cite{generic}; it enables to interpret entanglement in this
subclass of Gaussian states entirely in terms of the two-point
correlations between any pair of modes.

From a practical point of view, Gaussian resources have been widely
used to implement paradigmatic protocols of CV quantum
information, such as two-party and multiparty teleportation
\cite{Braunstein98,Furusawa98,network,naturusawa,pirandolareview},
and quantum key distribution \cite{Cry1,Cry2,gran}; they have been
proposed for achieving one-way quantum computation with CV generalizations
of cluster states \cite{menicucci}, and in the multiparty setting they have been
proven useful to solve Byzantine agreement \cite{sanpera}. In this
respect, one theoretical result of direct interest for the characterization
of entanglement in Gaussian states is the qualitative
and quantitative {\em equivalence} \cite{telepoppate} between the
presence of bipartite (multipartite) entanglement in two-mode
($N$-mode) fully symmetric Gaussian states shared as resources for a
two-party teleportation experiment \cite{Braunstein98,Furusawa98}
($N$-party teleportation network \cite{network,naturusawa}), and the
maximal {\em fidelity} of the protocol \cite{bfkjmo,hammerer},
optimized over local single-mode unitary operations performed on the
shared resource. In the special case of three-mode, pure GHZ/$W$
states, this optimal fidelity is a monotonically increasing function
of the residual contangle, providing the latter with a strong
operational interpretation. Based on this equivalence, one can
experimentally verify the promiscuous sharing structure of
tripartite Gaussian entanglement in such states in terms of the
success of two-party and three-party teleportation experiments
\cite{3mj}.

Gaussian states are currently considered key resources to realize
light-matter interfaced quantum communication networks. It has been
experimentally demonstrated how a
coherent state of light can be stored onto an atomic
memory \cite{memorypolzik}, or teleported to a distant atomic
ensemble via a hybrid light-matter two-mode entangled Gaussian
resource \cite{telepolzik}.

Gaussian states play a prominent role in many-body
physics, being ground and thermal states of harmonic lattice
Hamiltonians \cite{chain}. Entanglement entropy scaling in these
systems has been shown to follow an area law \cite{area,areanew}. In
this context, entanglement distribution can be understood by
resorting to the ``matrix-product'' framework, which results in an
insightful characterization of the long-range correlation properties
of some harmonic models \cite{gvbs}. Thermodynamical concepts have
also been applied to the characterization of Gaussian entanglement:
recently, a ``microcanonical'' measure over the second moments of
pure Gaussian states under an energy constraint has been introduced
\cite{typical}, and employed to investigate the statistical
properties of the bipartite entanglement in such states. Under that
measure, the distribution of entanglement concentrates around a
finite value at the thermodynamical limit and, in general, the
typical entanglement of Gaussian states with maximal energy $E$ is
{\em not} close to the maximum allowed by $E$.

A rather recent field of research concerns the investigation of
Gaussian states in a relativistic setting. The entanglement between
the modes of a free scalar field from the perspective of observers
in relative acceleration has been studied \cite{ahnkim,meivette}.
The loss of entanglement due to the Unruh effect has been
reinterpreted in the light of a redistribution of entanglement
between accessible and unaccessible causally disconnected modes
\cite{meivette}. Such studies are of relevance in the context of the
information loss paradox in black holes \cite{neibuchi}, in the
general framework of relativistic quantum information
\cite{peresterno}.

From a fundamental point of view, some important conceptual and
foundational problems include the operative interpretation and the
possible applications of the relative entropy of entanglement for
quantum states in infinite-dimensional Hilbert spaces
\cite{vedralrmp}, proving whether the Gaussian entanglement of
formation coincides with the true entanglement of formation or not,
elucidating the hierarchical structure of the distributed Gaussian
entanglement between $N$ modes, relative to all possible
multipartitions $k \leq N$, and extending the quantitative
investigation and exploitation of CV entanglement in a relativistic
setting to the domain of squeezed and photon-augmented non Gaussian states
\cite{illunongaussrelative}. Interest in non Gaussian entanglement
is not restricted to foundational questions. In certain cases non
Gaussian states may prove useful as resources in quantum information
protocols, for instance for teleportation with non Gaussian
mixed-state resources closely resembling Gaussian ones \cite{lund},
quantum information networks with superpositions of odd photon
number states \cite{molmernetwork}, and the experimental
measurability of some quantifiers of entanglement such as the
logarithmic negativity \cite{kitagawascozzese}. It has been recently
proven that there exist some particular classes of non Gaussian
squeezed states that can be produced and used as entangled resources
within the standard Braunstein-Kimble teleportation protocol
\cite{Braunstein98}, and allow a sharp enhancement of the fidelity of
teleportation, compared to Gaussian resources at fixed degree of
squeezing \cite{risorseilluminate}.

We have seen how the monogamy constraint establishes a natural
ordering and a hierarchy of entanglement of Gaussian states, that
goes beyond the frustration effects that arise, for instance, in
symmetric graphs \cite{frusta}. It is then natural to investigate
how and in what form monogamy constraints arise in non Gaussian
states of many-body CV systems, and, perhaps even more important,
what is the structure of distributed entanglement for hybrid
many-body systems composed of qubits, and/or qu$d$its, interacting
with single-mode or multi-mode fields \cite{tracey}. These
investigations may be of special interest for the understanding of
decoherence and entanglement degradation in different bath
configurations, and the development of possible protection schemes
\cite{illugiampiplastina}. From this point of view, some interesting
hints come from a recent study of the ground-state entanglement in
highly connected systems made of harmonic oscillators and spin-$1/2$
systems \cite{acin}. We may expect that this area of research has more
surprising, only apparently counterintuitive, results in store, besides
the recent finding that, in analogy with finite-dimensional systems,
independent oscillators can become entangled when coupled to a common
environment \cite{benattifloreanini}

A fundamental achievement would of course be the complete
understanding of entanglement sharing in a fully relativistic
setting for general interacting systems with a general tensor
product structure of individual Hilbert spaces of arbitrary
dimension. In a non relativistic framework, the investigation of the
structure of entanglement in hybrid CV-qubit systems is not only of
conceptual importance, but it is relevant for applications as well.
Here, we should at least mention a proposal for a quantum optical
implementation of hybrid quantum computation, where qubit degrees of
freedom for computation are combined with quantum continuous
variables for communication \cite{ibridovanloock}, and a suggested
scheme of hybrid quantum repeaters for long-distance distribution of
quantum entanglement based on dispersive interactions between
coherent light with large average photon number and single,
far-detuned atoms or semiconductor impurities in optical cavities
\cite{ripetitoreibrido}. A hybrid CV memory realized by indirect
interactions between different modes, mediated by qubits, has been
recently shown to have very appealing features compared to
pure-qubit quantum registers \cite{paternostro}.

It seems fitting to conclude this review by commenting on the
intriguing possibility of observing CV entanglement at the interface
between microscopic and macroscopic scales. In this context, it is
encouraging that the existence of optomechanical entanglement
between a macroscopic movable mirror and a cavity field has been
theoretically demonstrated and predicted to be quite robust in
realistic experimental situations, up to temperatures well in reach
of current cryogenic technologies \cite{vitali}. This examples
together with the few others mentioned above should suffice to
convince the reader that continuous-variable entanglement, together
with its applications in fundamental quantum mechanics and quantum
information, is a very active and lively field of research, where
more progress and new exciting developments may be expected in the
near future.

\ack We wold like to acknowledge the continuous exchanges and
discussions over the last three years with our friends, colleagues,
and collaborators at the Quantum Theory Group in Salerno, Lorenzo
Albano Farias, Fabio Dell'Anno, Silvio De Siena, Antonio Di Lisi,
Salvatore M. Giampaolo, and especially Alessio Serafini, for sharing
with us the joy of working together on some important topics covered
by the present review. Some of the results that have been reviewed
in the present work are the fruit of intense collaborations with
Tohya Hiroshima, Marie Ericsson, Ivette Fuentes-Schuller, and the
group of Claude Fabre at the LKB in Paris. We have greatly benefited
from general and specific discussions with Sougato Bose, Sam
Braunstein, Hans Briegel, Nicolas Cerf, Ignacio Cirac, Artur Ekert,
Jens Eisert, Jaromir Fiur\'a{\u s}ek, Jeff Kimble, Ole Krueger, Gerd Leuchs, Norbert
L\"utkenhaus, Klaus Moelmer, Martin Plenio, Eugene Polzik, Anna
Sanpera, Peter van Loock, Vlatko Vedral, Frank Verstraete, Ian Walmsley, Reinhard
Werner, Michael Wolf, and William K. Wootters. Stimulating exchanges
with Luigi Amico, Fabio Benatti, Ferdinando de Pasquale, Rosario
Fazio, Paolo Mataloni, Massimo Palma, Matteo Paris, Saverio
Pascazio, Mauro Paternostro, Marco Piani, Francesco Plastina,
Lorenza Viola, David Vitali, and Paolo Zanardi are gratefully
acknowledged. Last but not least, we thank Fabio Benatti and Roberto
Floreanini for the happy and collaborative atmosphere created at the
Workshop on Quantum Entanglement held in Trieste in June 2006.


\section*{References}


\providecommand{\newblock}{}

\end{document}